\documentclass[dvips,moor]{informs2}              

\long\def\hide#1{}

\usepackage{subfigure}
\usepackage{comment}
\usepackage{epsfig}
\usepackage{enumerate}
\usepackage{algorithmic}
\usepackage{enumitem}
\usepackage{setspace}
\usepackage{type1cm}
\usepackage{rotate}
\usepackage{tikz}
\usetikzlibrary{decorations.pathreplacing}
\usepackage{color}

\newcommand{\eProof}{\hspace*{.1in} \hfill
\begin{picture}(6,6)
\thicklines \put(0,0){\line(0,7){7}} \put(1,0){\line(0,7){7}}
\put(1.5,0){\line(0,7){7}} \put(2,0){\line(0,7){7}}
\put(3,0){\line(0,7){7}} \put(4.5,0){\line(0,7){7}}
\put(4,0){\line(0,7){7}} \put(5,0){\line(0,7){7}}
\end{picture} }

\usepackage{natbib}
 \NatBibNumeric
 \def\bibfont{\small}%
 \def\bibsep{\smallskipamount}%
 \def\bibhang{24pt}%
 \bibpunct[, ]{[}{]}{,}{n}{}{,}%

\newcommand{\CL}{black!30!red}
\newcommand{\CQ}{black!30!blue}
\newcommand{\CC}{black!60!green}
\newcommand{\cl}{red}
\newcommand{\cq}{blue}
\newcommand{\cc}{black!20!green}

\newcommand{\change}[1]{{\textcolor{black}{#1}}{}}
\newcommand{\red}[1]{{\textcolor{black}{#1}}{}}

\usepackage[colorlinks=true,bookmarks=true,urlcolor=blue,
     citecolor=blue,linkcolor=blue,bookmarksopen=false,draft=false]{hyperref}

\def\EMAIL#1{\href{mailto:#1}{#1}}

\TheoremsNumberedThrough     
\ECRepeatTheorems

\EquationsNumberedThrough    

\MANUSCRIPTNO{OPRE-2014-01-053} 

\begin{document}


\RUNAUTHOR{Gopalakrishnan, Doroudi, Ward, and Wierman}

\RUNTITLE{Routing and Staffing when Servers are Strategic}

\TITLE{Routing and Staffing when Servers are Strategic}

\ARTICLEAUTHORS{%
\AUTHOR{Ragavendran Gopalakrishnan}
\AFF{Xerox Research Centre India, Bangalore, Karnataka 560103,\\ \EMAIL{Ragavendran.Gopalakrishnan@xerox.com}}
\AUTHOR{Sherwin Doroudi}
\AFF{Tepper School of Business, Carnegie Mellon University, Pittsburgh, PA 15213,\\ \EMAIL{sdoroudi@andrew.cmu.edu}}
\AUTHOR{Amy R. Ward}
\AFF{Marshall School of Business, University of Southern California, Los Angeles, CA 90089,\\ \EMAIL{amyward@marshall.usc.edu}}
\AUTHOR{Adam Wierman}
\AFF{Department of Computing and Mathematical Sciences, California Institute of Technology, Pasadena, CA 91125,\\ \EMAIL{adamw@caltech.edu}}
} 

\ABSTRACT{%
Traditionally, research focusing on the design of routing and staffing policies for service systems has modeled servers as having fixed (possibly heterogeneous) service rates. However, service systems are generally staffed by people. Furthermore, people respond to workload incentives; that is, how hard a person works can depend both on how much work there is, and how the work is divided between the people responsible for it.  In a service system, the routing and staffing policies control such workload incentives; and so the rate servers work will be impacted by the system's routing and staffing policies.  This observation has consequences when modeling service system performance, and our objective in this paper is to investigate those consequences.

We do this in the context of the $M$/$M$/$N$ queue, which is the canonical model for large service systems. First, we present a model for ``strategic'' servers that choose their service rate in order to maximize a trade-off between an ``effort cost'', which captures the idea that servers exert more effort when working at a faster rate, and a ``value of idleness'', which assumes that servers value having idle time.  Next, we characterize the symmetric Nash equilibrium service rate under any routing policy that routes based on the server idle time (such as the longest idle server first policy).  \red{We find that the system must operate in a quality-driven regime, in which servers have idle time, in order for an equilibrium to exist. The implication is that to have an equilibrium solution the staffing must have a first-order term that strictly exceeds that of the common square-root staffing policy. Then, within the class of policies that admit an equilibrium, we (asymptotically) solve the problem of minimizing the total cost, when there are linear staffing costs and linear waiting costs.} Finally, we end by exploring the question of whether routing policies that are based on the service rate, instead of the server idle time, can improve system performance.
}%


\SUBJECTCLASS{Primary: Queues: applications, limit theorems; secondary: Games/group decisions: noncooperative}

\maketitle

%


\section{Introduction.}
\label{s.introduction}

There is a broad and deep literature studying the scheduling and staffing of service systems that bridges operations research, applied probability, and computer science.  This \red{literature} has had, and is continuing to have, a significant practical impact on the design of call centers (see, for example, the survey papers~\cite{GansKooleMandelbaum2003} and~\cite{AksinArmonyMehrotra07}), health care systems (see, for example, the recent book~\cite{HoppLovejoy2013}), and large-scale computing systems (see, for example, the recent book~\cite{Harchol2013}), among other areas.  Traditionally, this literature on scheduling and staffing has modeled the servers of the system as having fixed (possibly heterogeneous) service rates and then, given these rates, scheduling and staffing policies are proposed and analyzed.  However, in reality, \red{when the servers are \emph{people},} the rate a server chooses to work can be, and often is, impacted by the scheduling and staffing policies used by the system.

For example, if requests are always scheduled to the ``fastest'' server whenever that server is available, then this server may have the incentive to slow her rate to avoid being overloaded with work.  Similarly, if extra staff is always assigned to the division of a service system that is the busiest, then servers may have the incentive to reduce their service rates in order to ensure their division is assigned the extra staff. The previous two examples are simplistic; however, strategic behavior has been observed in practice in service systems.  For example, empirical data from call centers shows many calls that last near 0 seconds~\cite{GansKooleMandelbaum2003}.  This strategic behavior of the servers allowed them to obtain ``rest breaks'' by hanging up on customers -- a rather dramatic means of avoiding being overloaded with work.  For another example, academics are often guilty of strategic behavior when reviewing for journals.  It is rare for reviews to be submitted before an assigned deadline since, if someone is known for reviewing papers very quickly, then they are likely to be assigned more reviews by the editor.

Clearly, the strategic behavior illustrated by the preceding examples can have a significant impact on the performance provided by a service system.  One could implement a staffing or scheduling policy that is provably optimal under classical scheduling models, where servers are \red{nonstrategic}, and end up with far from optimal system performance as a result of undesirable strategic incentives created by the policy. Consequently, it is crucial for service systems to be designed in a manner that provides the proper incentives for such ``strategic servers''.

In practice, there are two approaches used for creating the proper incentives for strategic servers: one can either provide structured bonuses for employees depending on their job performance (performance-based payments) or one can provide incentives in how scheduling and staffing is performed that reward good job performance (incentive-aware scheduling).  While there has been considerable research on how to design performance-based payments in the operations management and economics communities; the incentives created by scheduling and staffing policies are much less understood.  In particular, \emph{the goal of this paper is to initiate the study of incentive-aware scheduling and staffing policies for strategic servers.}

The design of incentive-aware scheduling and staffing policies is important for a wide variety of service systems.  In particular, in many systems performance-based payments such as bonuses are simply not possible, e.g., in service systems staffed by volunteers such as academic reviewing.   Furthermore, many service systems do not use performance-based compensation schemes; for example, the 2005 benchmark survey on call center agent compensation in the U.S.\ shows that a large fraction of call centers pay a fixed hourly wage (and have no performance-based compensation)~\cite{compensationSurvey2005}.

Even when performance-based payments are possible, the incentives created by scheduling and staffing policies impact the performance of the service system, and thus impact the success of performance-based payments.  Further, since incentive-aware scheduling and staffing does not involve monetary payments (beyond a fixed employee salary), it may be less expensive to provide incentives through scheduling and staffing than through monetary bonuses.  Additionally, providing incentives through scheduling and staffing  eliminates many concerns about ``unfairness'' that stem from differential payments to employees.

Of course, the discussion above assumes that the incentives created by scheduling and staffing can be significant enough to impact the behavior.  A priori it is not clear if \red{they are}, since simply changing the scheduling and staffing policies may not provide strong enough incentives to strategic servers to \red{significantly change} service rates, and thus system performance. It is exactly this uncertainty that motivates the current paper, which seeks to understand the impact of the incentives created by scheduling and staffing, and then to design incentive-aware staffing and scheduling policies that provide near-optimal system performance without the use of monetary incentives.

\subsection{Contributions of this paper.}

This paper makes three main contributions. \red{We introduce a new model for the strategic behavior of servers in large service systems and, additionally, we initiate the study of staffing and routing in the context of strategic servers. Each of these contributions is described in the following.}

\emph{Modeling Strategic Servers \red{(Sections~\ref{Section:Model} and~\ref{Section:Model:MMN})}:}   The essential first step for an analysis of strategic servers is a model for server behavior that is simple enough to be analytically tractable and yet rich enough to capture the salient influences on how each server may choose her service rate.  Our model is motivated by work in labor economics that identifies two main factors that impact the utility of agents: effort cost and idleness.   More specifically, it is common in labor economics to model agents as having some ``effort cost'' function that models the decrease in utility which comes from an increase in effort~\cite{Cahuc2004}.  Additionally, it is a frequent empirical observation that agents in service systems engage in strategic behavior to increase the amount of idle time they have~\cite{GansKooleMandelbaum2003}.  The key feature of the form of the utility we propose in Section~\ref{Section:Model} is that it captures the inherent trade-off between idleness and effort.  In particular, a faster service rate would mean quicker completion of jobs and might result in a higher idle time, but it would also result in a higher effort cost.

\red{In Section~\ref{Section:Model:MMN} of this paper, we apply our model in the context of a $M$/$M$/$N$ system, analyzing the first order condition, and provide a necessary and sufficient condition for a solution to the first order condition to be a symmetric equilibrium service rate (Theorem~\ref{thm:conditionalsymeq}). In addition, we discuss the existence of solutions to the first order condition, and provide a sufficient condition for a unique solution (Theorem~\ref{lemma:MMN-SYM-FOC}).  These results are necessary in order to study staffing and routing decisions, as we do in Sections~\ref{Section:Staffing} and~\ref{Section:Scheduling}; however, it is important to note that the model is applicable more generally as well.}

\emph{Staffing Strategic Servers \red{(Section~\ref{Section:Staffing})}:}  The second piece of the paper  studies the impact strategic servers have on staffing policies in multi-server service systems. The decision of a staffing level for a service system has a crucial impact on the performance of the system.  As such, there is a large literature focusing on this question in the classical, \red{nonstrategic}, setting, and the optimal policy is well understood.  In particular, \red{the number of servers that must be staffed to ensure stability in a conventional $M$/$M$/$N$ queue with arrival rate $\lambda$ and fixed service rate $\mu$ should be strictly larger than the offered load, $\lambda/\mu$.}  However, when there are linear staffing and waiting costs, the economically optimal number of servers to staff is more. Specifically, the optimal policy employs the square root of the offered load more servers~\cite{BorstMandelbaumReiman04}.   \red{This results in efficient operation, because the system loading factor $\lambda/(N\mu)$ is close to one; and maintains quality of service, because the customer wait times are small (on the order of $1/\sqrt{\lambda}$).} Thus, this is often referred to as the Quality and Efficiency Driven (QED) regime or as square-root staffing.

Our contribution in this paper is to initiate the study of staffing strategic servers.  In the presence of strategic servers, the offered load depends on the arrival rate, the staffing, and the routing, through the servers' choice of their service rate.   \red{We show that an equilibrium service rate exists only if the number of servers staffed is order $\lambda$ more than the aforementioned square-root staffing (Theorem~\ref{lemma:FOCexistence_staffingInd}).  In particular, the system must operate in a quality-driven regime, in which the servers have idle time, instead of the quality-and-efficiency driven regime that arises under square-root staffing, in which servers do not have idle time.  Then, within the set of policies that admit an equilibrium service rate, we (asymptotically) solve the problem of minimizing the total cost, when there are linear staffing costs and linear waiting costs (Theorem~\ref{theorem:staffing_independent}).}

\emph{Routing to Strategic Servers \red{(Section~\ref{Section:Scheduling})}:}   The final piece of this paper studies the impact of strategic servers on the design of scheduling policies in multi-server service systems. When servers are not strategic, how to schedule (dispatch) jobs to servers in multi-server systems is well understood.  In particular, the most commonly proposed policies for this setting include Fastest Server First (FSF), which dispatches arriving jobs to the idle server with the fastest service rate; Longest Idle Server First (LISF), which dispatches jobs to the server that has been idle for the longest period of time; and Random, which dispatches the job to each idle server with equal probability. When strategic servers are not considered, FSF is the natural choice for reducing the mean response time (though it is not optimal in general~\cite{deVericourtZhou05,LinKumar84}).  However, in the context of strategic servers the story changes.  In particular, we prove that FSF has no symmetric equilibria when strategic servers are considered, even when there are just two servers.  Further, we prove that LISF, a commonly suggested policy for call centers due to its fairness properties, has the same, unique, symmetric equilibrium as random dispatching.  In fact, we prove that there is a large {\em policy-space collapse} -- all routing policies that are idle-time-order-based are equivalent in a very strong sense \red{ (Theorem~\ref{theorem:2-server-idlerandom})}.

With this in mind, one might suggest that Slowest Server First (SSF) would be a good dispatch policy, since it \red{could incentivize servers to work fast;} however, we prove that, like FSF, SSF has no symmetric equilibria \red{ (Theorem~\ref{theorem:FSF-noeq})}.  However, by ``softening'' \red{SSF's bias} toward slow servers, we are able to identify policies that are guaranteed to have a unique symmetric equilibrium and provide mean response times that are smaller than \red{that under LISF and Random (Theorem~\ref{theorem:unique-symeq})}.

A key message provided by the results described above is that scheduling policies must carefully balance two conflicting goals in the presence of strategic servers: \red{making efficient use of the service capacity} (e.g., by sending work to fast servers) while still incentivizing servers to work fast (e.g., by sending work to slow servers).  While these two goals are inherently in conflict, our results show that it is possible to balance them in a way that provides improved performance over Random.

\subsection{Related work.}
\label{Section:RelatedWork}

As we have already described, the question of how to route and staff in many-server systems when servers have fixed, \red{nonstrategic}, service rates is well-studied.  In general, this is a very difficult question, because the routing depends on the staffing and vice versa.  However, when all the servers serve at the same rate, the routing question is moot.  Then,~\cite{BorstMandelbaumReiman04} shows that square-root staffing, first introduced in~\cite{Erlang1948} and later formalized in~\cite{HalfinWhitt1981}, is economically optimal when both staffing and waiting costs are linear.  Furthermore, square root staffing is remarkably robust:  there is theoretical support for why it works so well for systems of moderate size~\cite{JaLeZw11}, and it continues to be economically optimal both when abandonment is added to the $M$/$M$/$N$ model~\cite{GarnettMandelbaumReiman2002} and when there is uncertainty in the arrival rate~\cite{KoArWa13}.  Hence, to study the joint routing and staffing question for more complex systems, that include heterogeneous servers that serve at different rates and heterogeneous customers, many authors have assumed square root staffing and show how to optimize the routing for various objective functions (see, for example,~\cite{Armony05,GurvichWhitt2007,Atar2005,Tezcan2008,TezcanDai2010}). In relation to this body of work, this paper shows that scheduling and routing results for classical many-server systems that assume fixed service rates must be revisited when servers exhibit strategic behavior.  \red{This is because they may not admit a symmetric equilibrium service rate in the case of square-root staffing (see Section~\ref{Section:Staffing}) or be feasible in the case of Fastest Server First routing (see Section~\ref{Section:Scheduling}).}

Importantly, the Fastest Server First routing policy mentioned earlier has already been recognized to be potentially problematic because it may be perceived as ``unfair''.  The issue from an operational standpoint is that there is strong indication in the \red{human resource management literature} that the perception of fairness affects employee performance~\cite{ColquittConlonWessonChristopher01,Cohen-CharashSpector01}. This has motivated  the analysis of ``fair'' routing policies that, for example, equalize the cumulative server idleness~\cite{Atar2011,Reed2012}, and the desire to find an optimal ``fair'' routing policy~\cite{ArmonyWard10,WardArmony13}.  Another approach is to formulate a model in which the servers choose their service rate in order to balance their desire for idle time (which is obtained by working faster) and the exertion required to serve faster.  This leads to a non-cooperative game for a $M$/$M$/$N$ queue in which the servers act as strategic players that selfishly maximize their utility.

Finally, the literature that is, perhaps, most closely related to the current paper is the literature on queueing games, which is surveyed in~\cite{HassinHaviv03}. The bulk of this literature focuses on the impact of customers acting strategically (e.g., deciding whether to join and which queue to join) on queueing performance. Still, there is a body of work within this literature that considers settings where servers can choose their service rate, e.g.,~\cite{Kalai92, Gilbert98, CachonHarker2002, Cachon2007}. However, in all of the aforementioned papers, there are two servers that derive utility from some monetary compensation per job or per unit of service that they provide, and there are no staffing decisions. In contrast, our work considers systems with more than two servers, and considers servers that derive utility from idle time (and have a cost of effort). The idea that servers value idle time is most similar to the setting in~\cite{GengHuhNag2013}, but that paper restricts its analysis to a two server model. Perhaps the closest previous work to the current paper in analysis spirit is~\cite{Allon2010}, which characterizes approximate equilibria in a market with many servers that compete on price and service level.  However, this is similar in theme to~\cite{Kalai92, CachonHarker2002} in the sense that they consider servers as competing firms in a market.  This contrasts with the current paper, where our focus is on competition between servers \emph{within the same firm}. 

\section{A model for strategic servers.}
\label{Section:Model}

The objective of this paper is to initiate an investigation into the effects of strategic servers on classical management decisions in service systems, e.g., staffing and routing.  We start by, in this section, describing formally our model for the behavior of a strategic server.

The term ``strategic server'' could be interpreted in many ways depending on the \red{server's goal.}  Thus, the key feature of the model is the utility function for a strategic server.  Our motivation comes from a service system staffed by people who are paid a fixed wage, independent of performance. In such settings, one may expect two key factors to have a first-order impact on the experience of the servers: the amount of effort they put forth and the amount of idle time they have.

Thus, a first-order model for the utility of a strategic server is to linearly combine the cost of effort with the idle time of the server.  This gives the following form for the utility of server $i$ in a service system with $N$ servers:
\begin{equation}\label{Equation:ServerUtilityGen}
U_i(\boldsymbol{\mu}) = I_i(\boldsymbol\mu) - c(\mu_i), \; i \in \{1,\ldots, N\},
\end{equation}
where $\boldsymbol{\mu}$ is a vector of the rate of work chosen by each server (i.e., the service rate vector), $I_i(\boldsymbol{\mu})$ is the time-average idle time experienced by server $i$ given the service rate vector $\boldsymbol{\mu}$, and $c(\mu_i)$ is the effort cost of server $i$.  We take $c$ to be an increasing, convex function which is the same for all servers. \red{ We assume that the strategic behavior of servers (choosing a utility-maximizing service rate) is independent of the state of the system and that the server has complete information about the steady state properties of the system when choosing a rate, i.e., they know the arrival rate, scheduling policy, staffing policy, etc., and thus can optimize $U_i(\boldsymbol{\mu})$.}

The key feature of the form of the utility in~\eqref{Equation:ServerUtilityGen} is that it captures the inherent \red{trade-off} between idleness and effort. \red{The idleness, and hence the utility, is a steady state quantity.} In particular, a faster service rate would mean quicker completion of jobs and might result in higher idle time \red{in steady state}, but it would also result in a higher effort cost.  This \red{trade-off} then creates a difficult challenge for staffing and routing \red{in} a service system. To increase throughput and decrease response times, one would like to route requests to the fastest servers, but by doing so the utility of servers decreases, making it less desirable to maintain a fast service rate. \red{Our model should be interpreted as providing insight into the \textit{systemic} incentives created by scheduling and staffing policies rather than the \textit{transitive} incentives created by the stochastic behavior of the system.}

Our focus in this paper will be to explore the consequences of strategic servers for staffing and routing in large service systems, specifically, in the $M$/$M$/$N$ setting.  However, the model is generic and can be studied in non-queueing contexts as well.

To quickly illustrate the issues created by strategic servers, a useful example to consider is that of \red{a} $M$/$M$/$1$ queue with a strategic server.

\noindent\hrulefill
\begin{example}[The M/M/1 queue with a strategic server]\label{ex:mm1}
In a classic $M$/$M$/$1$ system, jobs arrive at rate $\lambda$ into a queue with an infinite buffer, where they wait to obtain service from a single server having fixed service rate $\mu$.  When the server is strategic, instead of serving at a fixed rate $\mu$, the server chooses her service rate $\mu>\lambda$ in order to maximize the utility in~\eqref{Equation:ServerUtilityGen}.  To understand what service rate will emerge, recall that in a $M$/$M$/$1$ queue with $\mu>\lambda$ the steady state fraction of time that the server is idle is given by
$I(\mu) = 1-\frac{\lambda}{\mu}$.
Substituting this expression into~(\ref{Equation:ServerUtilityGen}) means that the utility of the server is given by \red{the following concave function:}
\begin{equation*}
U(\mu) = 1-\frac{\lambda}{\mu}-c(\mu).
\end{equation*}

We now have two possible scenarios. First, suppose that $c'(\lambda)<1/\lambda$, so that the cost function does not increase too fast. Then, $U(\mu)$ attains a maximum in $\left(\lambda,\infty\right)$ at a unique point $\mu^\star$, which is the optimal (utility maximizing) operating point for the strategic server. Thus, a stable operating point emerges, and the performance of this operating point can be derived explicitly when a specific form of a cost function is considered.

On the other hand, if $c'(\lambda)\geq 1/\lambda$, then $U(\mu)$ is strictly decreasing in $\left(\lambda,\infty\right)$ and hence does not attain a maximum in this interval. We interpret this case to mean that the server's inherent skill level (as indicated by the cost function) is such that the server must work extremely hard \red{just to} stabilize the system, and therefore should not have been hired in the first place.

For example, consider the class of cost functions $c(\mu)=c_E \mu^p$. If $c(\lambda)<\frac{1}{p}$, then $\mu^\star$ solves $\mu^\star c(\mu^\star)=\frac{\lambda}{p}$, which gives $\mu^\star=\left(\frac{\lambda}{c_E p}\right)^\frac{1}{p+1}>\lambda$.  On the other hand, if $c(\lambda)\geq\frac{1}{p}$, then $U(\mu)$ is strictly decreasing in $\left(\lambda,\infty\right)$ and hence does not attain a maximum in this interval.
\end{example}
\hrulefill

Before moving on to the analysis of the $M$/$M$/$N$ model with strategic servers, it is important to point out that \red{the model we study focuses on a linear trade-off between idleness and effort.  There are certainly many generalizations that are interesting to study in future work.  One particularly interesting generalization would be to consider a concave (and increasing) function of idle time in the utility function, since it is natural that the gain from improving idle time from 10\% to 20\% would be larger than the gain from improving idle time from 80\% to 90\%. A preliminary analysis highlights that the results in this paper would not qualitatively change in this context.\footnote{Specifically, if $g(I_i(\boldsymbol\mu))$ replaces $I_i(\boldsymbol\mu)$ in~\eqref{Equation:ServerUtilityGen}, all the results in Section~\ref{Section:Model:MMN} characterizing equilibria service rates are maintained so long as $g''' < 0$, except for Theorem~\ref{lemma:MMN-SYM-FOC}, whose sufficient condition would have to be adjusted to accommodate $g$. \change{In addition, our results could be made stronger depending on the specific form of $g$. For example, if $g$ is such that $\lim_{\mu_i\rightarrow \underline{\mu}_i+}U_i(\boldsymbol\mu)=-\infty$, then, a preliminary analysis reveals that it would not be necessary to impose the stability constraint $\mu_i > \lambda/N$ exogenously. Moreover, every solution to the symmetric first order condition~(\ref{eq:MMN-SYM-FOC}) would be a symmetric equilibrium (i.e., the sufficient condition of Theorem~\ref{thm:conditionalsymeq} as generalized for this case by Footnote 2 would automatically be satisfied).}}} 

\section{The \texorpdfstring{$M$/$M$/$N$}{M/M/N} queue with strategic servers.}
\label{Section:Model:MMN}

Our focus in this paper is on the staffing and routing decisions in large service systems, and so we adopt a classical model of this setting, the $M$/$M$/$N$, and adjust it by considering strategic servers, as \red{described} in Section~\ref{Section:Model}.  The analysis of staffing and routing policies is addressed in Sections~\ref{Section:Staffing} and~\ref{Section:Scheduling}, but before moving to such questions, we start by formally introducing the $M$/$M$/$N$ model, and performing some preliminary analysis that is useful both in the context of staffing and routing.

\subsection{Model and notation.}

In \red{a} $M$/$M$/$N$ queue, customers arrive to a service system having $N$ servers according to a Poisson process with rate $\lambda$.  Delayed customers (those that arrive to find all servers busy) are served according to the First In First Out (FIFO) discipline.  Each server is fully capable of handling any customer's service requirements.  The time required to serve each customer is independent and exponential, and has a mean of one time unit when the server works at rate one.  However, each server strategically chooses her service rate to maximize her own (steady state) utility, and so it is not a priori clear what the system service rates will be.

In this setting, the utility functions that the servers seek to maximize are given by
\red{\begin{equation}
\label{Equation:ServerUtility}
U_i(\boldsymbol{\mu};\lambda,N,\mathit{R}) = I_i(\boldsymbol\mu;\lambda,N,\mathit{R}) - c(\mu_i), \qquad i \in \{1,\ldots, N\},
\end{equation}
where $\boldsymbol\mu$ is the vector of service rates, $\lambda$ is the arrival rate, $N$ is the number of servers (staffing level), and $R$ is the routing policy. $I_i(\boldsymbol\mu;\lambda,N,\mathit{R})$ is the steady state fraction of time that server $i$ is idle.} $c(\mu)$ is an increasing, convex function with $c'''(\mu)\geq 0$, that represents the server effort cost.

Note that, as compared with~\eqref{Equation:ServerUtilityGen}, \red{we have emphasized the dependence on the arrival rate $\lambda$, staffing level $N$, and routing policy of the system, $\mathit{R}$. In the remainder of this article, we expose or suppress the dependence on these additional parameters as relevant to the discussion. In particular, note that the idle time fraction $I_i$ (and hence, the utility function $U_i$) in~(\ref{Equation:ServerUtility}) depends on how arriving customers are routed to the individual servers.}

There are a variety of routing policies that are feasible for the system manager.  In general, the system manager may use information about the order in which the servers became idle, the rates at which servers have been working, etc. This leads to the possibility of using simple policies such as Random, which chooses an idle server to route to uniformly at random, as well as more complex policies such as Longest/Shortest Idle Server First (LISF/SISF) and Fastest/Slowest Server First (FSF/SSF). We study the impact of this decision in detail in Section~\ref{Section:Scheduling}.

Given the routing policy chosen by the system manager and the form of the server utilities in~\eqref{Equation:ServerUtility}, the situation that emerges is a competition among the servers for the system idle time.  In particular, the routing policy yields a division of idle time among the servers, and both the division and the amount of idle time will depend on the service rates chosen by the servers.

As a result, the servers can be modeled as strategic players in a noncooperative game, and thus the operating point of the system is naturally modeled as an equilibrium of this game.  In particular, a Nash equilibrium of this game is a set of service rates $\boldsymbol{\mu}^\star$, such that,
\begin{equation} \label{Equation:NashDef}
U_i(\mu^\star_i,\boldsymbol{\mu}^\star_{-i};\mathit{R}) = \max_{\mu_i>\frac{\lambda}{N}} U_i(\mu_i,\boldsymbol{\mu}^\star_{-i};\mathit{R}),
\end{equation}
where $\boldsymbol{\mu}^\star_{-i}=(\mu^\star_1,\ldots,\mu^\star_{i-1},\mu^\star_{i+1},\ldots,\mu^\star_N)$ denotes the vector of service rates of all the servers except server $i$. Note that we exogenously impose the \red{(symmetric)} constraint that each server must work at a rate strictly greater than $\frac{\lambda}{N}$ in order to define a product action space that ensures the stability of the system.\footnote{One can imagine that servers, despite being strategic, would endogenously stabilize the system. \red{To test this, one could study a related game where the action sets of the servers are $(0,\infty)$. Then, the definition of the idle time $I_i(\boldsymbol\mu)$ must be extended into the range of $\boldsymbol\mu$ for which the system is overloaded; a natural way to do so is to define it to be zero in this range, which would ensure continuity at $\boldsymbol\mu$ for which the system is critically loaded. However, it is not differentiable there, which necessitates a careful piecewise analysis. A preliminary analysis indicates that in this scenario, no $\mu \in \left.\left(0,\frac{\lambda}{N}\right.\right]$ can ever be a symmetric equilibrium, and then, the necessary and sufficient condition of Theorem~\ref{thm:conditionalsymeq} would become $U(\mu^\star,\mu^\star)\geq \lim_{\mu_1\to 0+}U(\mu_1,\mu^\star)$, which is more demanding than~(\ref{eq:conditionalsymeq}) (e.g., it imposes a finite upper bound on $\mu^\star$), but not so much so that it disrupts the staffing results that rely on this theorem (e.g., Lemma~\ref{lemma:FOCisEquilibrium} still holds).}} \red{Such a constraint is necessary to allow steady state analysis, and does not eliminate any feasible symmetric equilibria.  We treat this bound as exogenously fixed, however in some situations a system manager may wish to impose quality standards on servers, which would correspond to imposing a larger lower bound (likely with correspondingly larger payments for servers). Investigating the impact of such quality standards is an interesting topic for future work.}

Our focus in this paper is on symmetric Nash equilibria.  With a slight abuse of notation, we say that $\mu^\star$ is a symmetric Nash equilibrium if $\boldsymbol{\mu}^\star=(\mu^\star,\ldots,\mu^\star)$ is a Nash equilibrium (solves~(\ref{Equation:NashDef})). \red{Throughout, the term ``equilibrium service rate'' means a symmetric Nash equilibrium service rate.}

We focus on symmetric Nash equilibria for two reasons.  First, because the agents we model intrinsically have the same skill level (as quantified by the effort cost functions), a symmetric equilibrium corresponds to a fair outcome.  As we have already discussed, this sort of fairness is often crucial in service organizations~\cite{ColquittConlonWessonChristopher01,Cohen-CharashSpector01,ArmonyWard10}. A second reason for focusing on symmetric equilibria is that analyzing symmetric equilibria is already technically challenging, and \red{it is not clear how to approach asymmetric equilibria in the contexts that we consider. Note that we do not rule out the existence of asymmetric equilibria; in fact, they likely exist, and it would be interesting to study whether they lead to better or worse system performance than their symmetric counterparts.}

\subsection{The \texorpdfstring{$M$/$M$/$N$}{M/M/N} queue with strategic servers and Random routing.}
\label{Section:Results:MMN}

\red{Before analyzing staffing and routing in detail, we first study the $M$/$M$/$N$ queue with strategic servers and Random routing.}  We focus on Random routing first because it is, perhaps, the most commonly studied policy in the classical literature on \red{nonstrategic} servers.  Further, this importance is magnified by a new ``policy-space collapse'' result included in Section~\ref{ssec:policyspacecollapse}, which shows that all idle-time-order-based routing policies (e.g., LISF and SISF) have equivalent steady state behavior, and thus have the same steady state behavior as Random routing. We stress that this result stands on its own in the classical, \red{nonstrategic} setting of a $M$/$M$/$N$ queue with heterogeneous service rates, but is also crucial \red{to analyze} routing to strategic servers (Section~\ref{Section:Scheduling}).

The key goal in analyzing a queueing system with strategic servers is to understand the equilibria service rates, i.e., show conditions that guarantee their existence and characterize the equilibria when they exist.  Theorems~\ref{thm:conditionalsymeq} and~\ref{lemma:MMN-SYM-FOC} of Section~\ref{Subsection:MMNEquilibrium} summarize these results for the $M$/$M$/$N$ queue with Random routing.  However, in order to obtain such results we must first characterize the idle time in a $M$/$M$/$N$ system in order to be able to understand the ``best responses'' for servers, and thus analyze their equilibrium behavior.  Such an analysis is the focus of Section~\ref{Subsection:MMNIdleTime}.

\subsubsection{The idle time of a tagged server.}\label{Subsection:MMNIdleTime}

In order to characterize the equilibria service rates, a key first step is to understand the idle time of \red{a} $M$/$M$/$N$ queue.  This is, of course, a well-studied model, and so one might expect to be able to use off-the-shelf results.  While this is true when the servers are homogeneous (i.e., all the server rates are the same), for heterogeneous systems, closed form expressions are challenging to obtain in general, and the resulting forms are quite complicated~\cite{Gumbel60}.

To characterize equilibria, we do need to understand the idle time of heterogeneous $M$/$M$/$N$ queues.  However, due to our focus on symmetric equilibria, we only need to understand a particular, mild, form of heterogeneity.  In particular, we need only understand the best response function for a ``deviating server'' when all other servers have the same service rate.  Given this limited form of heterogeneity, the form of the idle time function simplifies, but still remains quite complicated, as the following theorem shows.

\begin{theorem}\label{theorem:MMN-IDLE}
Consider a heterogenous $M$/$M$/$N$ system with Random routing and arrival rate $\lambda>0$, where $N-1$ servers operate at rate $\mu>\frac{\lambda}{N}$, and a tagged server operates at rate $\mu_1>\underline{\mu}_1=\left(\lambda-(N-1)\mu\right)^+$. The steady state probability that the tagged server is idle is given by:
\begin{equation}\label{eq:MMN-IDLE}
\red{I(\mu_1,\mu;\lambda,N)}=\left(1-\frac{\rho}{N}\right)\left(1-\frac{\rho}{N}\left(1-\frac{\mu}{\mu_1}\right)\left(1+\frac{ErlC(N,\rho)}{N-\left(\rho+1-\frac{\mu_1}{\mu}\right)}\right)\right)^{-1},
\end{equation}
where $\rho=\frac{\lambda}{\mu}$, and $ErlC(N,\rho)$ denotes the Erlang C formula, given by:
\begin{equation*}
ErlC(N,\rho)=\frac{\frac{\rho^N}{N!}\frac{N}{N-\rho}}{\sum_{j=0}^{N-1}\frac{\rho^j}{j!}+\frac{\rho^N}{N!}\frac{N}{N-\rho}}.
\end{equation*}
\end{theorem}

In order to understand this idle time function more, we derive expressions for the first two derivatives of \red{$I$} with respect to $\mu_1$ in the following theorem.  These results are crucial to the analysis of equilibrium behavior.

\begin{theorem}\label{theorem:MMN-DIDLE}
The first two partial derivatives of $I$ with respect to $\mu_1$ are given by

\begin{scriptsize}
\begin{equation}\label{eq:MMN-FPD}
\frac{\partial I}{\partial \mu_1} = \frac{I^2}{\mu_1^2}\frac{\lambda}{N-\rho}\left(1+\frac{ErlC(N,\rho)}{N-\left(\rho+1-\frac{\mu_1}{\mu}\right)}+\left(1-\frac{\mu_1}{\mu}\right)\frac{\mu_1}{\mu}\frac{ErlC(N,\rho)}{\left(N-\left(\rho+1-\frac{\mu_1}{\mu}\right)\right)^2}\right)
\end{equation}
\begin{equation}\label{eq:MMN-SPD}
\frac{\partial^2 I}{\partial \mu_1^2} = -\frac{2I^3}{\mu_1^3}\frac{\lambda}{N-\rho}\left(\left(1-\frac{\rho\ ErlC(N,\rho)}{\left(N-\left(\rho+1-\frac{\mu_1}{\mu}\right)\right)^2}\right)\left(1+\frac{ErlC(N,\rho)}{N-\left(\rho+1-\frac{\mu_1}{\mu}\right)}\right)+\left(N-\left(1-\frac{\mu_1}{\mu}\right)^2\right)\frac{\mu_1}{\mu}\frac{ErlC(N,\rho)}{\left(N-\left(\rho+1-\frac{\mu_1}{\mu}\right)\right)^3}\right)
\end{equation}
\end{scriptsize}
\end{theorem}

Importantly, it can be shown that the right hand side of~(\ref{eq:MMN-FPD}) is always positive, and therefore, the idle time is increasing in the service rate $\mu_1$, as expected. However, it is not clear through inspection of~(\ref{eq:MMN-SPD}) whether the second derivative is positive or negative. Our next theorem characterizes the second derivative, showing that the idle time could be convex at $\mu_1=\underline{\mu}_1$ to begin with, but if so, then as $\mu_1$ increases, it steadily becomes less convex, and is eventually concave.  This behavior adds considerable complication to the equilibrium analysis.

\begin{theorem}\label{theorem:MMN-IDLE-SHAPE}
The second derivative of the idle time satisfies the following properties:
\begin{enumerate}[label=(\alph*)]
\item There exists a threshold $\mu_1^\dagger\in[\underline{\mu}_1,\infty)$ such that $\frac{\partial^2 I}{\partial \mu_1^2}>0$ for $\underline{\mu}_1<\mu_1<\mu_1^\dagger$, and $\frac{\partial^2 I}{\partial \mu_1^2}<0$ for $\mu_1^\dagger<\mu_1<\infty$.
\item $\frac{\partial^2 I}{\partial \mu_1^2}>0 \Rightarrow \frac{\partial^3 I}{\partial \mu_1^3}<0$.
\end{enumerate}
\end{theorem}

We remark that it is possible that the threshold $\mu^\dagger$ could be greater than $\frac{\lambda}{N}$, so, restricting \red{the service rate of server $1$} to be greater than $\frac{\lambda}{N}$ does not necessarily simplify the analysis.

\subsubsection{Symmetric equilibrium analysis for a finite system.}\label{Subsection:MMNEquilibrium}

The properties of the idle time function derived in the previous section provide the key tools we need to characterize the symmetric equilibria service rates under Random routing for \red{a} $M$/$M$/$N$ system.

To characterize the symmetric equilibria, we consider the utility of a tagged server, without loss of generality, server $1$, under the mildly heterogeneous setup of Theorem~\ref{theorem:MMN-IDLE}.  We denote \red{it} by
\begin{equation}
\label{Equation:TaggedServerUtility}
\red{U(\mu_1,\mu;\lambda,N) = I(\mu_1,\mu;\lambda,N) - c(\mu_1)}
\end{equation}
For a symmetric equilibrium in $(\frac{\lambda}{N},\infty)$, we explore the first order and second order conditions for $U$ as a function of $\mu_1$ to have a maximum in $(\underline{\mu}_1,\infty)$.

The first order condition for an interior local maximum at $\mu_1$ is given by:
\begin{equation}\label{eq:MMN-FOC}
 \frac{\partial U}{\partial \mu_1}=0  \quad \Longrightarrow\quad  \frac{\partial I}{\partial \mu_1} = c'(\mu_1)
\end{equation}
Since we are interested in a symmetric equilibrium, we analyze the symmetric first order condition, obtained by plugging in $\mu_1=\mu$ in~(\ref{eq:MMN-FOC}):
\begin{equation}\label{eq:MMN-SYM-FOC}
 \left.\frac{\partial U}{\partial \mu_1}\right|_{\mu_1=\mu}=0 \quad
\Longrightarrow\quad  \frac{\lambda}{N^2\mu^2}\left(N-\frac{\lambda}{\mu} + ErlC\left(N,\frac{\lambda}{\mu}\right)\right) = c'(\mu)
\end{equation}

Now, suppose that $\mu^\star>\frac{\lambda}{N}$ satisfies the symmetric first order condition~(\ref{eq:MMN-SYM-FOC}). Then, $\mu_1=\mu^\star$ is a stationary point of $U(\mu_1,\mu^\star)$. It follows then, that \red{$\mu^\star$ will be a symmetric equilibrium for the servers (satisfying~(\ref{Equation:NashDef})) if and only if $U(\mu_1,\mu^\star)$ attains a global maximum at $\mu_1=\mu^\star$ in the interval $(\frac{\lambda}{N},\infty)$. While an obvious necessary condition for this is that $U(\mu^\star,\mu^\star) \geq U(\frac{\lambda}{N},\mu^\star)$, we show, perhaps surprisingly, that it is also sufficient, in the following theorem.}

\begin{theorem}\label{thm:conditionalsymeq}
\red{$\mu^\star>\frac{\lambda}{N}$ is a symmetric equilibrium if and only if it satisfies the symmetric first order condition~(\ref{eq:MMN-SYM-FOC}), and the inequality $U(\mu^\star,\mu^\star) \geq U(\frac{\lambda}{N},\mu^\star)$, i.e.,
\begin{equation}\label{eq:conditionalsymeq}
c(\mu) \leq c\left(\frac{\lambda}{N}\right) + \left(1-\frac{\rho}{N}\right)\left(1+\left(1-\frac{\rho}{N}+\frac{ErlC(N,\rho)}{N-1}\right)^{-1}\right)^{-1}.
\end{equation}}
\end{theorem}

Finally, we need to understand when the symmetric first order condition~(\ref{eq:MMN-SYM-FOC}) admits a feasible solution $\mu^\star>\frac{\lambda}{N}$. Towards that, \red{we present} sufficient conditions \change{for at least one feasible solution, as well as for a \textit{unique} feasible solution.}

\begin{theorem}\label{lemma:MMN-SYM-FOC}
\change{If $c'\left(\frac{\lambda}{N}\right)<\frac{1}{\lambda}$, then the symmetric first order condition~(\ref{eq:MMN-SYM-FOC}) has at least one solution for $\mu$ in $\left(\frac{\lambda}{N},\infty\right)$. In addition, if $2\frac{\lambda}{N}c'\left(\frac{\lambda}{N}\right)+\left(\frac{\lambda}{N}\right)^2 c''\left(\frac{\lambda}{N}\right)\geq 1$, then the symmetric first order condition~(\ref{eq:MMN-SYM-FOC}) has a unique solution for $\mu$ in $\left(\frac{\lambda}{N},\infty\right)$.}
\end{theorem}

\change{In the numerical results that follow, we see instances of zero, one, and two equilibria.\footnote{\change{In general, the symmetric first order condition~(\ref{eq:MMN-SYM-FOC}) can be rewritten as $$\mu^2c'(\mu) + \frac{\lambda}{N^2}\left(\rho - ErlC\left(N,\rho\right)\right) - \frac{\lambda}{N} = 0.$$ Note that, when the term $\rho-ErlC(N,\rho)$ is convex in $\mu$, it follows that the left hand side of the above equation is also convex in $\mu$, which implies that there are at most two symmetric equilibria.}} Interestingly, when more than one equilibrium exists, the equilibrium with the largest service rate, which leads to best system performance, also leads to highest server utility, and hence is also most preferred by the servers, as the following theorem shows.}

\begin{theorem}\label{thm:MMN-larger-eq-better}
\red{If the symmetric first order condition~(\ref{eq:MMN-SYM-FOC}) has two solutions, say $\mu_1^\star$ and $\mu_2^\star$, with $\mu_1^\star > \mu_2^\star > \frac{\lambda}{N}$, then $U(\mu_1^\star,\mu_1^\star) > U(\mu_2^\star,\mu_2^\star)$.}
\end{theorem}

\subsection{\red{Numerical examples.}}\label{subsection:equilibrium-numerics}

\red{Because of the complexity of the expression for the equilibrium service rate(s) given by the first order condition~(\ref{eq:MMN-SYM-FOC}) and the possibility of multiple equilibria, we discuss a few numerical examples here in order to provide intuition.  In addition, we point out some interesting characteristics that emerge as a consequence of strategic server behavior.}

\begin{figure}[bt]
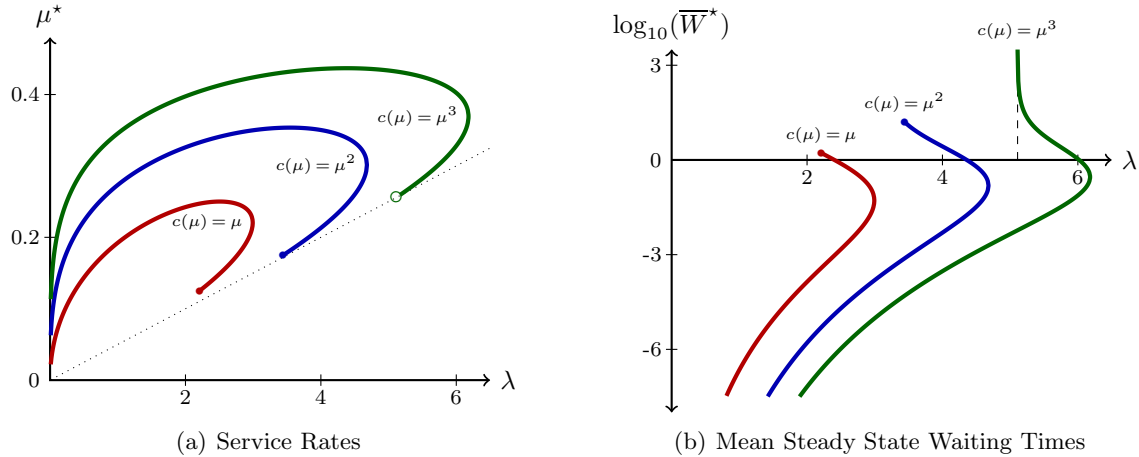

    \centering
    \subfigure[Service Rates]{\label{fig:1a}\input{fig1a.tikz}}
   \qquad
    \subfigure[Mean Steady State Waiting Times]{\label{fig:1b}\input{fig1b.tikz}}
    \caption{\red{Equilibrium behavior as a function of the arrival rate when the staffing level is fixed at $N=20$, for three different effort cost functions: linear, quadratic, and cubic. The dotted line in~(a) is $\mu=\lambda/N=\lambda/20$.\label{fig:1}}}
    \vspace{-0.2in}
\end{figure}

\begin{figure}[bt]
    \centering
    \subfigure[Service Rates]{\label{fig:2a}
    \begin{tikzpicture}[xscale=0.285,yscale=10]
\draw [<->,thick] (0,.48) -- (0,0) -- (20.5,0);
\node [above] at (0,.48) {{\small $\mu^{\star}$}};
\node [right] at (20.5,0) {{\small $N$}};
\draw (-.281,.2) -- (.281,.2);
\node [left] at (0,.2) {{\scriptsize 0.2}};
\draw (-.281,.4) -- (.281,.4);
\node [left] at (0,.4) {{\scriptsize 0.4}};
\node [left] at (0,0) {{\scriptsize 0}};
\draw (6,-.008) -- (6,.008);
\node [below] at (6,0) {{\scriptsize 6}};
\draw (12,-.008) -- (12,.008);
\node [below] at (12,0) {{\scriptsize 12}};
\draw (18,-.008) -- (18,.008);
\node [below] at (18,0) {{\scriptsize 18}};
\node [below] at (14, 0.234984) {{\tiny $c(\mu)=\mu$}};
\node [below] at (9.5, 0.331518) {{\tiny $c(\mu)=\mu^2$}};
\node [above] at (7, 0.445424) {{\tiny $c(\mu)=\mu^3$}};
\draw[dotted] (4.17, 0.479616) -- (4.27, 0.468384) -- (4.37, 0.457666) -- (4.47, 0.447427) -- (4.57, 0.437637) -- (4.67, 0.428266) -- (4.77, 0.419287) -- (4.87, 0.410678) -- (4.97, 0.402414) -- (5.07, 0.394477) -- (5.17, 0.386847) -- (5.27, 0.379507) -- (5.37, 0.372439) -- (5.47, 0.365631) -- (5.57, 0.359066) -- (5.67, 0.352734) -- (5.77, 0.34662) -- (5.87, 0.340716) -- (5.97, 0.335008) -- (6.07, 0.329489) -- (6.17, 0.324149) -- (6.27, 0.318979) -- (6.37, 0.313972) -- (6.47, 0.309119) -- (6.57, 0.304414) -- (6.67, 0.29985) -- (6.77, 0.295421) -- (6.87, 0.291121) -- (6.97, 0.286944) -- (7.07, 0.282885) -- (7.17, 0.27894) -- (7.27, 0.275103) -- (7.37, 0.27137) -- (7.47, 0.267738) -- (7.57, 0.264201) -- (7.67, 0.260756) -- (7.77, 0.2574) -- (7.87, 0.25413) -- (7.97, 0.250941) -- (8.07, 0.247831) -- (8.17, 0.244798) -- (8.27, 0.241838) -- (8.37, 0.238949) -- (8.47, 0.236128) -- (8.57, 0.233372) -- (8.67, 0.230681) -- (8.77, 0.22805) -- (8.87, 0.225479) -- (8.97, 0.222965) -- (9.07, 0.220507) -- (9.17, 0.218103) -- (9.27, 0.21575) -- (9.37, 0.213447) -- (9.47, 0.211193) -- (9.57, 0.208986) -- (9.67, 0.206825) -- (9.77, 0.204708) -- (9.87, 0.202634) -- (9.97, 0.200602) -- (10.07, 0.19861) -- (10.17, 0.196657) -- (10.27, 0.194742) -- (10.37, 0.192864) -- (10.47, 0.191022) -- (10.57, 0.189215) -- (10.67, 0.187441) -- (10.77, 0.185701) -- (10.87, 0.183993) -- (10.97, 0.182315) -- (11.07, 0.180668) -- (11.17, 0.179051) -- (11.27, 0.177462) -- (11.37, 0.175901) -- (11.47, 0.174368) -- (11.57, 0.172861) -- (11.67, 0.17138) -- (11.77, 0.169924) -- (11.87, 0.168492) -- (11.97, 0.167084) -- (12.07, 0.1657) -- (12.17, 0.164339) -- (12.27, 0.162999) -- (12.37, 0.161681) -- (12.47, 0.160385) -- (12.57, 0.159109) -- (12.67, 0.157853) -- (12.77, 0.156617) -- (12.87, 0.1554) -- (12.97, 0.154202) -- (13.07, 0.153022) -- (13.17, 0.15186) -- (13.27, 0.150716) -- (13.37, 0.149589) -- (13.47, 0.148478) -- (13.57, 0.147384) -- (13.67, 0.146306) -- (13.77, 0.145243) -- (13.87, 0.144196) -- (13.97, 0.143164) -- (14.07, 0.142146) -- (14.17, 0.141143) -- (14.27, 0.140154) -- (14.37, 0.139179) -- (14.47, 0.138217) -- (14.57, 0.137268) -- (14.67, 0.136333) -- (14.77, 0.13541) -- (14.87, 0.134499) -- (14.97, 0.133601) -- (15.07, 0.132714) -- (15.17, 0.131839) -- (15.27, 0.130976) -- (15.37, 0.130124) -- (15.47, 0.129282) -- (15.57, 0.128452) -- (15.67, 0.127632) -- (15.77, 0.126823) -- (15.87, 0.126024) -- (15.97, 0.125235) -- (16.07, 0.124456) -- (16.17, 0.123686) -- (16.27, 0.122926) -- (16.37, 0.122175) -- (16.47, 0.121433) -- (16.57, 0.1207) -- (16.67, 0.119976) -- (16.77, 0.119261) -- (16.87, 0.118554) -- (16.97, 0.117855) -- (17.07, 0.117165) -- (17.17, 0.116482) -- (17.27, 0.115808) -- (17.37, 0.115141) -- (17.47, 0.114482) -- (17.57, 0.11383) -- (17.67, 0.113186) -- (17.77, 0.112549) -- (17.87, 0.111919) -- (17.97, 0.111297) -- (18.07, 0.110681) -- (18.17, 0.110072) -- (18.27, 0.109469) -- (18.37, 0.108873) -- (18.47, 0.108284) -- (18.57, 0.107701) -- (18.67, 0.107124) -- (18.77, 0.106553) -- (18.87, 0.105988) -- (18.97, 0.10543) -- (19.07, 0.104877) -- (19.17, 0.10433) -- (19.27, 0.103788) -- (19.37, 0.103252) -- (19.47, 0.102722) -- (19.57, 0.102197) -- (19.67, 0.101678) -- (19.77, 0.101163) -- (19.87, 0.100654) -- (19.97, 0.10015) -- (20.07, 0.0996512);
\draw[->] (14, 0.244984) -- (15, 0.249518) -- (16, 0.250278) -- (17, 0.249323) -- (18, 0.247449) -- (19, 0.245056) -- (20, 0.242368);
\draw[->] (8, 0.331518) -- (9, 0.347964) -- (10, 0.353431) -- (11, 0.354539) -- (12, 0.353396) -- (13, 0.351005) -- (14, 0.347914) -- (15, 0.344442) -- (16, 0.340784) -- (17, 0.337057) -- (18, 0.333334) -- (19, 0.32966) -- (20, 0.326061);
\draw[->] (5, 0.410431) -- (6, 0.427034) -- (7, 0.435424) -- (8, 0.43851) -- (9, 0.438384) -- (10, 0.436346) -- (11, 0.433199) -- (12, 0.429439) -- (13, 0.425371) -- (14, 0.421183) -- (15, 0.416987) -- (16, 0.412849) -- (17, 0.408807) -- (18, 0.404883) -- (19, 0.401086) -- (20, 0.39742);
\draw (14, 0.178031) -- (15, 0.159256) -- (16, 0.145393);
\foreach \Point in {(15, 0.249518), (17, 0.249323), (18, 
  0.247449), (19, 0.245056), (15, 0.159256)} {\node at \Point {{\tiny \textcolor{\CL}{\textbullet}}};}
 \foreach \Point in {(8, 0.331518), (10, 0.353431), (11, 
  0.354539), (13, 0.351005), (14, 0.347914), (15, 
  0.344442), (16, 0.340784), (17, 0.337057), (18, 0.333334), (19, 
  0.32966)} {\node at \Point {{\tiny \textcolor{\CQ}{\textbullet}}};}
 \foreach \Point in {(5, 0.410431), (6, 0.427034), (8, 0.43851), (9, 
  0.438384), (10, 0.436346), (12, 0.429439), (13, 
  0.425371), (14, 0.421183), (15, 0.416987), (16, 0.412849), (17, 
  0.408807), (18, 0.404883), (19, 0.401086)} {\node at \Point {{\tiny \textcolor{\CC}{\textbullet}}};}
\node at (16, 0.250278) {{\scriptsize \textcolor{\CL}{$\diamond$}}};
\node at (16, 0.145393) {{\scriptsize \textcolor{\CL}{$\diamond$}}};
\node at (12, 0.353396) {{\scriptsize \textcolor{\CQ}{$\diamond$}}};
\node at (11, 0.433199) {{\scriptsize \textcolor{\CC}{$\diamond$}}};
\node at (14, 0.244984) {{\tiny \textcolor{\CL}{$\boldsymbol\times$}}};
\node at (14, 0.178031) {{\tiny \textcolor{\CL}{$\boldsymbol\times$}}};
\node at (9, 0.347964) {{\tiny \textcolor{\CQ}{$\boldsymbol\times$}}};
\node at (7, 0.435424) {{\tiny \textcolor{\CC}{$\boldsymbol\times$}}};
\end{tikzpicture}}
   \qquad
    \subfigure[Mean Steady State Waiting Times]{\label{fig:2b}
    \begin{tikzpicture}[xscale=0.3,yscale=0.47]
\draw [<->,thick] (0,3.5) -- (0,-8);
\draw [->,thick] (0,0) -- (20.5,0);
\node [above] at (0,3.5) {{\small $\log_{10}(\overline{W}^{\star})$}};
\node [right] at (20.5,0) {{\small $N$}};
\node [left] at (0,0) {{\scriptsize 0}};
\draw (-.281,3) -- (.281,3);
\node [left] at (0,3) {{\scriptsize 3}};
\draw (-.281,-3) -- (.281,-3);
\node [left] at (0,-3) {{\scriptsize -3}};
\draw (-.281,-6) -- (.281,-6);
\node [left] at (0,-6) {{\scriptsize -6}};
\draw (6,-.17) -- (6,.17);
\node [below] at (6,0) {{\scriptsize 6}};
\draw (12,-.17) -- (12,.17);
\node [below] at (12,0) {{\scriptsize 12}};
\draw (18,-.17) -- (18,.17);
\node [below] at (18,0) {{\scriptsize 18}};
\node [above] at (5., 1.25448) {{\tiny $c(\mu)=\mu^3$}};
\node [above] at (9, -0.1) {{\tiny $c(\mu)=\mu^2$}};
\node [right] at (13,-0.85) {{\tiny $c(\mu)=\mu$}};
\draw[->] (14., -1.49748) -- (15., -1.94885) -- (16., -2.35119) -- (17., -2.73514) -- (18., -3.11179) -- (19., -3.4865) -- (20., -3.86219);
\draw[->] (8., -0.252398) -- (9., -0.848671) -- (10., -1.31884) -- (11., -1.75422) -- (12., -2.17891) -- (13., -2.60302) -- (14., -3.03163) -- (15., -3.46754) -- (16., -3.91229) -- (17., -4.36676) -- (18., -4.83137) -- (19., -5.3063) -- (20., -5.79157);
\draw[->] (5., 1.25448) -- (6., -0.0693694) -- (7., -0.648825) -- (8., -1.13479) -- (9., -1.59339) -- (10., -2.04658) -- (11., -2.50411) -- (12., -2.97094) -- (13., -3.44966) -- (14., -3.94162) -- (15., -4.44745) -- (16., -4.96735) -- (17., -5.50126) -- (18., -6.04898) -- (19., -6.61021) -- (20., -7.18461);
\draw (14., -0.160741) -- (15., 0.0248052) -- (16., 0.15072);
\foreach \Point in {(15., -1.94885),  (17., -2.73514),
(18., -3.11179), (19., -3.4865), (15., 0.0248052)} {\node at \Point {{\tiny \textcolor{\CL}{\textbullet}}};}
 \foreach \Point in {(8., -0.252398),
 (10., -1.31884), (11., -1.75422),
(13., -2.60302), (14., -3.03163), (15., -3.46754), (16., -3.91229),
(17., -4.36676), (18., -4.83137), (19., -5.3063)} {\node at \Point {{\tiny \textcolor{\CQ}{\textbullet}}};}
 \foreach \Point in {(5., 1.25448), (6., -0.0693694),  (8., -1.13479),
(9., -1.59339), (10., -2.04658), (12., -2.97094),
(13., -3.44966), (14., -3.94162), (15., -4.44745), (16., -4.96735),
(17., -5.50126), (18., -6.04898), (19., -6.61021)} {\node at \Point {{\tiny \textcolor{\CC}{\textbullet}}};}
\node at (16., -2.35119) {{\scriptsize \textcolor{\CL}{$\diamond$}}};
\node at (16., 0.15072) {{\scriptsize \textcolor{\CL}{$\diamond$}}};
\node at (12., -2.17891) {{\scriptsize \textcolor{\CQ}{$\diamond$}}};
\node at (11., -2.50411) {{\scriptsize \textcolor{\CC}{$\diamond$}}};
\node at (14., -1.49748) {{\tiny \textcolor{\CL}{$\boldsymbol\times$}}};
\node at (14., -0.160741) {{\tiny \textcolor{\CL}{$\boldsymbol\times$}}};
\node at (9., -0.848671) {{\tiny \textcolor{\CQ}{$\boldsymbol\times$}}};
\node at (7., -0.648825) {{\tiny \textcolor{\CC}{$\boldsymbol\times$}}};
\end{tikzpicture} }
    \caption{\red{Equilibrium behavior as a function of the staffing level when the arrival rate is fixed at $\lambda=2$, for three different effort cost functions: linear, quadratic, and cubic. The dotted curve in~(a) is $\mu=\lambda/N=2/N$. The data points marked $\times$ and $\diamond$ correspond to $N^{ao,2}$ and $N^{opt,2}$ respectively.\label{fig:2}}}
    \vspace{-0.2in}
\end{figure}
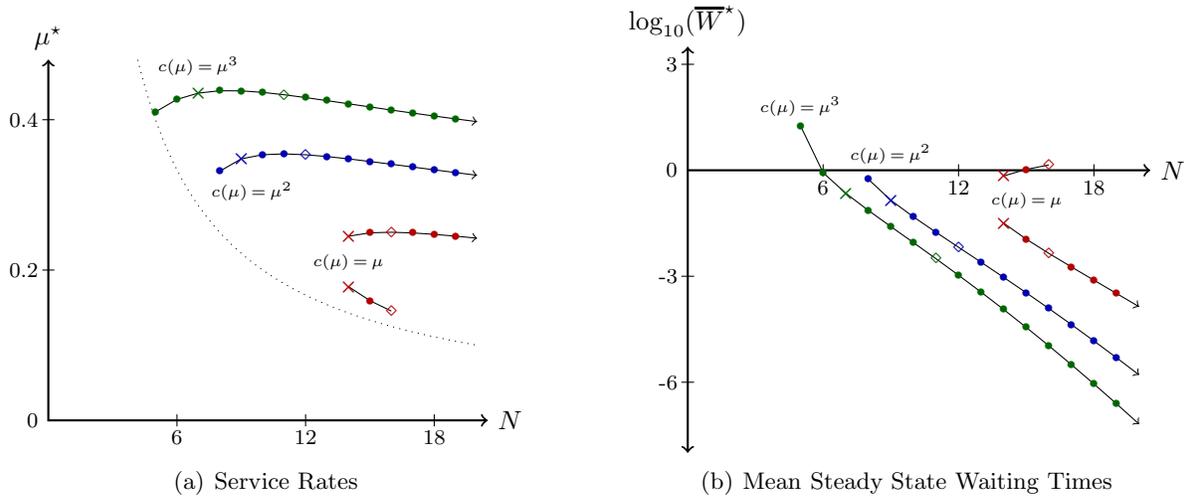

\red{We present two sets of graphs below: one that varies the arrival rate $\lambda$ while holding the staffing level fixed at $N=20$ (Figure~\ref{fig:1}), and one that varies the staffing level $N$ while holding the arrival rate fixed at $\lambda=2$ (Figure~\ref{fig:2}). In each set, we plot the following two equilibrium quantities: (a) service rates, and (b) mean steady state waiting times. Note that the graphs in Figure~\ref{fig:2} only show data points corresponding to integer values of $N$; the thin line through these points is only meant as a visual tool that helps bring out the pattern. Each of the four graphs shows data for three different effort cost functions: $c(\mu)=\mu$, $c(\mu)=\mu^2$, and $c(\mu)=\mu^3$, which are depicted in red, blue, and green respectively. The data points in Figure~\ref{fig:2} marked $\times$ and $\diamond$ correspond to special staffing levels $N^{ao,2}$ and $N^{opt,2}$ respectively, which are introduced later, in Section~\ref{Section:Staffing}.}

\red{The first observation we make is that there are at most two equilibria. Further, for large enough values of the minimum service rate $\frac{\lambda}{N}$, there is no equilibrium. (In Figure~\ref{fig:1a} where $N$ is fixed, this happens for large $\lambda$, and in Figure~\ref{fig:2a} where $\lambda$ is fixed, this happens for small $N$.) On the other hand, when the minimum service rate $\frac{\lambda}{N}$ is small enough, there is a unique equilibrium\change{; for this range, even if the symmetric first order condition~(\ref{eq:MMN-SYM-FOC}) has another solution greater than $\frac{\lambda}{N}$, it fails to satisfy~(\ref{eq:conditionalsymeq})}. \change{If an intermediate value of $\frac{\lambda}{N}$ is small enough for~(\ref{eq:MMN-SYM-FOC}) to have two feasible solutions, but not too small so that both solutions satisfy~(\ref{eq:conditionalsymeq}), then there are two equilibria.}}

\red{The second observation we make is that the two equilibria have very different behaviors.  As illustrated in Figure~\ref{fig:1a}, the larger equilibrium service rate first increases and then decreases while the corresponding mean steady state waiting time in Figure~\ref{fig:1b} steadily increases. In contrast, as the smaller equilibrium service rate increases, the corresponding mean steady state waiting time decreases. The relationship between the equilibrium service rates and waiting times is similarly inconsistent in Figure~\ref{fig:2}. \change{This behavior is not consistent with results from classical, nonstrategic models, and could serve as a starting point to explaining empiric observations that are also not consistent with classical, nonstrategic models. For example, the non-monotonicity of service rate in workload is consistent with behavior observed in a hospital setting in~\cite{KcTer2009}.}} 

\section{Staffing strategic servers.} \label{Section:Staffing}

One of the most studied questions for the design of service systems is staffing. Specifically, how many servers should be used for a given arrival rate.  \red{In the classical, nonstrategic setting,} this question is well understood.  In particular, as mentioned in the introduction, square-root staffing is known to be optimal when there are linear staffing and waiting costs~\cite{BorstMandelbaumReiman04}.

In contrast, there is no previous work studying staffing in the context of strategic servers. {\em The goal of this section is to initiate the study of the impact that strategic servers have on staffing.}  \red{To get a feeling for the issues involved, consider a system with arrival rate $\lambda$ and two possible staffing policies:  $N_1 = \lambda$ and $N_2=2 \lambda$, where $N_i$ is the number of servers staffed under policy $i$ given arrival rate $\lambda$.  Under $N_1$, if the servers work at any rate slightly larger than 1, then they will have almost no idle time, and so they will have incentive to work harder.  However, if servers are added, so that the provisioning is as in $N_2$, then servers will have plentiful idle time when working at rate 1, and thus not have incentive to work harder.  Thus, the staffing level has a fundamental impact on the incentives of the servers.}

The above highlights that one should expect significant differences in staffing when strategic servers are considered. \red{In particular, the key issue is that the staffing level itself creates incentives for the servers to speed up or slow down, because it influences the balance between effort and idle time.} Thus, the policies that are optimal in the nonstrategic setting are likely suboptimal in the strategic setting, and vice versa.

\red{The goal of the analysis in this section is to find the staffing level that  minimizes costs when the system manager incurs linear staffing and waiting costs, within the class of policies that admit a symmetric equilibrium service rate.}  However, the analysis in the previous section highlights that determining the exact optimal policy is difficult, since we only have an implicit characterization of a symmetric equilibrium service rate in~\eqref{eq:MMN-SYM-FOC}. As a result, we focus our attention on the setting where $\lambda$ is large, and look for an asymptotically optimal policy.

As expected, the asymptotically optimal staffing policy we design for the case of strategic servers differs considerably from the optimal policies in the nonstrategic setting. In particular, \red{in order for a symmetric equilibrium service rate to exist, the staffing level must be} {\em order $\lambda$  larger} than the optimal staffing in the classical, nonstrategic setting.  Then, the system operates in a {\em quality-driven (QD)} regime instead of the {\em quality-and-efficiency-driven (QED)} regime that results from square-root staffing.  This is intuitive given that the servers value their idle time, and in the QD regime they have idle time but in the QED regime their idle time is negligible.

The remainder of this section is organized as follows.  We first introduce the cost structure and define asymptotic optimality in Section~\ref{Subsection:CostFunction}. Then, in Section~\ref{Subsection:IStaffing}, we provide \red{a simple approximation of a symmetric equilibrium service rate and} an asymptotically optimal staffing policy. Finally, in Section~\ref{subsection:overstaffing}, \red{we compare our asymptotically optimal staffing policy for strategic servers with the square-root staffing policy that is asymptotically optimal in the nonstrategic setting.}

\subsection{Preliminaries.} \label{Subsection:CostFunction}

Our focus in this section is on \red{a} $M$/$M$/$N$ queue with strategic servers, as introduced in Section~\ref{Section:Model:MMN}.  We assume Random routing throughout this section.  \red{It follows that our results hold for any ``idle-time-order-based'' routing policy (as explained in the beginning of Section~\ref{Section:Results:MMN} and validated by Theorem~\ref{theorem:2-server-idlerandom}).} The cost structure we assume is consistent with the one in~\cite{BorstMandelbaumReiman04}, under which square-root staffing is asymptotically optimal when servers are not strategic.  \red{  In their cost structure, there are linear staffing and waiting costs.   One difference in our setting is that there may be multiple equilibrium service rates. In light of Theorem~\ref{thm:MMN-larger-eq-better}, we focus on the \change{largest} symmetric equilibrium service rate, and assume $\overline{W}^\star$ denotes the mean steady state waiting time in a $M$/$M$/$N$ queue with arrival rate $\lambda$, and strategic servers that serve at the largest symmetric equilibrium service rate (when there is more than one equilibrium).\footnote{\red{Note that the staffing policy we derive in this section (Theorem~\ref{theorem:staffing_independent}) will be asymptotically optimal regardless of which equilibrium service rate the servers choose.}}  Then, the total system cost is}
\begin{equation*}
C^\star(N,\lambda) = c_S N + \overline{w} \lambda \overline{W}^\star,
\end{equation*}
where $c_S$ is the per-unit staffing cost and $\overline{w}$ is the per-unit waiting cost. The $\star$ superscript \red{indicates} that the mean steady state waiting time, and hence, the cost function, depends on the \change{(largest)} symmetric equilibrium service rate $\mu^\star$, which in turn depends on $N$ and $\lambda$.

\red{The function $C^\star(N,\lambda)$ is well-defined only if a symmetric equilibrium service rate, under which the system is stable, exists.  Furthermore, we would like to rule out having an unboundedly large symmetric equilibrium service rate because then the server utility~(\ref{Equation:ServerUtilityGen}) will be large and negative -- and it is hard to imagine servers wanting to participate in such a game.
\begin{definition} \label{definition:admissible}
A staffing policy $N^\lambda$ is \emph{admissible} if the following two properties hold:
    \begin{enumerate}
    \item[(i)] There exists a symmetric equilibrium $\mu^{\star,\lambda}$ under which the system is stable ($\lambda < \mu^{\star,\lambda} N^\lambda$) for all large enough $\lambda$.
    \item[(ii)] There exists a sequence of symmetric equilibria $\{\mu^{\star,\lambda}, \lambda >0\}$ for which $\limsup_{\lambda \rightarrow \infty} \mu^{\star,\lambda} < \infty$.
    \end{enumerate}
\end{definition}
If the requirement (ii) in the above definition is not satisfied, then the server utility will approach $-\infty$ as the service rates become unboundedly large.  The servers will not want to participate in such a game.  As long as the requirement (ii) is satisfied, we can assume the server payment is sufficient to ensure that the servers have positive utility.
}

\red{We let $\Pi$ denote the set of admissible staffing policies.}  We would like to solve for
\begin{equation}\label{eq:N-opt-lambda}
\red{N^{opt,\lambda} = \argmin_{N \in \Pi} C^\star(N,\lambda).}
\end{equation}
However, given the difficulty of deriving \red{$N^{{opt,\lambda}}$} directly, we instead characterize the first order growth term of \red{$N^{opt,\lambda}$} in terms of $\lambda$.  To do this, we consider a sequence of systems, indexed by the arrival rate $\lambda$, and let $\lambda$ become large.

Our convention when we wish to refer to any process or quantity associated with the system having arrival rate $\lambda$ is to superscript the appropriate symbol by $\lambda$.  In particular, $N^\lambda$ denotes the staffing level in the system having arrival rate $\lambda$, and $\mu^{\star,\lambda}$ denotes an equilibrium service rate (assuming existence) in the system with arrival rate $\lambda$ and staffing level $N^\lambda$.  \red{We assume} $\overline{W}^{\star,\lambda}$ equals the mean steady state waiting time in a $M$/$M$/$N^\lambda$ queue with arrival rate $\lambda$ \red{ when the servers work at the largest equilibrium service rate.  The associated cost is}
\begin{equation} \label{eq:cost}
C^{\star,\lambda}(N^\lambda) = c_S N^\lambda + \overline{w} \lambda \overline{W}^{\star,\lambda}.
\end{equation}

\red{Given this setup, we would like to find an admissible staffing policy $N^\lambda$ that has close to the minimum cost $C^{\star,\lambda}(N^{opt,\lambda})$.}
\begin{definition}
\red{A staffing policy $N^\lambda$ is \emph{asymptotically optimal} if it is admissible ($N^\lambda \in \Pi$) and
 \[
\lim_{\lambda \rightarrow \infty} \frac{C^{\star,\lambda}(N^{\lambda})}{C^{\star,\lambda}(N^{opt,\lambda})} = 1.
\]}
\end{definition}

\red{In what follows, we use the $o$ and $\omega$ notations to denote the limiting behavior of functions. Formally, for any two real-valued functions $f(x),g(x)$ that take nonzero values for sufficiently large $x$, we say that $f(x) = o(g(x))$ (equivalently, $g(x)=\omega(f(x))$) if $\lim_{x\rightarrow\infty}\frac{f(x)}{g(x)}=0.$ In other words, $f$ is dominated by $g$ asymptotically (equivalently, $g$ dominates $f$ asymptotically).}

\subsection{\red{An asymptotically optimal staffing policy.}} \label{Subsection:IStaffing}

The class of policies we study are those that staff independently of the equilibrium service rates\red{, which are endogenously determined according to the analysis in Section~\ref{Section:Results:MMN}.}  More specifically, these are policies that choose $N^\lambda$ purely as a function of $\lambda$. Initially, it is unclear what functional form an asymptotically optimal staffing policy can take in the strategic server setting.  Thus, to begin, it is important to rule out policies that \textit{cannot} be asymptotically optimal.  The following proposition does this, and highlights that asymptotically optimal policies must be \red{asymptotically} linear in $\lambda$.

\begin{proposition} \label{proposition:nonaoInd}
Suppose $N^\lambda = f(\lambda) + o(f(\lambda)) \mbox{ for some function } f.$
If either $f(\lambda) = o(\lambda)$ or $f(\lambda) = \omega(\lambda)$, then the staffing policy $N^\lambda$ cannot be asymptotically optimal.
\end{proposition}

Intuitively, if $f(\lambda) = o(\lambda)$, understaffing forces the servers to work too hard, \red{their service rates growing unboundedly (and hence their utilities approaching $-\infty$) as $\lambda$ becomes large.} On the other hand, the servers may prefer to have $f(\lambda) = \omega(\lambda)$ because the overstaffing allows them to be lazier; however, the overstaffing is too expensive for the system manager.

Proposition~\ref{proposition:nonaoInd} implies that to find a staffing policy that is asymptotically optimal, we need only search within the class of policies that have the following form:
\begin{equation} \label{eq:staffingInd}
N^\lambda = \frac{1}{a}\lambda + o(\lambda), \mbox{ for } a \in (0,\infty).
\end{equation}
However, before we can search for the cost-minimizing $a$, we must ensure that the staffing~(\ref{eq:staffingInd}) guarantees the existence of a symmetric equilibria $\mu^{\star,\lambda}$ for all large enough $\lambda$.  It turns out that this is only true when $a$ satisfies certain conditions.  After providing these conditions (see \red{Theorem}~\ref{lemma:FOCexistence_staffingInd} in the following), we evaluate the cost function as $\lambda$ becomes large to find the $a^\star$ (defined in~(\ref{eq:optaindependentstaffing})) under which~(\ref{eq:staffingInd}) is an asymptotically optimal staffing policy (see Theorem~\ref{theorem:staffing_independent}).

\subsubsection*{Equilibrium characterization.} The challenge in characterizing equilibria comes from the complexity of the first order condition derived in Section~\ref{Section:Model:MMN}.  This complexity drives our focus on the large $\lambda$ regime.

The first order condition for a symmetric equilibrium~(\ref{eq:MMN-SYM-FOC}) is equivalently written as
\begin{equation} \label{eq:FOCstaffingInd}
\frac{\lambda}{N^\lambda}\left( \mu\left(1+\frac{ErlC(N^\lambda,\lambda/\mu)}{N^\lambda} \right) - \frac{\lambda}{N^\lambda} \right) = \mu^3 c'(\mu).
\end{equation}
Under the staffing policy~(\ref{eq:staffingInd}), when the limit $\lambda \rightarrow \infty$ is taken, this becomes
\begin{equation} \label{eq:limiting-FOC-one}
 a(\mu-a) = \mu^3 c'(\mu).
   \end{equation}
   \red{Since $\mu^3 c'(\mu) >0$, it follows that any solution $\mu$ has $\mu>a$.  Therefore, under the optimistic assumption that a symmetric equilibrium solution $\mu^{\star,\lambda}$ converging to the aforementioned solution $\mu$ exists, it follows that
\[
\lambda/ \mu^{\star,\lambda} < \lambda/a
\]
for all large enough $\lambda$.
In words,
the presence of strategic servers that value their idle time forces the system manager to staff order $\lambda$ more servers than the offered load $\lambda/ \mu^{\star,\lambda}$.  In particular, since the growth rate of $N^\lambda$ is $\lambda/a$, {\it the system will operate in the quality-driven regime}.
}

\red{The properties of the equation~(\ref{eq:limiting-FOC-one}) are easier to see when it is rewritten as}
 \begin{equation}  \label{eq:limiting_mu_Ind}
 \frac{1}{a} = \frac{\mu^2}{a^2} c'(\mu)+\frac{1}{\mu}.
 \end{equation}
Note that the left-hand side of~(\ref{eq:limiting_mu_Ind}) is a constant function and the right-hand side is a convex function.  These functions either cross at exactly two points, at exactly one point, or never intersect, depending on $a$. That information then can be used to show whether or not there exists a solution to the first order condition~(\ref{eq:FOCstaffingInd}), depending on the value of $a$ in the staffing policy~(\ref{eq:staffingInd}).
\begin{theorem} \label{lemma:FOCexistence_staffingInd}  The following holds for all large enough $\lambda$.
\begin{enumerate}
\item[(i)] Suppose $a>0$ is such that there exists $\mu_2 > \mu_1 >0$ that solve~(\ref{eq:limiting_mu_Ind}).  Then, there exist two solutions that solve~(\ref{eq:FOCstaffingInd}).
\item[(ii)]  Suppose $a>0$ is such that there exists exactly one $\mu_1>0$ that solves~(\ref{eq:limiting_mu_Ind}).
\begin{enumerate}
\item[(a)] Suppose $N^\lambda - \frac{\lambda}{a} \geq 0$.  Then, there exist two solutions that solve~(\ref{eq:FOCstaffingInd}).
\item[(b)] Otherwise, if $N^\lambda - \frac{\red{\lambda}}{a} < -3$, then there does not exist a solution $\mu^\lambda$ to~(\ref{eq:FOCstaffingInd}).
\end{enumerate}
\end{enumerate}
Furthermore, for any $\epsilon >0$, if $\mu^\lambda$ solves~(\ref{eq:FOCstaffingInd}), then $|\mu^\lambda - \mu| < \epsilon$ for some $\mu$ that solves~(\ref{eq:limiting_mu_Ind}).
\end{theorem}
\red{We are not sure if there exists a solution in the case of $N^\lambda - \frac{1}{a}\lambda \in [-3,0)$; however, given that we are focusing on a large $\lambda$ asymptotic regime, the range [-3,0) is vanishingly small.}

\red{Moving forward, once the existence of a solution to the first order condition~(\ref{eq:FOCstaffingInd}) is established, to conclude that solution is a
symmetric equilibrium service rate also requires verifying the condition~(\ref{eq:conditionalsymeq}) in Theorem~\ref{thm:conditionalsymeq}.  This can be done for any staffing policy~(\ref{eq:staffingInd}) under which the system operates in the quality driven-regime.}
\begin{lemma} \label{lemma:FOCisEquilibrium}
For any staffing policy $N^\lambda$ and associated $\mu^\lambda$ that satisfies the first order condition~(\ref{eq:FOCstaffingInd}), if
\[
\liminf_{\lambda \rightarrow \infty} \frac{N^\lambda \mu^\lambda}{\lambda} = d >1 \mbox{ and } \limsup_{\lambda \rightarrow \infty} \mu^\lambda < \infty,
\]
then $\mu^{\star,\lambda} = \mu^\lambda$ is a symmetric equilibrium for all large enough $\lambda$.
\end{lemma}

\red{Under the conditions for the existence of a solution to the first order condition~(\ref{eq:FOCstaffingInd}) in Theorem~\ref{lemma:FOCexistence_staffingInd}, it is also true that  the conditions of Lemma~\ref{lemma:FOCisEquilibrium} are satisfied.}  In particular, there exists a bounded sequence $\{ \mu^\lambda \}$ having
\[
\liminf_{\lambda \rightarrow \infty} \frac{N^\lambda \mu^\lambda}{\lambda} = \liminf_{\lambda \rightarrow \infty} \frac{\mu^\lambda}{a} + \mu^\lambda \frac{o(\lambda)}{\lambda} >1.
\]
This then guarantees that, for all large enough $\lambda$, there exists a solution $\mu^{\star,\lambda}$ to~(\ref{eq:FOCstaffingInd}) that is a symmetric equilibrium, under the conditions of \red{Theorem}~\ref{lemma:FOCexistence_staffingInd}.

There are either multiple symmetric equilibria for each $\lambda$ or 0, because from \red{Theorem}~\ref{lemma:FOCexistence_staffingInd} there are either multiple or zero solutions to the first order condition~(\ref{eq:FOCstaffingInd}).  These symmetric equilibria will be close when there exists exactly one $\mu$ that solves~(\ref{eq:limiting_mu_Ind}); however, they may not be close when there exist two $\mu$ that solve~(\ref{eq:limiting_mu_Ind}).  We show in the following that this does not affect what staffing policy should be asymptotically optimal.

\subsubsection*{Optimal staffing.}  Given the characterization of symmetric equilibria under a staffing policy~(\ref{eq:staffingInd}), we can now move to the task of \red{optimizing the staffing level, i.e., optimizing $a$. The first step is to characterize the associated cost, which is done in the following proposition.}

\begin{proposition} \label{proposition:staffing_independent}
Suppose $a>0$ is such that there exists  $\mu>0$ that solves~(\ref{eq:limiting_mu_Ind}).  Then, under the staffing policy~(\ref{eq:staffingInd}),
\[
\frac{C^{\star,\lambda}(N^\lambda)}{\lambda} \rightarrow \frac{1}{a}c_S, \mbox{ as } \lambda \rightarrow \infty.
\]
\end{proposition}

Proposition~\ref{proposition:staffing_independent} implies that to minimize costs within the class of staffing policies that satisfy~(\ref{eq:staffingInd}), the maximum $a$ under which there exists at least one solution to~(\ref{eq:limiting_mu_Ind}) should be used.  That is, we should choose $a$ to be
\begin{equation}
\change{a^\star := \sup{\mathcal{A}}, \mbox{ where } \mathcal{A} := \left\{ a>0: \mbox{ there exists at least one solution $\mu>0$ to~(\ref{eq:limiting_mu_Ind})} \right\}.} \label{eq:optaindependentstaffing}
\end{equation}

\begin{lemma} \label{lemma:Afinite}
$a^\star \in \mathcal{A}$ is finite.
\end{lemma}

Importantly, this $a^\star$ is not only optimal among \red{the class of staffing policies that satisfy~(\ref{eq:staffingInd})}, it is asymptotically optimal among all \red{admissible} staffing policies.  In particular, the following theorem shows that as $\lambda$ becomes unboundedly large, no other \red{admissible} staffing policy can \red{asymptotically} achieve strictly lower cost than the one in~(\ref{eq:staffingInd}) with $a=a^\star$.

\begin{theorem} \label{theorem:staffing_independent}
If $N^{{ao},\lambda}$ satisfies~(\ref{eq:staffingInd}) with $a = a^\star$, then \red{$N^{{ao},\lambda}$ is admissible and asymptotically optimal.}
Furthermore,
\[
\lim_{\lambda \rightarrow \infty} \frac{C^{\star,\lambda} (N^{{ao},\lambda})}{\lambda} = \lim_{\lambda \rightarrow \infty} \frac{C^{\star,\lambda} (N^{opt,\lambda})}{\lambda} = c_S \frac{1}{a^\star}.
\]
\end{theorem}

\red{Note that an inspection of the proof of Theorem~\ref{theorem:staffing_independent} shows that it holds regardless of which equilibrium service rate is used to define $\overline{W}^{\star,\lambda}$.  Hence, even though we have defined $\overline{W}^{\star,\lambda}$ to be the mean steady state waiting time when the servers serve at the largest equilibrium service rate, this is not necessary.  The staffing policy $N^{{ao},\lambda}$ in Theorem~\ref{theorem:staffing_independent} will be asymptotically optimal regardless of which equilibrium service rate the servers choose.}

Though the above theorem characterizes an asymptotically optimal staffing level, because the definition of $a^\star$ is implicit, it is difficult to develop intuition.  To highlight the structure more clearly, the following lemma characterizes $a^\star$ for a specific class of effort cost functions.

\begin{lemma} \label{lemma:ex_polynomial_cost}
Suppose $c(\mu)=c_E \mu^p$ for some $c_E \in [1,\infty)$ and $p\geq 1$.
Then,
\[
a^\star = \left[ \frac{(p+1)}{(p+2)} \left( \frac{1}{c_E p (p+2)} \right)^{\frac{1}{p+1}}  \right] ^{(p+1)/p} < \mu^\star = \left( \frac{p+1}{c_E p (p+2)^2} \right)^{\frac{1}{p}} <1,
\]
and $a^\star$ and $\mu^\star$ are both increasing in $p$.
Furthermore,
\[
\mbox{ if } a \left\{ \begin{array}{l} < \\ > \\ = \end{array} \right\} a^\star, \mbox{ then } \left\{
 \begin{array}{l}\mbox{there are 2 non-negative solutions to~(\ref{eq:limiting_mu_Ind})} \\ \mbox{there is no non-negative solution to~(\ref{eq:limiting_mu_Ind})} \\ \mbox{there is exactly one solution to~(\ref{eq:limiting_mu_Ind})} \end{array} \right..
\]
\end{lemma}

There are several interesting relationships between the effort cost function and the staffing level that follow from Lemma~\ref{lemma:ex_polynomial_cost}.   First, for fixed $p$,
\[
a^\star(p) \downarrow 0 \mbox{ as } c_E \rightarrow \infty.
\]
In words, the system manager must staff more and more servers as effort becomes more costly.  Second, for fixed $c_E$, since $a^\star(p)$ is increasing in $p$, the system manager can staff less servers when the cost function becomes ``more convex''.  The lower staffing level forces the servers to work at a higher service rate since since $\mu^\star(p)$ is also increasing in $p$.  We will revisit this idea that convexity is helpful to the system manager in the next section.

\subsection{Contrasting staffing policies for strategic and nonstrategic servers.} \label{subsection:overstaffing}
One of the most crucial observations that the previous section makes about the impact of strategic servers on staffing is that \red{the strategic behavior leads the system to a quality-driven regime.  In this section, we explore this issue in more detail, by comparing to the optimal staffing rule that arises when servers are not strategic, and then attempting to implement that staffing rule.}

\subsubsection*{Nonstrategic servers.} Recall that, for the conventional $M$/$M$/$N$ queue (without strategic servers), square-root staffing minimizes costs as $\lambda$ becomes large (see equation (1), Proposition 6.2, and Example 6.3 in~\cite{BorstMandelbaumReiman04}). So, we can define
    \[
    C^\lambda_\mu(N) = c_SN + \overline{w} \lambda \overline{W}_\mu^\lambda
    \]
to be the cost associated with staffing $N$ nonstrategic servers that work at the fixed service rate $\mu$.  Further,
    \[
    N_\mu^{{opt},\lambda} = \argmin_{N > \frac{\lambda}{\mu}} C_\mu^\lambda(N)
    \]
is the staffing level that minimizes expected cost when the system arrival rate is $\lambda$ and the service rate is fixed to be $\mu$.
So, the staffing rule
    \begin{equation} \label{eq:BMRstaffing}
    N^{{BMR},\lambda}_\mu = \frac{\lambda}{\mu} + y^\star \sqrt{\frac{\lambda}{\mu}}
    \end{equation}
is asymptotically optimal in the sense that
    \[
    \lim_{\lambda \rightarrow \infty} \frac{C_\mu^\lambda(N_\mu^{{BMR},\lambda})}{C_\mu^\lambda(N_{\red{\mu}}^{opt,\lambda})} = 1.
    \]
\red{Here, $y^\star := \argmin_{y >0} \left\{ c_S y + \frac{w\alpha(y)}{y} \right\}$, where $\alpha(y)=\left(1+\frac{y}{h(-y)}\right)^{-1}$ with $h(\cdot)$ being the hazard rate function of the standard normal distribution, namely, $h(x):=\frac{\phi(x)}{1-\Phi(x)}$ with $\phi(x)= \frac{1}{\sqrt{2\pi}}e^{-\frac{x^2}{2}}$ and $\Phi(x)=\int_{-\infty}^{x}\phi(t)dt$. The staffing rule~(\ref{eq:BMRstaffing}) is the famous square-root safety staffing rule.}

\subsubsection*{Contrasting strategic and nonstrategic servers.} \red{In order to compare the case of strategic servers to the case of nonstrategic servers, it is natural to fix $\mu$ in~(\ref{eq:BMRstaffing}) to the limiting service rate that results from using the optimal staffing rule $N^{{ao},\lambda}$ defined in Theorem~\ref{theorem:staffing_independent}.   We see that $N^{{ao},\lambda}$ staffs order $\lambda$ more servers than $N^{{BMR},\lambda}_{\mu^\star}$, where $\mu^\star$ solves~(\ref{eq:limiting_mu_Ind}) for $a = a^\star$, because any solution to~(\ref{eq:limiting_mu_Ind}) has $a > \mu$.  When the effort cost function is $c(\mu) = c_E \mu^p$ for $p \geq 1$, we know from Lemma~\ref{lemma:ex_polynomial_cost} and Theorem~\ref{lemma:FOCexistence_staffingInd} (since the $a^\star$ is unique) that
\[
\mu^{\star,\lambda} \rightarrow \mu^\star \mbox{ as } \lambda \rightarrow \infty,
\]
where $\mu^\star$ is as given in Lemma~\ref{lemma:ex_polynomial_cost}.  Then, the difference in the staffing levels is
\begin{eqnarray*}
 N^{{ao},\lambda} - N_{\mu^\star}^{{BMR},\lambda}  =  \left( \frac{1}{a^\star} - \frac{1}{\mu^\star} \right) \lambda + o(\lambda)  =  \frac{1}{a^\star} \left( \frac{1}{p+2} \right) \lambda + o(\lambda).
\end{eqnarray*}
Since $a^\star = a^\star(p)$ is increasing in $p$ from Lemma~\ref{lemma:ex_polynomial_cost}, we see that the difference $ N^{{ao},\lambda} - N_{\mu^\star}^{BMR,\lambda}$ decreases to 0 as the cost function becomes ``more convex''.  This is consistent with our observation at the end of the previous subsection that convexity is helpful to the system manager.}

\red{It is natural to wonder if a system manager can force the servers to work harder by adopting the staffing policy suggested by the analysis of nonstrategic servers, i.e.,
\begin{equation} \label{eq:BMRstrategic}
N^{\star,{BMR},\lambda} = \frac{\lambda}{\mu^{\star,\lambda}} + y^\star \sqrt{\frac{\lambda}{\mu^{\star,\lambda}}}.
\end{equation}
The interpretation of this staffing rule requires care, because the offered load $\lambda / \mu^{\star,\lambda}$ is itself a function of the staffing level (and the arrival rate) through an equilibrium service rate $\mu^{\star,\lambda}$.  The superscript $\star$ emphasizes this dependence.}

\red{The first question concerns whether or not the staffing policy~(\ref{eq:BMRstrategic}) is even possible in practice, because the staffing level depends on an equilibrium service rate and vice versa.  More specifically, for a given staffing level, the servers relatively quickly arrive at an equilibrium service
rate.  Then, when system demand grows, the system manager increases the staffing, and the servers again arrive at an equilibrium service rate.  In other words, there are two games, one played on a faster time scale (that is the servers settling to an equilibrium service rate), and one played on a slower time scale (that is the servers responding to added capacity).}

\red{To analyze the staffing policy~(\ref{eq:BMRstrategic}), note that the
 first order condition for a symmetric equilibrium~(\ref{eq:MMN-SYM-FOC}) is equivalently written as
\[
\frac{\lambda/\mu}{\left( N^{\star,BMR,\lambda} \right)^2} \left( N^{\star,BMR,\lambda} - \frac{\lambda}{\mu} + ErlC\left( N^{\star,BMR,\lambda}, \frac{\lambda}{\mu} \right) \right) = \mu c'(\mu).
\]
Then, if $\mu^\lambda$ is a solution to the first order condition under the staffing $N^{\star,BMR,\lambda}$ from~(\ref{eq:BMRstrategic}), from substituting
$N^{\star,BMR,\lambda}$ into the above expression, $\mu^\lambda$ must satisfy
\[
  \frac{\lambda/\mu^\lambda}{\left( \lambda/\mu^\lambda + y^\star \sqrt{\lambda/\mu^\lambda} \right)^2} \left( y^\star \sqrt{\lambda/\mu^\lambda} + ErlC\left( \lambda/\mu^\lambda + y^\star \sqrt{\lambda/\mu^\lambda}, \frac{\lambda}{\mu^\lambda} \right) \right) = \mu^\lambda c'(\mu^\lambda).
\]
As $\lambda$ becomes large, since $ErlC\left( \lambda/\mu + y \sqrt{\lambda/\mu}, \frac{\lambda}{\mu} \right)$ is bounded above by 1, the left-hand side of the above expression has limit 0. Furthermore, the right-hand side of the above equation is non-negative and increasing as a function of $\mu$. Hence any sequence of solutions $\mu^\lambda$ to the first order condition has the limiting behavior
\[
\mu^\lambda \rightarrow 0, \mbox{ as } \lambda \rightarrow \infty,
\]
which cannot be a symmetric equilibrium service rate because we require the servers to work fast enough to stabilize the system.}

\red{One possibility is to expand the definition of an equilibrium service rate in~(\ref{Equation:ServerUtilityGen}) to allow the servers to work exactly at the lower bound $\lambda/N$.  In fact, the system manager may now be tempted to push the servers to work even faster.  However, faster service cannot be mandated for free -- there must be a trade-off; for example, the service quality may suffer or the salaries should be higher.}

\subsection{\red{Numerical examples.}}\label{subsection:staffing-numerics}

\red{In order to understand how well our asymptotically optimal staffing policy $N^{ao,\lambda}$ performs in comparison with the optimal policy $N^{opt,\lambda}$ for finite $\lambda$, and how fast the corresponding system cost converges to the optimal cost, we present some results from numerical analysis in this section.}

\begin{figure}[ht]
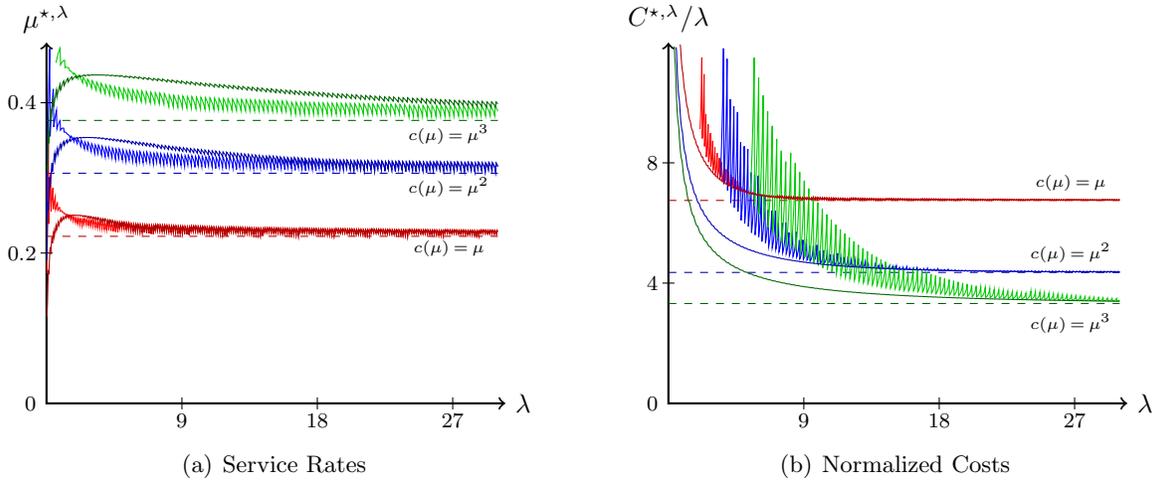

    \centering
    \subfigure[Service Rates]{\label{fig:3a}
    \input{fig3a.tikz}}
   \qquad
    \subfigure[Normalized Costs]{\label{fig:3b}
    \input{fig3b.tikz}}
    \caption{Equilibrium behavior as a function of the arrival rate for the optimal and asymptotically optimal staffing policies, for three different effort cost functions: linear, quadratic, and cubic.\label{fig:3}}
\end{figure}

\red{We consider two staffing policies: (i) $N^{opt,\lambda}$ (defined in~(\ref{eq:N-opt-lambda})), and (ii) $N^{ao,\lambda}$ (defined in Theorem~\ref{theorem:staffing_independent} and~(\ref{eq:optaindependentstaffing})) where we ignore the $o(\lambda)$ term of~(\ref{eq:staffingInd}). For each, we first round up the staffing level if necessary, and then plot the following two equilibrium quantities as a function of the arrival rate $\lambda$: (a) service rates $\mu^{\star,\lambda}$ (if there is more than one, we pick the largest), and (b) normalized costs $C^{\star,\lambda}/\lambda$. \change{We calculate $N^{opt,\lambda}$ numerically, by iterating over the staffing levels that admit equilibria (and we choose the lowest cost when there are multiple equilibria).} These plots are shown in Figure~\ref{fig:3} for three different effort cost functions: $c(\mu)=\mu$, $c(\mu)=\mu^2$, and $c(\mu)=\mu^3$, which are depicted in red, blue, and green respectively. For each color, the curve with the darker shade corresponds to $N^{opt,\lambda}$ and the curve with the lighter shade corresponds to $N^{ao,\lambda}$. The horizontal dashed lines correspond to the limiting values as $\lambda\to\infty$.}

\red{An immediate first observation is the jaggedness of the curves, which is a direct result of the discreteness of the staffing levels $N^{opt,\lambda}$ and $N^{ao,\lambda}$. In particular, as the arrival rate $\lambda$ increases, the equilibrium service rate $\mu^{\star,\lambda}$ decreases (respectively, the equilibrium normalized cost $C^{\star,\lambda}/\lambda$ increases) smoothly until the staffing policy adds an additional server, which causes a sharp increase (respectively, decrease). The jaggedness is especially pronounced for smaller $\lambda$, resulting in a complex pre-limit behavior that necessitates asymptotic analysis in order to obtain analytic results.}

\red{However, despite the jaggedness, the plots illustrate clearly that both the equilibrium service rates and normalized costs of the optimal policy $N^{ao,\lambda}$ converge quickly to those of the optimal policy $N^{opt,\lambda}$, highlighting that our asymptotic results are predictive at realistically sized systems. } 

\section{Routing to strategic servers.}
\label{Section:Scheduling}

Thus far we have focused our discussion on staffing, assuming that jobs are routed randomly to servers when there is a choice.  Of course, the decision of how to route jobs to servers is another crucial aspect of the design of service systems.  As such, the analysis of routing policies has received considerable attention in the queueing literature, when servers are not strategic.  In this section, we begin to investigate the impact of strategic servers on the design of routing policies.

In the classical literature studying routing when servers are nonstrategic, a wide variety of policies have been considered.  These include ``rate-based policies'' such as Fastest Server First (FSF) and Slowest Server First (SSF); as well as \red{``idle-time-order-based policies''} such as Longest Idle Server First (LISF) and Shortest Idle Server First (SISF).  Among these routing policies, FSF is a natural choice to minimize the mean response time (although, as noted in the Introduction, it is not optimal in general).  This leads to the question:  how does FSF perform when servers are strategic?  In particular, does it perform better than the Random routing that we have so far studied?

\red{Before studying optimal routing to improve performance, we must first answer the following even more fundamental question:} what routing policies admit symmetric equilibria? This is a very challenging goal, as can be seen by the complexity of the analysis for the $M$/$M$/$N$ under Random routing.  This \red{section} provides a first step towards that goal.

The results in this section focus on two broad classes of routing policies \textit{idle-time-order-based policies} and \textit{rate-based policies}, which are introduced in turn in the following.

\subsection{Idle-time-order-based policies.}

Informally, idle-time-order-based policies are those routing policies that use only the rank ordering of when servers \red{last} became idle in order to determine how to route incoming jobs. To describe the class of idle-time-order-based policies precisely, let $\mathcal{I}(t)$ be the set of servers idle at time $t>0$, and, when $\mathcal{I}(t)\not=\emptyset$, let $\boldsymbol{s}(t)=(s_1,\ldots,s_{|\mathcal{I}(t)|})$ denote the ordered vector of idle servers at time $t$, where server $s_j$ became idle before server $s_k$ whenever $j<k$.  For $n\geq 1$, let $\mathcal{P}_n=\Delta(\{1,\ldots,n\})$ denote the set of all probability distributions over the set $\{1,\ldots,n\}$. An idle-time-order-based routing policy is defined by a collection of probability distributions $\boldsymbol{p} = \{p^S\}_{S\in 2^{\{1,2,\ldots,N\}}\backslash\emptyset}$, such that $p^S\in\mathcal{P}_{|S|}$, for all $S\in 2^{\{1,2,\ldots,N\}}\backslash\emptyset$. Under this policy, at time $t$, the next job in queue is assigned to idle server $s_j$ with probability $p^{\mathcal{I}(t)}(j)$. Examples of idle-time-order-based routing policies are as follows.
\begin{enumerate}
\item \red{\textit{Random.}}  An arriving customer that finds more than one server idle is equally likely to be routed to any of those servers.  Then, $p^S = (1/|S|,\ldots,1/|S|)$ for all $S\in 2^{\{1,2,\ldots,N\}}\backslash\emptyset$.
\item \red{\textit{Weighted Random.}}  Each such arriving customer is routed to one of the idle servers with probabilities that may depend on the order in which the servers became idle.  For example, if
    \begin{equation*}
    p^S(j) = \frac{|S|+1-j}{\sum_{n=1}^{|S|}n}, \; j \in S, \mbox{ for } s_j \in S, \mbox{ for all } S\in 2^{\{1,2,\ldots,N\}}\backslash\emptyset,
    \end{equation*}
    then the probabilities are decreasing according to the order in which the servers became idle. Note that $\sum_j p^S(j) = \frac{|S|(|S|+1) - \frac{1}{2}|S|(|S|+1)}{\frac{1}{2}|S|(|S|+1)}=1$.
\item \red{\textit{Longest Idle Server First (Shortest Idle Server First).}}  Each such arriving customer is routed to the server that has idled the longest (idled the shortest).  Then, $p^S = (1,0,\ldots,0)$ ($p^S = (0,\ldots,0,1)$) for all $S \subseteq \{1,2,\ldots,N\}$.
\end{enumerate}

\subsubsection{Policy-space collapse.}\label{ssec:policyspacecollapse}

Surprisingly, it turns out that all idle-time-order-based policies are ``equivalent'' in a very strong sense --- they all lead to the same steady state probabilities, resulting in a remarkable \textit{policy-space collapse} result, which we discuss in the following.

Fix $\mathit{R}$ to be some idle-time-order-based routing policy, defined through the collection of probability distributions $\boldsymbol{p} = \{p^S\}_{\emptyset\not=S\subseteq\{1,2,\ldots,N\}}$.  The states of the associated continuous time Markov chain are defined as follows:
 \begin{itemize}
\item State $B$ is the state where all servers are busy, but there are no jobs waiting in the queue.
\item State $\boldsymbol s=(s_1,s_2,\ldots,s_{|\mathcal I|})$ is the ordered vector  of idle servers $\mathcal{I}$. When $\mathcal{I}=\emptyset$, we identify the empty vector $\boldsymbol{s}$ with state $B$.
\item State $m$ ($m\geq 0$) is the state where all servers are busy and there are $m$ jobs waiting in the queue (i.e., there are $N+m$ jobs in the system). We identify state $0$ with state $B$.
\end{itemize}

When all servers are busy, there is no routing, and so the system behaves exactly as \red{a} $M$/$M$/$1$ queue with arrival rate $\lambda$ and service rate $\mu_1+\cdots+\mu_N$.  Then, from the local balance equations, the associated steady state probabilities
$\pi_B$ and $\pi_m$ for $m=0,1,2,\ldots$, must satisfy
\begin{equation} \label{eq:known}
\pi_m=(\lambda/\mu)^m\pi_B \mbox{ where } \mu=\DS\sum_{j=1}^N\mu_j.
\end{equation}

One can anticipate that the remaining steady state probabilities satisfy
\begin{equation}\label{eq:claim}
\pi_{\boldsymbol s}=\pi_B\prod_{s\in\mathcal{I}}\frac{\mu_s}{\lambda}\quad\mbox{ for all }\boldsymbol s=(s_1,s_2,\ldots,s_{|\mathcal{I}|})\text{ with }|\mathcal{I}|>0,
\end{equation}
and the following theorem verifies this by establishing that the detailed balance equations are satisfied.

\begin{theorem}\label{theorem:2-server-idlerandom}
All idle-time-order-based policies have the steady state probabilities that are uniquely determined by~(\ref{eq:known})-(\ref{eq:claim}), together with the normalization constraint that their sum is one.
\end{theorem}

Theorem~\ref{theorem:2-server-idlerandom} is remarkable because there is no dependence on the collection of probability distributions $\boldsymbol{p}$ that define $\mathit{R}$.
Therefore,  it follows that all idle-time-order-based routing policies result in the same steady state probabilities. Note that, concurrently, a similar result has been discovered independently in the context of loss systems~\cite{HajiRoss13}.

In relation to our server game, it follows from Theorem~\ref{theorem:2-server-idlerandom} that all idle-time-order-based policies have the same equilibrium behavior as Random.  This is because an equilibrium service rate depends on the routing policy through the server idle time vector $(I_1(\boldsymbol\mu;\mathit{R}),\ldots,I_N(\boldsymbol\mu;\mathit{R}))$, \red{which can be found from the steady state probabilities in~(\ref{eq:known})-(\ref{eq:claim}).} As a consequence, if there exists (does not exist) an equilibrium service rate under Random, then there exists (does not exist) an equilibrium service rate under any idle-time-order-based policy. In summary, it is not possible to achieve better performance than under Random by employing any idle-time-order-based policy.

\subsection{Rate-based policies.}

Informally, a rate-based policy is one that makes routing decisions using only information about the rates of the servers. As before, let $\mathcal{I}(t)$ denote the set of idle servers at time $t$. In a rate-based routing policy, jobs are assigned to idle servers only based on their service rates. We consider a parameterized class of rate-based routing policies that we term $r$\textit{-routing policies} ($r\in\mathbb{R}$). Under an $r$-routing policy, at time $t$, the next job in queue is assigned to idle server $i\in\mathcal{I}(t)$ with probability
\begin{equation*}
p_i(\boldsymbol\mu,t;r)=\frac{\mu_i^r}{\displaystyle\sum_{j\in\mathcal{I}(t)}\mu_j^r}
\end{equation*}
Notice that for special values of the parameter $r$, we recover well-known policies. For example, setting $r=0$ results in Random; as $r\rightarrow\infty$, it approaches FSF; and as $r\rightarrow -\infty$, it approaches SSF.

In order to understand the performance \red{of} rate-based policies, the first step is to perform an equilibrium analysis, i.e., we need to understand what the steady state idle times look like under any $r$-routing policy. The following proposition provides us with the required expressions.

\begin{proposition}\label{proposition:idletime}
Consider a heterogeneous $M$/$M$/$2$ system under an $r$-routing policy, with arrival rate $\lambda>0$ and servers $1$ and $2$ operating at rates $\mu_1$ and $\mu_2$ respectively. The steady state probability that server $1$ is idle is given by:
\begin{equation*}
I^r_1(\mu_1,\mu_2)=\frac{\mu_1(\mu_1+\mu_2-\lambda)\left[(\lambda+\mu_2)^2+\mu_1\mu_2+\frac{\mu_2^r}{\mu_1^r+\mu_2^r}(\lambda\mu_1+\lambda\mu_2)\right]}{\mu_1\mu_2(\mu_1+\mu_2)^2+(\lambda\mu_1+\lambda\mu_2)\left[\mu_1^2+2\mu_1\mu_2-\frac{\mu_1^r}{\mu_1^r+\mu_2^r}(\mu_1^2-\mu_2^2)\right]+(\lambda\mu_1)^2+(\lambda\mu_2)^2},
\end{equation*}
and the steady state probability that server $2$ is idle is given by $I^r_2(\mu_1,\mu_2)=I^r_1(\mu_2,\mu_1)$.
\end{proposition}

Note that we restrict ourselves to a $2$-server system for this analysis. This is due to the fact that there are no closed form expressions known for the resulting Markov chains for systems with more than 3 servers.  It may be possible to extend these results to $3$ servers using results from~\cite{Mokaddis98}; but, the expressions are intimidating, to say the least. However, the analysis for two servers is already enough to highlight important structure about the impact of strategic servers on policy design.

In particular, our first result concerns the FSF and SSF routing policies, which can be obtained in the limit when $r\to\infty$ and $r\to-\infty$ respectively. Recall that FSF is asymptotically optimal in the nonstrategic setting. Intuitively, however, it penalizes the servers that work the fastest by sending them more and more jobs. In a strategic setting, this might incentivize servers to decrease their service rate, which is not good for the performance of the system. One may wonder if by doing the opposite, that is, using the SSF policy, servers can be incentivized to increase their service rate. However, \red{our next theorem (Theorem~\ref{theorem:FSF-noeq})} shows that neither of these policies is useful if we are interested in symmetric equilibria.

Recall that our model for strategic servers already assumes an increasing, convex effort cost function with $c'''(\mu)\geq 0$. For the rest of this section, in addition, we assume that $c'(\frac{\lambda}{2})<\frac{1}{\lambda}$. (Recall that this is identical to the sufficient condition $c'(\frac{\lambda}{N})<\frac{1}{\lambda}$ which we introduced in Section~\ref{Section:Results:MMN}, on substituting $N=2$.)\footnote{The sufficient condition $c'(\frac{\lambda}{2})<\frac{1}{\lambda}$ might seem rather strong, but it can be shown that it is necessary for the symmetric first order condition to have a unique solution. This is because, if $c'(\frac{\lambda}{2})>\frac{1}{\lambda}$, then the function $\varphi(\mu)$, defined in~(\ref{eq:varphi}), ceases to be monotonic, and as a result, for any given $r$, the first order condition $\varphi(\mu)=r$ could have more than one solution.}

\begin{theorem}\label{theorem:FSF-noeq}
Consider \red{a} $M$/$M$/$2$ queue with strategic servers. Then, FSF and SSF do not admit a symmetric equilibrium.
\end{theorem}

Moving beyond FSF and SSF, we continue our equilibrium analysis (for a finite $r$) by using the first order conditions to show that whenever an $r$-routing policy admits a symmetric equilibrium, it is unique. Furthermore, we provide an expression for the corresponding symmetric equilibrium service rate in terms of $r$, which brings out a useful monotonicity property.
\begin{theorem}\label{theorem:unique-symeq}
Consider \red{a} $M$/$M$/$2$ queue with strategic servers. Then, any $r$-routing policy that admits a symmetric equilibrium, admits a unique symmetric equilibrium, given by $\mu^\star=\varphi^{-1}(r)$, where $\varphi\colon(\frac{\lambda}{2},\infty)\to\mathbb{R}$ is the function defined by
\begin{equation}\label{eq:varphi}
\varphi(\mu)=\frac{4(\lambda+\mu)}{\lambda(\lambda-2\mu)}\left(\mu(\lambda+2\mu)c'(\mu)-\lambda\right).
\end{equation}
Furthermore, among all such policies, $\mu^\star$ is decreasing in $r$, and therefore, $\mathbb{E}[T]$, the mean response time \red{(a.k.a.\ sojourn time)} at symmetric equilibrium is increasing in $r$.
\end{theorem}
In light of the inverse relationship between $r$ and $\mu^\star$ that is established by this theorem, the system manager would ideally choose the smallest $r$ such that the corresponding $r$-routing policy admits a symmetric equilibrium, which is in line with the intuition that a bias towards SSF (the limiting $r$-routing policy as $r\to-\infty$) incentivizes servers to work harder. However, there is a hard limit on how small an $r$ can be chosen (concurrently, how large an equilibrium service rate $\mu^\star$ can be achieved) so that there exists a symmetric equilibrium, as evidenced by our next theorem.

\begin{theorem}\label{theorem:boundedinterval}
Consider \red{a} $M$/$M$/$2$ queue with strategic servers. Then, there exists $\overline{\mu},\underline{r}\in\mathbb{R}$, with $\underline{r}=\varphi(\overline{\mu})$, such that no service rate $\mu>\overline{\mu}$ can be a symmetric equilibrium under any $r$-routing policy, and no $r$-routing policy with $r<\underline{r}$ admits a symmetric equilibrium.
\end{theorem}

The proof of this theorem is constructive and we do exhibit an $\underline{r}$, however, it is not clear whether this is tight, that is, whether there exists a symmetric equilibrium for all $r$-routing policies with $r\geq\underline{r}$. We provide a partial answer to this question of what $r$-routing policies do admit symmetric equilibria in the following theorem.

\begin{theorem}\label{theorem:guarantee-symeq}
Consider \red{a} $M$/$M$/$2$ queue with strategic servers. Then, there exists a unique symmetric equilibrium under any $r$-routing policy with $r\in\{-2,-1,0,1\}$.
\end{theorem}

\begin{figure}
\centering
    \input{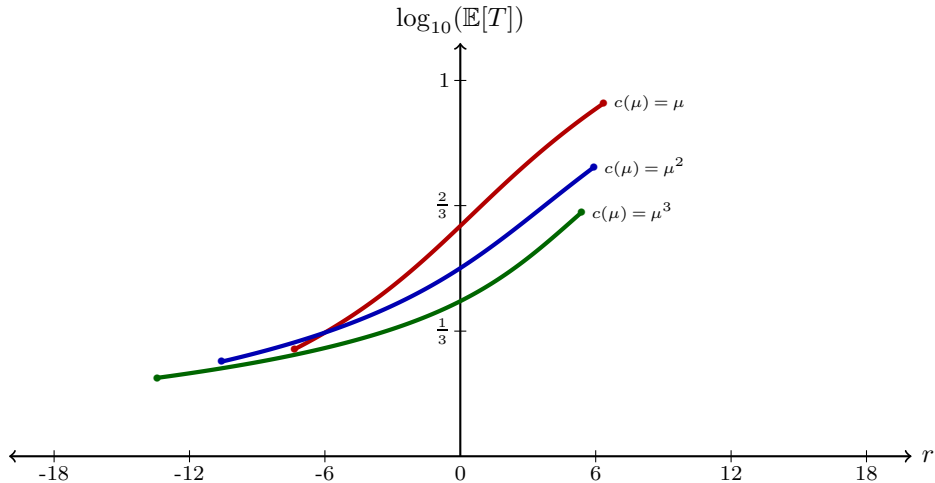}
    \caption{\red{Equilibrium mean response time (a.k.a.\ sojourn time) as a function of the policy parameter, $r$, when the arrival rate is $\lambda=\frac{1}{4}$, for three different effort cost functions: linear, quadratic, and cubic.}\label{fig:mama}}
\end{figure}

Notice that we show equilibrium existence for four integral values of $r$. It is challenging to show that all $r$-routing policies in the interval $[-2,1]$ admit a symmetric equilibrium. This theorem provides an upper bound on the $\underline{r}$ of the previous theorem, that is, $\underline{r}\leq-2$. Therefore, if the specific cost function $c$ is unknown, then the system manager can guarantee better performance than Random ($r=0$), by setting $r=-2$. If the specific cost function is known, the system manager may be able to employ a lower $r$ to obtain even better performance. \red{For example, consider a 2-server system with $\lambda=1/4$ and one of three different effort cost functions: $c(\mu)=\mu$, $c(\mu)=\mu^2$, and $c(\mu)=\mu^3$. Figure~\ref{fig:mama} shows the corresponding equilibrium mean response times (in red, blue, and green, respectively). It is worth noting that the more convex the effort cost function, larger the range of $r$ (and smaller the minimum value of $r$) for which a symmetric equilibrium exists.}

\section{Concluding remarks.}\label{Section:Conclusion}
The rate at which each server works in a service system has important consequences for service system design.  However, traditional models of large service systems do not capture the fact that human servers respond to incentives created by scheduling and staffing policies, because traditional models assume each server works at a given fixed service rate.  In this paper, we initiate the study of a class of strategic servers that seek to optimize a utility function which values idle time and includes an effort cost.

Our focus is on the analysis of staffing and routing policies for \red{a} $M$/$M$/$N$ queue with strategic servers, and our results highlight that strategic servers have a dramatic impact on the optimal policies in both cases.  In particular, policies that are optimal in the classical, nonstrategic setting can perform quite poorly when servers act strategically.

For example, a consequence of the strategic server behavior is that the cost-minimizing staffing level is order $\lambda$ larger than square-root staffing, the cost minimizing staffing level for systems with fixed service rate.  \red{In particular, any system with strategic servers operates in the quality-driven regime at equilibrium (as opposed to the quality-and-efficiency-driven regime that arises under square-root staffing), in which the servers all enjoy non-negligible idle time.}

\red{The intuitive reason square-root staffing is not feasible in the context of strategic servers is that the servers do not value their idleness enough in comparison to their effort cost.  This causes the servers to work too slowly, making idle time scarce. } In the economics literature~\cite{Brock68,Lynn91}, it is common to assume that scarce goods are more highly valued.  If we assume that the servers valued their idle time more heavily as the idle time becomes scarcer, then the servers would work faster in order to make sure they achieved some.  This suggests the following interesting direction for future research:  what is the relationship between the assumed value of idle time in~(\ref{Equation:ServerUtility}) and the resulting cost minimizing staffing policy? Another situation in which servers may not care about idle time becoming scarce is when their compensation depends on their service volume (which is increasing in their service rate).  Then, it is reasonable to expect the servers prefer to have negligible idle time.  It would be interesting to be able to identify a class of compensation schemes under which that is the case.

The aforementioned two future research directions become even more interesting when the class of routing policies is expanded to include rate-based policies.  This paper solves the \emph{joint} routing and staffing problem within the class of idle-time-order-based policies.  Section~\ref{Section:Scheduling} suggests that by expanding the class of routing policies to also include rate-based policies we should be able to achieve better system performance (although it is clear that the analysis becomes much more difficult).  The richer question also aspires to understand the relationship between the server idle time value, the compensation scheme, the (potentially) rate-based routing policy, and the number of strategic servers to staff.

Finally, it is important to note that we have focused on symmetric equilibrium service rates.  We have not proven that asymmetric equilibria do not exist. Thus, it is natural to wonder if there are routing and staffing policies that result in an asymmetric equilibrium.  Potentially, there could be one group of servers that have low effort costs but negligible idle time and another group of servers that enjoy plentiful idle time but have high effort costs.  The question of asymmetric equilibria becomes even more interesting when the servers have different utility functions.  For example, more experienced servers likely have lower effort costs than new hires.  Also, different servers can value their idle time differently.  How do we design routing and staffing policies that are respectful of such considerations? 


\ACKNOWLEDGMENT{This work was supported by NSF grant \#CCF-1101470, AFOSR grant \#FA9550-12-1-0359, and ONR grant \#N00014-09-1-0751. An important source of inspiration for this work came from discussions with Mor Armony regarding how to fairly assign workload to employees. We would also like to thank Bert Zwart for an illuminating discussion regarding asymptotically optimal staffing policies.}


\bibliographystyle{ormsv080} 
\bibliography{OR_server_game} 

\begin{thebibliography}{43}
\expandafter\ifx\csname natexlab\endcsname\relax\def\natexlab#1{#1}\fi
\expandafter\ifx\csname url\endcsname\relax
  \def\url#1{{\tt #1}}\fi
\expandafter\ifx\csname urlprefix\endcsname\relax\def\urlprefix{URL }\fi
\expandafter\ifx\csname urlstyle\endcsname\relax
  \expandafter\ifx\csname doi\endcsname\relax
  \def\doi#1{doi:\discretionary{}{}{}#1}\fi \else
  \expandafter\ifx\csname doi\endcsname\relax
  \def\doi{doi:\discretionary{}{}{}\begingroup \urlstyle{rm}\Url}\fi \fi

\bibitem[{Aksin et~al.(2007)Aksin, Armony, and
  Mehrotra}]{AksinArmonyMehrotra07}
Aksin, Z., M.~Armony, V.~Mehrotra. 2007.
\newblock The modern call-center: {A} multi-disciplinary perspective on
  operations management research.
\newblock {\it {Prod. Oper. Manag.}\/} {\bf 16}(6) 665--688.

\bibitem[{Allon and Gurvich(2010)}]{Allon2010}
Allon, G., I.~Gurvich. 2010.
\newblock Pricing and dimensioning competing large-scale service providers.
\newblock {\it {M\&SOM}\/} {\bf 12}(3) 449--469.

\bibitem[{Anton(2005)}]{compensationSurvey2005}
Anton, J. 2005.
\newblock One-minute survey report \#488: \textup{A}gent compensation \&
  advancement.
\newblock Document Tracking Number SRV488-080305.

\bibitem[{Armony(2005)}]{Armony05}
Armony, M. 2005.
\newblock Dynamic routing in large-scale service systems with heterogeneous
  servers.
\newblock {\it {Queueing Syst. Theory Appl.}\/} {\bf 51}(3-4) 287--329.

\bibitem[{Armony and Ward(2010)}]{ArmonyWard10}
Armony, M., A.~R. Ward. 2010.
\newblock Fair dynamic routing in large-scale heterogeneous-server systems.
\newblock {\it {Oper. Res.}\/} {\bf 58} 624--637.

\bibitem[{Atar(2005)}]{Atar2005}
Atar, R. 2005.
\newblock Scheduling control for queueing systems with many servers:
  \textup{A}symptotic optimality in heavy traffic.
\newblock {\it {Ann. Appl. Probab.}\/} {\bf 15}(4) 2606--2650.

\bibitem[{Atar et~al.(2011)Atar, Shaki, and Shwartz}]{Atar2011}
Atar, R., Y.~Y. Shaki, A.~Shwartz. 2011.
\newblock A blind policy for equalizing cumulative idleness.
\newblock {\it Queueing Syst.\/} {\bf 67}(4) 275--293.

\bibitem[{Borst et~al.(2004)Borst, Mandelbaum, and
  Reiman}]{BorstMandelbaumReiman04}
Borst, S., A.~Mandelbaum, M.~I. Reiman. 2004.
\newblock Dimensioning large call centers.
\newblock {\it {Oper. Res.}\/} {\bf 52}(1) 17--34.

\bibitem[{Brock(1968)}]{Brock68}
Brock, T.~C. 1968.
\newblock Implications of commodity theory for value change.
\newblock {\it Psychological Foundations of Attitudes\/}  243--275.

\bibitem[{Cachon and Harker(2002)}]{CachonHarker2002}
Cachon, G.~P., P.~T. Harker. 2002.
\newblock Competition and outsourcing with scale economies.
\newblock {\it Manage. Sci.\/} {\bf 48}(10) 1314--1333.

\bibitem[{Cachon and Zhang(2007)}]{Cachon2007}
Cachon, G.~P., F.~Zhang. 2007.
\newblock Obtaining fast service in a queueing system via performance-based
  allocation of demand.
\newblock {\it Manage. Sci.\/} {\bf 53}(3) 408--420.

\bibitem[{Cahuc and Zylberberg(2004)}]{Cahuc2004}
Cahuc, P., A.~Zylberberg. 2004.
\newblock {\it {Labor Economics}\/}.
\newblock MIT Press.

\bibitem[{Cheng et~al.(2004)Cheng, Reeves, Vorobeychik, and
  Wellman}]{Cheng2004}
Cheng, S.-F., D.~M. Reeves, Y.~Vorobeychik, M.~P. Wellman. 2004.
\newblock Notes on equilibria in symmetric games.
\newblock {\it International Workshop On Game Theoretic And Decision Theoretic
  Agents (GTDT)\/}. 71--78.

\bibitem[{Cohen-Charash and Spector(2001)}]{Cohen-CharashSpector01}
Cohen-Charash, Y., P.~E. Spector. 2001.
\newblock {The role of justice in organizations: {A} meta-analysis}.
\newblock {\it {Organ. Behav. and Hum. Dec.}\/} {\bf 86}(2) 278--321.

\bibitem[{Colquitt et~al.(2001)Colquitt, Conlon, Wesson, Porter, and
  Ng}]{ColquittConlonWessonChristopher01}
Colquitt, J.~A., D.~E. Conlon, M.~J. Wesson, C.~O. L.~H. Porter, K.~Y. Ng.
  2001.
\newblock {Justice at the millennium: A meta-analytic review of 25 years of
  organizational justice research}.
\newblock {\it {J. Appl. Psychol.}\/} {\bf 86}(3) 425--445.

\bibitem[{de~V\'{e}ricourt and Zhou(2005)}]{deVericourtZhou05}
de~V\'{e}ricourt, F., Y.-P. Zhou. 2005.
\newblock Managing response time in a call-routing problem with service
  failure.
\newblock {\it {Oper. Res.}\/} {\bf 53}(6) 968--981.

\bibitem[{Erlang(1948)}]{Erlang1948}
Erlang, A.~K. 1948.
\newblock On the rational determination of the number of circuits.
\newblock E.~Brockmeyer, H.~L. Halstrom, A.~Jensen, eds., {\it {The Life and
  Works of A.~K. Erlang}\/}. {The Copenhagen Telephone Company}, 216--221.

\bibitem[{Gans et~al.(2003)Gans, Koole, and
  Mandelbaum}]{GansKooleMandelbaum2003}
Gans, N., G.~Koole, A.~Mandelbaum. 2003.
\newblock Telephone call centers: \textup{T}utorial, review, and research
  prospects.
\newblock {\it {M\&SOM}\/} {\bf 5}(2) 79--141.

\bibitem[{Garnett et~al.(2002)Garnett, Mandelbaum, and
  Reiman}]{GarnettMandelbaumReiman2002}
Garnett, O., A.~Mandelbaum, M.~Reiman. 2002.
\newblock Designing a call center with impatient customers.
\newblock {\it {M\&SOM}\/} {\bf 4}(3) 208--227.

\bibitem[{Geng et~al.(2013)Geng, Huh, and Nagarajan}]{GengHuhNag2013}
Geng, X., W.~T. Huh, M.~Nagarajan. 2013.
\newblock Strategic and fair routing policies in a decentralized service
  system.
\newblock Working paper.

\bibitem[{Gilbert and Weng(1998)}]{Gilbert98}
Gilbert, S.~M., Z.~K. Weng. 1998.
\newblock Incentive effects favor nonconsolidating queues in a service system:
  \textup{T}he principal-agent perspective.
\newblock {\it Manage. Sci.\/} {\bf 44}(12) 1662--1669.

\bibitem[{Gumbel(1960)}]{Gumbel60}
Gumbel, H. 1960.
\newblock Waiting lines with heterogeneous servers.
\newblock {\it {Oper. Res.}\/} {\bf 8}(4) 504--511.

\bibitem[{Gurvich and Whitt(2007)}]{GurvichWhitt2007}
Gurvich, I., W.~Whitt. 2007.
\newblock Scheduling flexible servers with convex delay costs in many-server
  service systems.
\newblock {\it {M\&SOM}\/} {\bf 11}(2) 237--253.

\bibitem[{Haji and Ross(2013)}]{HajiRoss13}
Haji, B., S.~M. Ross. 2013.
\newblock A queueing loss model with heterogenous skill based servers under
  idle time ordering policies.
\newblock Working paper.

\bibitem[{Halfin and Whitt(1981)}]{HalfinWhitt1981}
Halfin, S., W.~Whitt. 1981.
\newblock Heavy-traffic limits for queues with many exponential servers.
\newblock {\it {Oper. Res.}\/} {\bf 29}(3) 567--588.

\bibitem[{Harchol-Balter(2013)}]{Harchol2013}
Harchol-Balter, M. 2013.
\newblock {\it {Performance Modeling and Design of Computer Systems: Queueing
  Theory in Action}\/}.
\newblock Cambridge University Press.

\bibitem[{Harel(1988)}]{Harel1988}
Harel, A. 1988.
\newblock Sharp bounds and simple approximations for the {E}rlang delay and
  loss formulas.
\newblock {\it {Manage. Sci.}\/} {\bf 34}(8) 959--972.

\bibitem[{Hassin and Haviv(2003)}]{HassinHaviv03}
Hassin, R., M.~Haviv. 2003.
\newblock {\it {To Queue or Not to Queue: Equilibrium Behavior in Queueing
  Systems}\/}.
\newblock Kluwer.

\bibitem[{Hopp and Lovejoy(2013)}]{HoppLovejoy2013}
Hopp, W., W.~Lovejoy. 2013.
\newblock {\it {Hospital Operations: Principles of High Efficiency Health
  Care}\/}.
\newblock Financial Times Press.

\bibitem[{Janssen et~al.(2011)Janssen, van \textup{L}eeuwaarden, and
  Zwart}]{JaLeZw11}
Janssen, A.~J.~E.~M., J.~S.H. van \textup{L}eeuwaarden, B.~Zwart. 2011.
\newblock Refining square-root safety staffing by expanding \textup{E}rlang
  \textup{C}.
\newblock {\it Oper. Res.\/} {\bf 59}(6) 1512--1522.

\bibitem[{Kalai et~al.(1992)Kalai, Kamien, and Rubinovitch}]{Kalai92}
Kalai, E., M.~I. Kamien, M.~Rubinovitch. 1992.
\newblock Optimal service speeds in a competitive environment.
\newblock {\it Manage. Sci.\/} {\bf 38}(8) 1154--1163.

\bibitem[{Kc and Terwiesch(2009)}]{KcTer2009}
Kc, D.~S., C.~Terwiesch. 2009.
\newblock Impact of workload on service time and patient safety: \textup{A}n
  econometric analysis of hospital operations.
\newblock {\it {Manage. Sci.}\/} {\bf 55}(9) 1486--1498.

\bibitem[{Kocaga et~al.(2013)Kocaga, Armony, and Ward}]{KoArWa13}
Kocaga, L., M.~Armony, A.~R. Ward. 2013.
\newblock Staffing call centers with uncertain arrival rates and co-sourcing.
\newblock Working paper.

\bibitem[{Krishnamoorthi(1963)}]{Krishnamoorthi1963}
Krishnamoorthi, B. 1963.
\newblock On \textup{P}oisson queue with two heterogeneous servers.
\newblock {\it {Oper. Res.}\/} {\bf 11}(3) 321--330.

\bibitem[{Lin and Kumar(1984)}]{LinKumar84}
Lin, W., P.~Kumar. 1984.
\newblock Optimal control of a queueing system with two heterogeneous servers.
\newblock {\it {IEEE Trans. Autom. Contr.}\/} {\bf 29}(8) 696--703.

\bibitem[{Lynn(1991)}]{Lynn91}
Lynn, M. 1991.
\newblock Scarcity effects on value: \textup{A} quantitative review of the
  commodity theory literature.
\newblock {\it Psychol. Market.\/} {\bf 8}(1) 43--57.

\bibitem[{Mokaddis et~al.(1998)Mokaddis, Matta, and El~Genaidy}]{Mokaddis98}
Mokaddis, G.~S., C.~H. Matta, M.~M. El~Genaidy. 1998.
\newblock On \textup{P}oisson queue with three heterogeneous servers.
\newblock {\it International Journal of Information and Management Sciences\/}
  {\bf 9} 53--60.

\bibitem[{Reed and Shaki(2013)}]{Reed2012}
Reed, J., Y.~Shaki. 2013.
\newblock A fair policy for the \textup{G/GI/N} queue with multiple server
  pools.
\newblock Preprint.

\bibitem[{Saaty(1960)}]{Saaty1960}
Saaty, T.~L. 1960.
\newblock Time-dependent solution of the many-merver \textup{P}oisson queue.
\newblock {\it {Oper. Res.}\/} {\bf 8}(6) 755--772.

\bibitem[{Tezcan(2008)}]{Tezcan2008}
Tezcan, T. 2008.
\newblock Optimal control of distributed parallel server systems under the
  \textup{H}alfin and \textup{W}hitt regime.
\newblock {\it {Math. Oper. Res.}\/} {\bf 33} 51--90.

\bibitem[{Tezcan and Dai(2010)}]{TezcanDai2010}
Tezcan, T., J.~Dai. 2010.
\newblock Dynamic control of \textup{N}-systems with many servers:
  \textup{A}symptotic optimality of a static priority policy in heavy traffic.
\newblock {\it {Oper. Res.}\/} {\bf 58}(1) 94--110.

\bibitem[{Ward and Armony(2013)}]{WardArmony13}
Ward, A.~R., M.~Armony. 2013.
\newblock Blind fair routing in large-scale service systems with heterogeneous
  customers and servers.
\newblock {\it {Oper. Res.}\/} {\bf 61} 228--243.

\bibitem[{Whitt(2002)}]{WhittNotes}
Whitt, W. 2002.
\newblock \textup{IEOR} 6707: \textup{A}dvanced \textup{T}opics in
  \textup{Q}ueueing \textup{T}heory: \textup{F}ocus on \textup{C}ustomer
  \textup{C}ontact \textup{C}enters.
\newblock Homework 1e Solutions, see
  \url{http://www.columbia.edu/~ww2040/ErlangBandCFormulas.pdf}.

\end{thebibliography}


\ECSwitch


\ECHead{Routing and Staffing when Servers are Strategic:  Technical Appendix}

\noindent Ragavendran Gopalakrishnan, Sherwin Doroudi, Amy R. Ward, and Adam Wierman

\vspace{0.25in}

In this technical appendix, we provide proofs for the results stated in the main body of the manuscript titled:  ``Routing and Staffing when Servers are Strategic''.  The proofs of these results are in the order in which they appear in the main body.

\subsection*{PROOFS FROM SECTION~\ref{Section:Model:MMN}}

\subsubsection*{Proof of Theorem~\ref{theorem:MMN-IDLE}.}

The starting point of this proof is the expression for the steady state probabilities of a general heterogeneous $M$/$M$/$N$ system with Random routing, which was derived in~\cite{Gumbel60}. Before stating this more general result, we first set up the required notation. Let $\mu_1,\mu_2,\ldots,\mu_N$ denote the service rates of the $N$ servers, and let $\rho_j=\frac{\lambda}{\mu_j}$, $1\leq j\leq N$. We assume that $\sum_{j=1}^N\rho_j^{-1}> 1$ for stability. Let $(a_1,a_2,\ldots,a_k)$ denote the state of the system when there are $k$ jobs in the system ($0<k<N$) and the busy servers are $\{a_1,a_2,\ldots,a_k\}$, where $1\leq a_1<a_2<\cdots<a_k\leq N$. Let $P(a_1,a_2,\ldots,a_k)$ denote the steady state probability of the system being in state $(a_1,a_2,\ldots,a_k)$. Also, let $P_k$ denote the steady state probability of $k$ jobs in the system. Then,
\begin{equation}\label{eq:steady_state_gen}
P(a_1,a_2,\ldots,a_k) = \frac{(N-k)!\ P_0\ \rho_{a_1}\rho_{a_2}\cdots\rho_{a_k}}{N!},
\end{equation}
where $P_0$, the steady state probability that the system is empty, is given by:
\begin{equation}\label{eq:P0_gen}
P_0 = \frac{N!\ C_N^N}{D_N},
\end{equation}
where, for $1\leq j\leq N$,
\begin{equation}\label{eq:Cjn_gen}
\begin{split}
C_j^N &= \text{sum of combinations of }j\ \rho_i^{-1}\text{ values from }N\ \rho_i^{-1}\text{ values}\\
&=\sum_{a_1=1}^{N-j+1}\sum_{a_2=a_1+1}^{N-j+2}\cdots\sum_{a_{j-1}=a_{j-2}+1}^{N-j+j-1}\sum_{a_j=a_{j-1}+1}^{N}\rho_{a_1}^{-1}\rho_{a_2}^{-1}\cdots\rho_{a_j}^{-1},
\end{split}
\end{equation}
and
\begin{equation}\label{eq:Dn_gen}
D_N = \sum_{j=1}^N j!\ C_j^N + \frac{C_1^N}{C_1^N-1}.
\end{equation}
Note that,
\begin{equation*}
C_N^N = \prod_{i=1}^N \rho_i^{-1}\qquad\text{and}\qquad C_1^N = \sum_{i=1}^N \rho_i^{-1}.
\end{equation*}
Also, by convention, we write $C_0^N=1$. The steady state probability that a tagged server, say server 1, is idle is obtained by summing up the steady state probabilities of every state in which server 1 is idle:
\begin{equation}\label{eq:idle_gen}
I(\mu_1,\mu_2,\ldots,\mu_N;\lambda,N) = P_0 + \sum_{k=1}^{N-1}\sum_{2\leq a_1\leq\cdots\leq a_k\leq N}P(a_1,a_2,\ldots,a_k)
\end{equation}

We now simplify the expressions above for our special system where the tagged server works at a rate $\mu_1$ and all other servers work at rate $\mu$. Without loss of generality, we pick server 1 to be the tagged server, and we set $\mu_2=\mu_3=\cdots=\mu_N=\mu$, and therefore, $\rho_2=\rho_3=\cdots=\rho_N=\rho=\frac{\lambda}{\mu}$. Then,~(\ref{eq:steady_state_gen}) simplifies to:
\begin{equation}\label{eq:steady_state_sym}
P(a_1,a_2,\ldots,a_k) = \frac{(N-k)!\ P_0\ \rho^k}{N!}, 2\leq a_1\leq \cdots\leq a_k\leq N
\end{equation}
In order to simplify~(\ref{eq:Cjn_gen}), we observe that
\begin{equation*}
C_j^N = \rho_1^{-1} C_{j-1}^{N-1} + C_j^{N-1}
\end{equation*}
where the terms $C_{j-1}^{N-1}$ and $C_j^{N-1}$ are obtained by applying~(\ref{eq:Cjn_gen}) to a homogeneous $M$/$M$/$(N-1)$ system with arrival rate $\lambda$ and all servers operating at rate $\mu$. This results in:
\begin{equation}\label{eq:Cjn_sym}
C_j^N = \frac{\rho}{N\rho_1}\left(j \binom{N}{j} \rho^{-j}\right) + \frac{1}{N}\left((N-j) \binom{N}{j} \rho^{-j}\right)
\end{equation}
The corresponding special cases are given by: $C_0^N=1$, $C_1^N=\rho_1^{-1}+(N-1)\rho$, and $C_N^N=\frac{\rho}{\rho_1}\rho^{-N}$. We then simplify~(\ref{eq:Dn_gen}) by substituting for $C_j^N$ from~(\ref{eq:Cjn_sym}), to obtain:
\begin{equation}\label{eq:Dn_sym}
\begin{split}
D_N &= \left(\frac{N!}{\rho^N}\left(\frac{\rho}{\rho_1}+\frac{\rho}{N}\left(1-\frac{\rho}{\rho_1}\right)\right)\sum_{j=0}^{N-1}\frac{\rho^j}{j!}+\frac{\rho}{\rho_1}-1\right) + \left(1+\frac{1}{C_1^N-1}\right)\\
&= \frac{\rho}{\rho_1}\left(\frac{N!}{\rho^N}\left(1-\frac{\rho}{N}\left(1-\frac{\rho_1}{\rho}\right)\right)\sum_{j=0}^{N-1}\frac{\rho^j}{j!} + 1 + \frac{\rho_1}{\rho}\frac{\rho}{N-\left(\rho+1-\frac{\rho}{\rho_1}\right)}\right)\\
\end{split}
\end{equation}
Next, we simplify~(\ref{eq:P0_gen}) by substituting for $D_N$ from~(\ref{eq:Dn_sym}), to obtain:
\begin{equation*}
P_0 = \left(\left(1-\frac{\rho}{N}\left(1-\frac{\rho_1}{\rho}\right)\right)\sum_{j=0}^{N-1}\frac{\rho^j}{j!} + \frac{\rho^N}{N!}\left(1 + \frac{\rho_1}{N-\left(\rho+1-\frac{\rho}{\rho_1}\right)}\right)\right)^{-1}
\end{equation*}
To express $P_0$ in terms of $ErlC(N,\rho)$, the Erlang C formula, we add and subtract the term $\frac{N}{N-\rho}\frac{\rho^N}{N!}$ within, to obtain:

\begin{footnotesize}
\begin{equation*}
P_0 = \left(\left(1-\frac{\rho}{N}\left(1-\frac{\rho_1}{\rho}\right)\right)\left(\sum_{j=0}^{N-1}\frac{\rho^j}{j!} + \frac{N}{N-\rho}\frac{\rho^N}{N!}\right) + \frac{\rho^N}{N!}\left(1 + \frac{\rho_1}{N-\left(\rho+1-\frac{\rho}{\rho_1}\right)} - \frac{N}{N-\rho}\left(1 - \frac{\rho}{N}\left(1-\frac{\rho_1}{\rho}\right)\right)\right)\right)^{-1}
\end{equation*}
\end{footnotesize}

\noindent which reduces to:
\begin{equation}\label{eq:P0_sym}
\begin{split}
P_0 &= \left(\left(1-\frac{\rho}{N}\left(1-\frac{\rho_1}{\rho}\right)\right)\left(\sum_{j=0}^{N-1}\frac{\rho^j}{j!} + \frac{N}{N-\rho}\frac{\rho^N}{N!}\right) - \frac{\rho}{N}\left(1-\frac{\rho_1}{\rho}\right)\frac{\frac{N}{N-\rho}\frac{\rho^N}{N!}}{N-\left(\rho+1-\frac{\rho}{\rho_1}\right)}\right)^{-1}\\
&= \left(\sum_{j=0}^{N-1}\frac{\rho^j}{j!} + \frac{N}{N-\rho}\frac{\rho^N}{N!}\right)^{-1} \left(1-\frac{\rho}{N}\left(1-\frac{\rho_1}{\rho}\right)\left(1+\frac{ErlC(N,\rho)}{N-\left(\rho+1-\frac{\rho}{\rho_1}\right)}\right)\right)^{-1}
\end{split}
\end{equation}
Finally,~(\ref{eq:idle_gen}) simplifies to:
\begin{equation*}
I(\mu_1,\mu,\mu,\ldots,\mu;\lambda,N) = P_0 + \sum_{k=1}^{N-1}\binom{N-1}{k}P(2,3,\ldots,k+1)
\end{equation*}
Substituting for $P_0$ from~(\ref{eq:P0_sym}) and $P(2,3,\ldots,k+1)$ from~(\ref{eq:steady_state_sym}), we get:
\begin{equation*}
\begin{split}
I(\mu_1,\mu;\lambda,N) &= P_0 + \sum_{k=1}^{N-1}\binom{N-1}{k}\left(\frac{(N-k)!\ P_0\ \rho^k}{N!}\right)\\
&= \left(1-\frac{\rho}{N}\right)\left(\sum_{k=0}^{N-1}\frac{\rho^k}{k!}+\frac{N}{N-\rho}\frac{\rho^N}{N!}\right)P_0\\
&= \left(1-\frac{\rho}{N}\right)\left(1-\frac{\rho}{N}\left(1-\frac{\rho_1}{\rho}\right)\left(1+\frac{ErlC(N,\rho)}{N-\left(\rho+1-\frac{\rho}{\rho_1}\right)}\right)\right)^{-1}\\
&= \left(1-\frac{\rho}{N}\right)\left(1-\frac{\rho}{N}\left(1-\frac{\mu}{\mu_1}\right)\left(1+\frac{ErlC(N,\rho)}{N-\left(\rho+1-\frac{\mu_1}{\mu}\right)}\right)\right)^{-1},
\end{split}
\end{equation*}
as desired.
\eProof

\subsubsection*{Proof of Theorem~\ref{theorem:MMN-DIDLE}.}

We start with the expression for $I$ from~(\ref{eq:MMN-IDLE}), and take its first partial derivative with respect to $\mu_1$:

\begin{scriptsize}
\begin{equation*}
\begin{split}
\frac{\partial I}{\partial \mu_1} &= -\left(1-\frac{\rho}{N}\right)\left(1-\frac{\rho}{N}\left(1-\frac{\mu}{\mu_1}\right)\left(1+\frac{ErlC(N,\rho)}{N-\left(\rho+1-\frac{\mu_1}{\mu}\right)}\right)\right)^{-2}\frac{\partial}{\partial\mu_1}\left(1-\frac{\rho}{N}\left(1-\frac{\mu}{\mu_1}\right)\left(1+\frac{ErlC(N,\rho)}{N-\left(\rho+1-\frac{\mu_1}{\mu}\right)}\right)\right)\\
&= -\frac{N}{N-\rho}I^2\frac{\partial}{\partial\mu_1}\left(1-\frac{\rho}{N}\left(1-\frac{\mu}{\mu_1}\right)\left(1+\frac{ErlC(N,\rho)}{N-\left(\rho+1-\frac{\mu_1}{\mu}\right)}\right)\right) = \frac{\rho}{N-\rho}I^2\frac{\partial}{\partial\mu_1}\left(\left(1-\frac{\mu}{\mu_1}\right)\left(1+\frac{ErlC(N,\rho)}{N-\left(\rho+1-\frac{\mu_1}{\mu}\right)}\right)\right)
\end{split}
\end{equation*}
\end{scriptsize}

\noindent Applying the product rule, and simplifying the expression, we get~(\ref{eq:MMN-FPD}). Next, for convenience, we rewrite~(\ref{eq:MMN-FPD}) as:

\begin{scriptsize}
\begin{equation}\label{eq:MMN-SPD-INTER1}
\frac{N-\rho}{\lambda}\frac{\partial I}{\partial \mu_1} = \frac{I^2}{\mu_1^2}\left(1+\frac{ErlC(N,\rho)}{N-\left(\rho+1-\frac{\mu_1}{\mu}\right)}+\left(1-\frac{\mu_1}{\mu}\right)\frac{\mu_1}{\mu}\frac{ErlC(N,\rho)}{\left(N-\left(\rho+1-\frac{\mu_1}{\mu}\right)\right)^2}\right)
\end{equation}
\end{scriptsize}

\noindent Differentiating this equation once more with respect to $\mu_1$ by applying the product rule, we get:

\begin{scriptsize}
\begin{equation*}
\begin{split}
\frac{N-\rho}{\lambda}\frac{\partial^2I}{\partial\mu_1^2} &= \left(\frac{2I}{\mu_1^2}\frac{\partial I}{\partial\mu_1} - \frac{2I^2}{\mu_1^3}\right) \left(1+\frac{ErlC(N,\rho)}{N-\left(\rho+1-\frac{\mu_1}{\mu}\right)}+\left(1-\frac{\mu_1}{\mu}\right)\frac{\mu_1}{\mu}\frac{ErlC(N,\rho)}{\left(N-\left(\rho+1-\frac{\mu_1}{\mu}\right)\right)^2}\right)\\
&\qquad\qquad\ + \frac{I^2}{\mu_1^2}\frac{\partial}{\partial\mu_1} \left(1+\frac{ErlC(N,\rho)}{N-\left(\rho+1-\frac{\mu_1}{\mu}\right)}+\left(1-\frac{\mu_1}{\mu}\right)\frac{\mu_1}{\mu}\frac{ErlC(N,\rho)}{\left(N-\left(\rho+1-\frac{\mu_1}{\mu}\right)\right)^2}\right)\\
&= \left(\frac{2I}{\mu_1^2}\frac{\partial I}{\partial\mu_1} - \frac{2I^2}{\mu_1^3}\right)\frac{\mu_1^2}{I^2}\frac{N-\rho}{\lambda}\frac{\partial I}{\partial\mu_1} + \frac{I^2}{\mu_1^2}\frac{\partial}{\partial\mu_1}
\left(1+\frac{ErlC(N,\rho)}{N-\left(\rho+1-\frac{\mu_1}{\mu}\right)}+\left(1-\frac{\mu_1}{\mu}\right)\frac{\mu_1}{\mu}\frac{ErlC(N,\rho)}{\left(N-\left(\rho+1-\frac{\mu_1}{\mu}\right)\right)^2}\right)
\end{split}
\end{equation*}
\end{scriptsize}

\noindent Applying the product rule for the second term, and simplifying the expression, we get:

\begin{scriptsize}
\begin{equation*}
\frac{\partial^2I}{\partial\mu_1^2} = \frac{2}{I}\left(\frac{\partial I}{\partial\mu_1}\right)^2 - \frac{2}{\mu_1}\left(\frac{\partial I}{\partial\mu_1}\right) - \frac{2I^2}{\mu_1\mu^2}\frac{\lambda}{N-\rho}\frac{ErlC(N,\rho)}{\left(N-\left(\rho+1-\frac{\mu_1}{\mu}\right)\right)^2}\left(1+\left(1-\frac{\mu_1}{\mu}\right)\frac{1}{N-\left(\rho+1-\frac{\mu_1}{\mu}\right)}\right)
\end{equation*}
\end{scriptsize}

\noindent The expression in~(\ref{eq:MMN-SPD}) is then obtained by substituting for $\frac{\partial I}{\partial\mu_1}$ from~(\ref{eq:MMN-FPD}), and carefully going through some incredibly messy (but straightforward) algebra.
\eProof

\subsubsection*{Proof of Theorem~\ref{theorem:MMN-IDLE-SHAPE}.}

In order to prove this theorem, we make the transformation
\begin{equation}\label{eq:transformation}
t = \rho+1-\frac{\mu_1}{\mu}
\end{equation}
For example, when $\mu_1=\underline{\mu}_1=\left(\lambda-(N-1)\mu\right)^+$, $t=\overline{t}=\min\left(\rho+1,N\right)$. Using this transformation, the $\frac{I}{\mu_1}$ term that appears in the beginning of the expression for the second derivative of the idle time~(\ref{eq:MMN-SPD}) can be written in terms of $t$ as follows.
\begin{equation*}
\frac{I}{\mu_1} = \frac{\left(N-\rho\right)\left(N-t\right)}{\mu g(t)}
\end{equation*}
where
\begin{equation*}
g(t) = N\left(N-t\right)\left(\rho+1-t\right)-\rho\left(\rho-t\right)\left(N-t+ErlC(N,\rho)\right)
\end{equation*}
Note that $g(t)>0$, since $I>0$, $N>\rho$, and from stability, $N>t$. Substituting this in~(\ref{eq:MMN-SPD}), and using~(\ref{eq:transformation}) to complete the transformation, we get the following expression for the second derivative of the idle time in terms of $t$.
\begin{equation*}
\frac{\partial^2I}{\partial\mu_1^2} = H(t) = -\frac{2\lambda\left(N-\rho\right)^2 f(t)}{\mu^3 g^3(t)}
\end{equation*}
where we use the notation $g^3(t)$ to denote $\left(g(t)\right)^3$, and
\begin{equation*}
f(t) = \left(\left(N-t\right)^2-\rho ErlC(N,\rho)\right)\left(N-t+ErlC(N,\rho)\right)+\left(N-\left(\rho-t\right)^2\right)\left(\rho+1-t\right)ErlC(N,\rho)
\end{equation*}
In order to prove the theorem, we now need to show that
\begin{enumerate}[label=(\alph*)]
\item There exists a threshold $t^\dagger\in\left(\left.-\infty,\overline{t}\right]\right.$ such that $H(t)<0$ for $-\infty<t<t^\dagger$, and $H(t)>0$ for $t^\dagger<t<\overline{t}$.
\item $H(t)>0\Rightarrow H'(t)>0$.
\end{enumerate}
To show these statements, we prove the following three properties of $f$ and $g$.
\begin{itemize}
\item $f(t)$ is a decreasing function of $t$.
\item $g(t)$ is a decreasing function of $t$.
\item $f(0)>0$.
\end{itemize}
In what follows, for convenience, we denote $ErlC(N,\rho)$ simply by $C$. Differentiating $f(t)$, we get
\begin{equation*}
\begin{split}
f'(t) &= -\left((N-t)^2-\rho C\right)-2(N-t)(N-t+C)-\left(N-(\rho-t)^2\right)C+2(\rho-t)(\rho+1-t)C\\
&= -3\left((N-t)^2+\left(-(\rho-t)^2+(N-\rho)\right)C\right)\\
&= -3\left((N-t)^2(1-C)+\left(\left((N-t)^2-(\rho-t)^2\right)+(N-\rho)\right)C\right)\\
&= -3\left((N-t)^2(1-C)+\left((N-t+\rho-t)(N-\rho)+(N-\rho)\right)C\right)\\
&= -3\left((N-t)^2(1-C)+(N-t+\rho+1-t)(N-\rho)C\right)\\
&< 0
\end{split}
\end{equation*}
The last step follows by noting that $N-t>0$, $\rho+1-t\geq 0$, $N-\rho>0$, and $0<ErlC(N,\rho)<1$ when $0<\rho<N$. This shows that $f(t)$ is a decreasing function of $t$. Next, differentiating $g(t)$, we get
\begin{equation*}
\begin{split}
g'(t) &= -N(N-t)-N(\rho+1-t)+\rho(\rho-t)+\rho(N-t+C)\\
&= -N(N-t+\rho+1-t)+\rho(\rho+1-t)+\rho(N-t)-\rho(1-C)\\
&= -(N-\rho)(N-t+\rho+1-t)-\rho(1-C)\\
&< 0
\end{split}
\end{equation*}
The last step follows by noting that $N-t>0$, $\rho+1-t\geq 0$, $N-\rho>0$, and $0<ErlC(N,\rho)<1$ when $0<\rho<N$. This shows that $g(t)$ is a decreasing function of $t$. Finally, evaluating $f(0)$, we get
\begin{equation*}
\begin{split}
f(0) &= (N^2-\rho C)(N+C)+(N-\rho^2)(\rho+1)C\\
&= N^3-\rho^3 C+N^2C-\rho^2C+NC-\rho C^2\\
&= (N^3-\rho^3)+\rho^3(1-C)+(N^2-\rho^2)C+(N-\rho)C+\rho C(1-C)\\
&> 0
\end{split}
\end{equation*}
The last step follows by noting that $N-\rho>0$, and $0<ErlC(N,\rho)<1$ when $0<\rho<N$.

We are now ready to prove the statements (a-b).
\begin{enumerate}[label=(\alph*)]
\item First, note that because $f(t)$ is decreasing and $f(0)>0$, there exists a threshold $t^\dagger\in\left(\left.0,\overline{t}\right]\right.$ such that $f(t)>0$ for $-\infty<t<t^\dagger$, and $f(t)<0$ for $t^\dagger<t<\overline{t}$. \red{(Note that if $f(\overline{t})>0$, then we let $t^\dagger=\overline{t}$ so that $f(t)<0$ in an empty interval.)} Next, since $g(t)>0$ for all $t\in \left(\left.-\infty,\overline{t}\right]\right.$, the sign of $H(t)$ is simply the opposite of the sign of $f(t)$. Statement (a) now follows directly.
\item Statement (b) is equivalent to showing that $f(t)<0\Rightarrow H'(t)>0$. Differentiating $H(t)$, we get
    \begin{equation*}
    \begin{split}
    H'(t) &= - \frac{2\lambda(N-\rho)^2}{\mu^3}\left(\frac{g^3(t)f'(t)-3f(t)g^2(t)g'(t)}{g^6(t)}\right)\\
    &= - \frac{2\lambda(N-\rho)^2}{\mu^3}\left(\frac{g(t)f'(t)-3f(t)g'(t)}{g^4(t)}\right)
    \end{split}
    \end{equation*}
    Since $g(t)>0$, $f'(t)<0$, and $g'(t)<0$, it follows that $H'(t)>0$ whenever $f(t)<0$.
\end{enumerate}
This concludes the proof.
\eProof

\subsubsection*{Proof of Theorem~\ref{thm:conditionalsymeq}.}

\red{The ``only if'' direction is straightforward.  Briefly, it follows from the fact that, by definition, any symmetric equilibrium $\mu^*>\frac{\lambda}{N}$ must be an interior global maximizer of $U(\mu_1,\mu^\star)$ in the interval $\mu_1\in(\frac{\lambda}{N},\infty)$. }

\red{The ``if'' direction requires more care.  We first show that the utility function $U(\mu_1,\mu^\star)$ inherits the properties of the idle time function $I(\mu_1,\mu^\star)$ as laid out in Theorem~\ref{theorem:MMN-IDLE-SHAPE}, and then consider the two cases when it is either increasing or decreasing at $\mu_1=\frac{\lambda}{N}$.}

\red{Recall that $U(\mu_1,\mu^\star)=I(\mu_1,\mu^\star)-c(\mu_1)$. Let $\mu_1^\dagger\in\left[\left.\underline{\mu}_1,\infty\right)\right.$ be the threshold of Theorem~\ref{theorem:MMN-IDLE-SHAPE}. We subdivide the interval $(\underline{\mu}_1,\infty)$ as follows, in order to analyze $U(\mu_1,\mu^\star)$.}

\begin{itemize}
\item \red{Consider the interval $(\underline{\mu}_1,\mu_1^\dagger)$, where, from Theorem~\ref{theorem:MMN-IDLE-SHAPE}, we know that $I'''(\mu_1,\mu^\star)<0$. Therefore, $U'''(\mu_1,\mu^\star)=I'''(\mu_1,\mu^\star)-c'''(\mu_1)<0$. This means that $U''(\mu_1,\mu^\star)$ is decreasing in this interval.  (Note that this interval could be empty, i.e., it is possible that $\mu_1^\dagger=\underline{\mu}_1$.)}
\item \red{Consider the interval $(\mu_1^\dagger,\infty)$, where, from Theorem~\ref{theorem:MMN-IDLE-SHAPE}, we know that $I''(\mu_1,\mu^\star)<0$. Therefore, $U''(\mu_1,\mu^\star)=I''(\mu_1,\mu^\star)-c''(\mu_1)<0$. This means that $U(\mu_1,\mu^\star)$ is concave in this interval.}
\end{itemize}

\red{Thus, the utility function $U(\mu_1,\mu^\star)$, like the idle time function $I(\mu_1,\mu^\star)$, may start out as a convex function at $\mu_1=\underline{\mu}_1$, but it eventually becomes concave, and stays concave thereafter. Moreover, because the cost function $c$ is increasing and convex, $\lim_{\mu_1\rightarrow\infty}U(\mu_1,\mu^\star)=-\infty$, which implies that $U(\mu_1,\mu^\star)$ must eventually be \textit{decreasing} concave.}

\red{We now consider two possibilities for the behavior of $U(\mu_1,\mu^\star)$ in the interval $(\frac{\lambda}{N},\infty)$:}

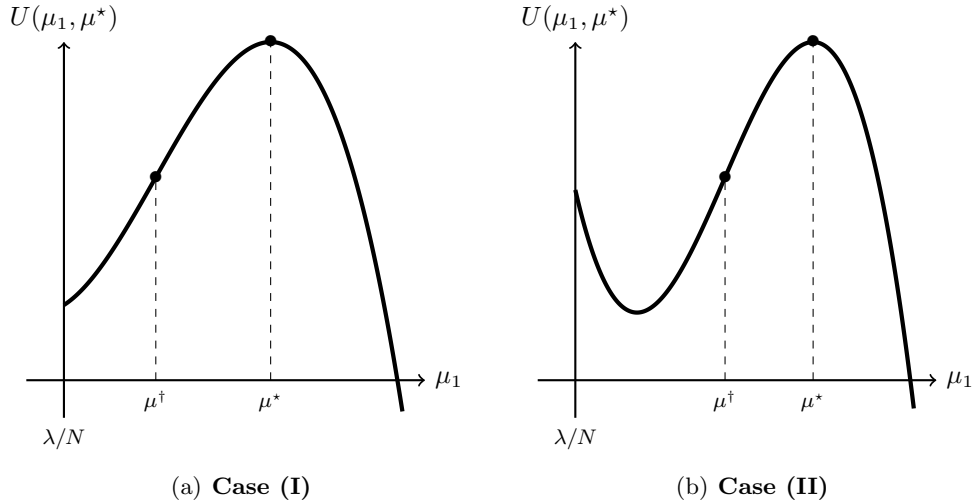
\begin{figure}[ht]
\centering
\subfigure[\red{\textbf{Case (I)}}\label{fig:conditionalsymeq-1}]{
\begin{tikzpicture}
\draw[thick,->] (0,-.5) -- (0,4.5);
\draw[thick,->] (-.5,0) -- (4.8,0);
\node[above] at (0,4.5) {{\small $U(\mu_1,\mu^\star)$}};
\node[right] at (4.8,0) {{\small $\mu_1$}};
\draw[dashed] (1.22137, 0) -- (1.22137, 2.7);
\node at (1.22137, 2.7) {{\textbullet}};
\draw[dashed] (2.74809, 0) -- (2.74809, 4.5);
\node at (2.74809, 4.5) {{\textbullet}};
\node[below] at (0,-.5) {{\scriptsize $\lambda/N$}};
\node[below] at (1.22137, 0) {{\scriptsize $\mu^\dagger$}};
\node[below] at (2.74809, -.038) {{\scriptsize $\mu^\star$}};
\draw[ultra thick] (0., 1.0008) -- (0.01, 1.00726) -- (0.02, 1.0139) -- (0.03, 1.02073) -- (0.04, 1.02773) -- (0.05, 1.03492) -- (0.06, 1.04228) -- (0.07, 1.04982) -- (0.08, 1.05753) -- (0.09, 1.06542) -- (0.1, 1.07348) -- (0.11, 1.08171) -- (0.12, 1.09011) -- (0.13, 1.09867) -- (0.14, 1.1074) -- (0.15, 1.1163) -- (0.16, 1.12535) -- (0.17, 1.13457) -- (0.18, 1.14395) -- (0.19, 1.15348) -- (0.2, 1.16318) -- (0.21, 1.17302) -- (0.22, 1.18302) -- (0.23, 1.19318) -- (0.24, 1.20348) -- (0.25, 1.21393) -- (0.26, 1.22453) -- (0.27, 1.23528) -- (0.28, 1.24617) -- (0.29, 1.2572) -- (0.3, 1.26837) -- (0.31, 1.27969) -- (0.32, 1.29114) -- (0.33, 1.30273) -- (0.34, 1.31445) -- (0.35, 1.32631) -- (0.36, 1.3383) -- (0.37, 1.35042) -- (0.38, 1.36267) -- (0.39, 1.37505) -- (0.4, 1.38755) -- (0.41, 1.40018) -- (0.42, 1.41293) -- (0.43, 1.4258) -- (0.44, 1.43879) -- (0.45, 1.45191) -- (0.46, 1.46513) -- (0.47, 1.47848) -- (0.48, 1.49194) -- (0.49, 1.50551) -- (0.5, 1.51919) -- (0.51, 1.53298) -- (0.52, 1.54688) -- (0.53, 1.56089) -- (0.54, 1.575) -- (0.55, 1.58921) -- (0.56, 1.60353) -- (0.57, 1.61794) -- (0.58, 1.63246) -- (0.59, 1.64707) -- (0.6, 1.66178) -- (0.61, 1.67658) -- (0.62, 1.69147) -- (0.63, 1.70646) -- (0.64, 1.72154) -- (0.65, 1.7367) -- (0.66, 1.75195) -- (0.67, 1.76729) -- (0.68, 1.78271) -- (0.69, 1.79821) -- (0.7, 1.81379) -- (0.71, 1.82946) -- (0.72, 1.8452) -- (0.73, 1.86101) -- (0.74, 1.8769) -- (0.75, 1.89286) -- (0.76, 1.9089) -- (0.77, 1.925) -- (0.78, 1.94118) -- (0.79, 1.95742) -- (0.8, 1.97372) -- (0.81, 1.99009) -- (0.82, 2.00652) -- (0.83, 2.02302) -- (0.84, 2.03957) -- (0.85, 2.05618) -- (0.86, 2.07285) -- (0.87, 2.08957) -- (0.88, 2.10634) -- (0.89, 2.12317) -- (0.9, 2.14004) -- (0.91, 2.15697) -- (0.92, 2.17394) -- (0.93, 2.19096) -- (0.94, 2.20802) -- (0.95, 2.22513) -- (0.96, 2.24228) -- (0.97, 2.25946) -- (0.98, 2.27669) -- (0.99, 2.29395) -- (1., 2.31124) -- (1.01, 2.32857) -- (1.02, 2.34594) -- (1.03, 2.36333) -- (1.04, 2.38075) -- (1.05, 2.3982) -- (1.06, 2.41567) -- (1.07, 2.43317) -- (1.08, 2.45069) -- (1.09, 2.46824) -- (1.1, 2.4858) -- (1.11, 2.50338) -- (1.12, 2.52098) -- (1.13, 2.5386) -- (1.14, 2.55623) -- (1.15, 2.57387) -- (1.16, 2.59152) -- (1.17, 2.60918) -- (1.18, 2.62685) -- (1.19, 2.64452) -- (1.2, 2.6622) -- (1.21, 2.67989) -- (1.22, 2.69757) -- (1.23, 2.71525) -- (1.24, 2.73294) -- (1.25, 2.75062) -- (1.26, 2.7683) -- (1.27, 2.78597) -- (1.28, 2.80363) -- (1.29, 2.82128) -- (1.3, 2.83893) -- (1.31, 2.85656) -- (1.32, 2.87418) -- (1.33, 2.89178) -- (1.34, 2.90937) -- (1.35, 2.92694) -- (1.36, 2.94449) -- (1.37, 2.96201) -- (1.38, 2.97952) -- (1.39, 2.997) -- (1.4, 3.01446) -- (1.41, 3.03189) -- (1.42, 3.04929) -- (1.43, 3.06666) -- (1.44, 3.084) -- (1.45, 3.1013) -- (1.46, 3.11857) -- (1.47, 3.13581) -- (1.48, 3.153) -- (1.49, 3.17016) -- (1.5, 3.18728) -- (1.51, 3.20435) -- (1.52, 3.22138) -- (1.53, 3.23837) -- (1.54, 3.25531) -- (1.55, 3.2722) -- (1.56, 3.28904) -- (1.57, 3.30583) -- (1.58, 3.32256) -- (1.59, 3.33925) -- (1.6, 3.35587) -- (1.61, 3.37244) -- (1.62, 3.38895) -- (1.63, 3.4054) -- (1.64, 3.42179) -- (1.65, 3.43811) -- (1.66, 3.45437) -- (1.67, 3.47056) -- (1.68, 3.48668) -- (1.69, 3.50274) -- (1.7, 3.51872) -- (1.71, 3.53463) -- (1.72, 3.55047) -- (1.73, 3.56623) -- (1.74, 3.58191) -- (1.75, 3.59751) -- (1.76, 3.61304) -- (1.77, 3.62848) -- (1.78, 3.64384) -- (1.79, 3.65912) -- (1.8, 3.6743) -- (1.81, 3.6894) -- (1.82, 3.70442) -- (1.83, 3.71934) -- (1.84, 3.73416) -- (1.85, 3.7489) -- (1.86, 3.76354) -- (1.87, 3.77808) -- (1.88, 3.79252) -- (1.89, 3.80687) -- (1.9, 3.82111) -- (1.91, 3.83525) -- (1.92, 3.84928) -- (1.93, 3.86321) -- (1.94, 3.87703) -- (1.95, 3.89074) -- (1.96, 3.90434) -- (1.97, 3.91783) -- (1.98, 3.93121) -- (1.99, 3.94447) -- (2., 3.95761) -- (2.01, 3.97064) -- (2.02, 3.98355) -- (2.03, 3.99633) -- (2.04, 4.00899) -- (2.05, 4.02153) -- (2.06, 4.03394) -- (2.07, 4.04623) -- (2.08, 4.05838) -- (2.09, 4.07041) -- (2.1, 4.0823) -- (2.11, 4.09407) -- (2.12, 4.10569) -- (2.13, 4.11718) -- (2.14, 4.12853) -- (2.15, 4.13975) -- (2.16, 4.15082) -- (2.17, 4.16175) -- (2.18, 4.17253) -- (2.19, 4.18317) -- (2.2, 4.19366) -- (2.21, 4.20401) -- (2.22, 4.2142) -- (2.23, 4.22424) -- (2.24, 4.23413) -- (2.25, 4.24387) -- (2.26, 4.25345) -- (2.27, 4.26287) -- (2.28, 4.27213) -- (2.29, 4.28123) -- (2.3, 4.29017) -- (2.31, 4.29895) -- (2.32, 4.30756) -- (2.33, 4.316) -- (2.34, 4.32428) -- (2.35, 4.33238) -- (2.36, 4.34032) -- (2.37, 4.34808) -- (2.38, 4.35566) -- (2.39, 4.36308) -- (2.4, 4.37031) -- (2.41, 4.37737) -- (2.42, 4.38424) -- (2.43, 4.39093) -- (2.44, 4.39744) -- (2.45, 4.40377) -- (2.46, 4.40991) -- (2.47, 4.41586) -- (2.48, 4.42162) -- (2.49, 4.42719) -- (2.5, 4.43257) -- (2.51, 4.43775) -- (2.52, 4.44274) -- (2.53, 4.44753) -- (2.54, 4.45212) -- (2.55, 4.45651) -- (2.56, 4.4607) -- (2.57, 4.46469) -- (2.58, 4.46847) -- (2.59, 4.47205) -- (2.6, 4.47542) -- (2.61, 4.47858) -- (2.62, 4.48153) -- (2.63, 4.48426) -- (2.64, 4.48679) -- (2.65, 4.48909) -- (2.66, 4.49118) -- (2.67, 4.49306) -- (2.68, 4.49471) -- (2.69, 4.49614) -- (2.7, 4.49735) -- (2.71, 4.49833) -- (2.72, 4.49909) -- (2.73, 4.49962) -- (2.74, 4.49992) -- (2.75, 4.5) -- (2.76, 4.49984) -- (2.77, 4.49944) -- (2.78, 4.49881) -- (2.79, 4.49795) -- (2.8, 4.49684) -- (2.81, 4.4955) -- (2.82, 4.49392) -- (2.83, 4.49209) -- (2.84, 4.49002) -- (2.85, 4.4877) -- (2.86, 4.48514) -- (2.87, 4.48233) -- (2.88, 4.47926) -- (2.89, 4.47595) -- (2.9, 4.47238) -- (2.91, 4.46856) -- (2.92, 4.46448) -- (2.93, 4.46015) -- (2.94, 4.45555) -- (2.95, 4.4507) -- (2.96, 4.44558) -- (2.97, 4.44019) -- (2.98, 4.43455) -- (2.99, 4.42863) -- (3., 4.42245) -- (3.01, 4.416) -- (3.02, 4.40927) -- (3.03, 4.40228) -- (3.04, 4.395) -- (3.05, 4.38746) -- (3.06, 4.37963) -- (3.07, 4.37153) -- (3.08, 4.36314) -- (3.09, 4.35448) -- (3.1, 4.34553) -- (3.11, 4.33629) -- (3.12, 4.32677) -- (3.13, 4.31696) -- (3.14, 4.30686) -- (3.15, 4.29647) -- (3.16, 4.28579) -- (3.17, 4.27481) -- (3.18, 4.26354) -- (3.19, 4.25197) -- (3.2, 4.2401) -- (3.21, 4.22793) -- (3.22, 4.21546) -- (3.23, 4.20268) -- (3.24, 4.1896) -- (3.25, 4.17622) -- (3.26, 4.16252) -- (3.27, 4.14852) -- (3.28, 4.13421) -- (3.29, 4.11958) -- (3.3, 4.10464) -- (3.31, 4.08939) -- (3.32, 4.07381) -- (3.33, 4.05792) -- (3.34, 4.04171) -- (3.35, 4.02518) -- (3.36, 4.00832) -- (3.37, 3.99114) -- (3.38, 3.97364) -- (3.39, 3.95581) -- (3.4, 3.93764) -- (3.41, 3.91915) -- (3.42, 3.90032) -- (3.43, 3.88117) -- (3.44, 3.86167) -- (3.45, 3.84184) -- (3.46, 3.82167) -- (3.47, 3.80116) -- (3.48, 3.78031) -- (3.49, 3.75912) -- (3.5, 3.73759) -- (3.51, 3.7157) -- (3.52, 3.69347) -- (3.53, 3.67089) -- (3.54, 3.64797) -- (3.55, 3.62468) -- (3.56, 3.60105) -- (3.57, 3.57706) -- (3.58, 3.55272) -- (3.59, 3.52801) -- (3.6, 3.50295) -- (3.61, 3.47753) -- (3.62, 3.45174) -- (3.63, 3.42559) -- (3.64, 3.39907) -- (3.65, 3.37219) -- (3.66, 3.34494) -- (3.67, 3.31732) -- (3.68, 3.28933) -- (3.69, 3.26096) -- (3.7, 3.23222) -- (3.71, 3.2031) -- (3.72, 3.17361) -- (3.73, 3.14374) -- (3.74, 3.11348) -- (3.75, 3.08285) -- (3.76, 3.05183) -- (3.77, 3.02042) -- (3.78, 2.98863) -- (3.79, 2.95645) -- (3.8, 2.92388) -- (3.81, 2.89092) -- (3.82, 2.85756) -- (3.83, 2.82382) -- (3.84, 2.78967) -- (3.85, 2.75513) -- (3.86, 2.72019) -- (3.87, 2.68485) -- (3.88, 2.6491) -- (3.89, 2.61296) -- (3.9, 2.57641) -- (3.91, 2.53945) -- (3.92, 2.50208) -- (3.93, 2.46431) -- (3.94, 2.42612) -- (3.95, 2.38752) -- (3.96, 2.34851) -- (3.97, 2.30908) -- (3.98, 2.26924) -- (3.99, 2.22897) -- (4., 2.18829) -- (4.01, 2.14718) -- (4.02, 2.10565) -- (4.03, 2.0637) -- (4.04, 2.02132) -- (4.05, 1.97851) -- (4.06, 1.93528) -- (4.07, 1.89161) -- (4.08, 1.84751) -- (4.09, 1.80297) -- (4.1, 1.75801) -- (4.11, 1.7126) -- (4.12, 1.66676) -- (4.13, 1.62047) -- (4.14, 1.57375) -- (4.15, 1.52658) -- (4.16, 1.47897) -- (4.17, 1.43091) -- (4.18, 1.3824) -- (4.19, 1.33345) -- (4.2, 1.28404) -- (4.21, 1.23418) -- (4.22, 1.18387) -- (4.23, 1.13311) -- (4.24, 1.08188) -- (4.25, 1.0302) -- (4.26, 0.978063) -- (4.27, 0.925461) -- (4.28, 0.872398) -- (4.29, 0.818869) -- (4.3, 0.764876) -- (4.31, 0.710415) -- (4.32, 0.655485) -- (4.33, 0.600086) -- (4.34, 0.544214) -- (4.35, 0.487869) -- (4.36, 0.43105) -- (4.37, 0.373754) -- (4.38, 0.315981) -- (4.39, 0.257728) -- (4.4, 0.198994) -- (4.41, 0.139778) -- (4.42, 0.0800783) -- (4.43, 0.019893) -- (4.44, -0.0407791) -- (4.45, -0.10194) -- (4.46, -0.16359) -- (4.47, -0.225732) -- (4.48, -0.288367) -- (4.49, -0.351496) -- (4.5, -0.415122);
\end{tikzpicture} 
}
\subfigure[\red{\textbf{Case (II)}}\label{fig:conditionalsymeq-2}]{
\begin{tikzpicture}
\draw[thick,->] (0,-.5) -- (0,4.5);
\draw[thick,->] (-.5,0) -- (4.8,0);
\node[above] at (0,4.5) {{\small $U(\mu_1,\mu^\star)$}};
\node[right] at (4.8,0) {{\small $\mu_1$}};
\draw[dashed] (1.9904, 0) -- (1.9904, 2.7);
\node at (1.9904, 2.7) {{\textbullet}};
\draw[dashed] (3.16122, 0) -- (3.16122, 4.5);
\node at (3.16122, 4.5) {{\textbullet}};
\node[below] at (0,-.5) {{\scriptsize $\lambda/N$}};
\node[below] at (1.9904, 0) {{\scriptsize $\mu^\dagger$}};
\node[below] at (3.16122, -.038) {{\scriptsize $\mu^\star$}};
\draw[ultra thick] (0., 2.5317) -- (0.01, 2.48845) -- (0.02, 2.44587) -- (0.03, 2.40394) -- (0.04, 2.36268) -- (0.05, 2.32208) -- (0.06, 2.28212) -- (0.07, 2.24282) -- (0.08, 2.20416) -- (0.09, 2.16615) -- (0.1, 2.12878) -- (0.11, 2.09204) -- (0.12, 2.05593) -- (0.13, 2.02045) -- (0.14, 1.9856) -- (0.15, 1.95137) -- (0.16, 1.91777) -- (0.17, 1.88477) -- (0.18, 1.85239) -- (0.19, 1.82062) -- (0.2, 1.78945) -- (0.21, 1.75889) -- (0.22, 1.72893) -- (0.23, 1.69956) -- (0.24, 1.67078) -- (0.25, 1.64259) -- (0.26, 1.61499) -- (0.27, 1.58797) -- (0.28, 1.56153) -- (0.29, 1.53566) -- (0.3, 1.51037) -- (0.31, 1.48564) -- (0.32, 1.46148) -- (0.33, 1.43789) -- (0.34, 1.41485) -- (0.35, 1.39236) -- (0.36, 1.37043) -- (0.37, 1.34905) -- (0.38, 1.32821) -- (0.39, 1.30792) -- (0.4, 1.28816) -- (0.41, 1.26894) -- (0.42, 1.25024) -- (0.43, 1.23208) -- (0.44, 1.21444) -- (0.45, 1.19733) -- (0.46, 1.18073) -- (0.47, 1.16465) -- (0.48, 1.14908) -- (0.49, 1.13401) -- (0.5, 1.11946) -- (0.51, 1.1054) -- (0.52, 1.09184) -- (0.53, 1.07878) -- (0.54, 1.0662) -- (0.55, 1.05412) -- (0.56, 1.04252) -- (0.57, 1.0314) -- (0.58, 1.02076) -- (0.59, 1.01059) -- (0.6, 1.0009) -- (0.61, 0.991671) -- (0.62, 0.982909) -- (0.63, 0.974607) -- (0.64, 0.966763) -- (0.65, 0.959373) -- (0.66, 0.952434) -- (0.67, 0.945943) -- (0.68, 0.939896) -- (0.69, 0.93429) -- (0.7, 0.929121) -- (0.71, 0.924387) -- (0.72, 0.920083) -- (0.73, 0.916207) -- (0.74, 0.912755) -- (0.75, 0.909723) -- (0.76, 0.907109) -- (0.77, 0.904909) -- (0.78, 0.90312) -- (0.79, 0.901737) -- (0.8, 0.900759) -- (0.81, 0.900181) -- (0.82, 0.9) -- (0.83, 0.900213) -- (0.84, 0.900817) -- (0.85, 0.901807) -- (0.86, 0.903181) -- (0.87, 0.904936) -- (0.88, 0.907067) -- (0.89, 0.909572) -- (0.9, 0.912448) -- (0.91, 0.91569) -- (0.92, 0.919296) -- (0.93, 0.923261) -- (0.94, 0.927584) -- (0.95, 0.93226) -- (0.96, 0.937286) -- (0.97, 0.942659) -- (0.98, 0.948374) -- (0.99, 0.95443) -- (1., 0.960823) -- (1.01, 0.967549) -- (1.02, 0.974604) -- (1.03, 0.981986) -- (1.04, 0.989692) -- (1.05, 0.997717) -- (1.06, 1.00606) -- (1.07, 1.01471) -- (1.08, 1.02368) -- (1.09, 1.03295) -- (1.1, 1.04252) -- (1.11, 1.05239) -- (1.12, 1.06256) -- (1.13, 1.07302) -- (1.14, 1.08378) -- (1.15, 1.09481) -- (1.16, 1.10613) -- (1.17, 1.11773) -- (1.18, 1.12961) -- (1.19, 1.14176) -- (1.2, 1.15417) -- (1.21, 1.16686) -- (1.22, 1.1798) -- (1.23, 1.19301) -- (1.24, 1.20647) -- (1.25, 1.22018) -- (1.26, 1.23415) -- (1.27, 1.24836) -- (1.28, 1.26281) -- (1.29, 1.2775) -- (1.3, 1.29242) -- (1.31, 1.30758) -- (1.32, 1.32297) -- (1.33, 1.33858) -- (1.34, 1.35441) -- (1.35, 1.37047) -- (1.36, 1.38674) -- (1.37, 1.40322) -- (1.38, 1.41991) -- (1.39, 1.4368) -- (1.4, 1.4539) -- (1.41, 1.47119) -- (1.42, 1.48868) -- (1.43, 1.50637) -- (1.44, 1.52424) -- (1.45, 1.5423) -- (1.46, 1.56053) -- (1.47, 1.57895) -- (1.48, 1.59754) -- (1.49, 1.61631) -- (1.5, 1.63524) -- (1.51, 1.65433) -- (1.52, 1.67359) -- (1.53, 1.69301) -- (1.54, 1.71258) -- (1.55, 1.73231) -- (1.56, 1.75218) -- (1.57, 1.77219) -- (1.58, 1.79235) -- (1.59, 1.81265) -- (1.6, 1.83308) -- (1.61, 1.85364) -- (1.62, 1.87433) -- (1.63, 1.89514) -- (1.64, 1.91608) -- (1.65, 1.93713) -- (1.66, 1.9583) -- (1.67, 1.97958) -- (1.68, 2.00097) -- (1.69, 2.02246) -- (1.7, 2.04405) -- (1.71, 2.06574) -- (1.72, 2.08753) -- (1.73, 2.1094) -- (1.74, 2.13137) -- (1.75, 2.15341) -- (1.76, 2.17554) -- (1.77, 2.19775) -- (1.78, 2.22003) -- (1.79, 2.24238) -- (1.8, 2.2648) -- (1.81, 2.28728) -- (1.82, 2.30982) -- (1.83, 2.33242) -- (1.84, 2.35508) -- (1.85, 2.37778) -- (1.86, 2.40053) -- (1.87, 2.42333) -- (1.88, 2.44617) -- (1.89, 2.46904) -- (1.9, 2.49195) -- (1.91, 2.51489) -- (1.92, 2.53785) -- (1.93, 2.56084) -- (1.94, 2.58385) -- (1.95, 2.60687) -- (1.96, 2.62991) -- (1.97, 2.65296) -- (1.98, 2.67602) -- (1.99, 2.69908) -- (2., 2.72214) -- (2.01, 2.7452) -- (2.02, 2.76825) -- (2.03, 2.79129) -- (2.04, 2.81431) -- (2.05, 2.83732) -- (2.06, 2.86032) -- (2.07, 2.88328) -- (2.08, 2.90622) -- (2.09, 2.92913) -- (2.1, 2.95201) -- (2.11, 2.97485) -- (2.12, 2.99765) -- (2.13, 3.0204) -- (2.14, 3.04311) -- (2.15, 3.06577) -- (2.16, 3.08838) -- (2.17, 3.11092) -- (2.18, 3.13341) -- (2.19, 3.15583) -- (2.2, 3.17819) -- (2.21, 3.20048) -- (2.22, 3.22269) -- (2.23, 3.24482) -- (2.24, 3.26688) -- (2.25, 3.28885) -- (2.26, 3.31073) -- (2.27, 3.33252) -- (2.28, 3.35422) -- (2.29, 3.37582) -- (2.3, 3.39732) -- (2.31, 3.41872) -- (2.32, 3.44) -- (2.33, 3.46118) -- (2.34, 3.48224) -- (2.35, 3.50319) -- (2.36, 3.52401) -- (2.37, 3.54471) -- (2.38, 3.56529) -- (2.39, 3.58573) -- (2.4, 3.60603) -- (2.41, 3.6262) -- (2.42, 3.64623) -- (2.43, 3.66611) -- (2.44, 3.68585) -- (2.45, 3.70543) -- (2.46, 3.72486) -- (2.47, 3.74413) -- (2.48, 3.76324) -- (2.49, 3.78219) -- (2.5, 3.80097) -- (2.51, 3.81957) -- (2.52, 3.838) -- (2.53, 3.85625) -- (2.54, 3.87433) -- (2.55, 3.89221) -- (2.56, 3.90991) -- (2.57, 3.92742) -- (2.58, 3.94473) -- (2.59, 3.96184) -- (2.6, 3.97875) -- (2.61, 3.99546) -- (2.62, 4.01196) -- (2.63, 4.02824) -- (2.64, 4.04431) -- (2.65, 4.06016) -- (2.66, 4.07579) -- (2.67, 4.0912) -- (2.68, 4.10638) -- (2.69, 4.12132) -- (2.7, 4.13603) -- (2.71, 4.1505) -- (2.72, 4.16473) -- (2.73, 4.17871) -- (2.74, 4.19244) -- (2.75, 4.20592) -- (2.76, 4.21915) -- (2.77, 4.23212) -- (2.78, 4.24482) -- (2.79, 4.25726) -- (2.8, 4.26943) -- (2.81, 4.28133) -- (2.82, 4.29295) -- (2.83, 4.30429) -- (2.84, 4.31535) -- (2.85, 4.32613) -- (2.86, 4.33661) -- (2.87, 4.34681) -- (2.88, 4.3567) -- (2.89, 4.3663) -- (2.9, 4.37559) -- (2.91, 4.38458) -- (2.92, 4.39326) -- (2.93, 4.40163) -- (2.94, 4.40968) -- (2.95, 4.41741) -- (2.96, 4.42482) -- (2.97, 4.4319) -- (2.98, 4.43865) -- (2.99, 4.44507) -- (3., 4.45115) -- (3.01, 4.4569) -- (3.02, 4.4623) -- (3.03, 4.46735) -- (3.04, 4.47206) -- (3.05, 4.47641) -- (3.06, 4.4804) -- (3.07, 4.48404) -- (3.08, 4.48731) -- (3.09, 4.49021) -- (3.1, 4.49275) -- (3.11, 4.49491) -- (3.12, 4.49669) -- (3.13, 4.4981) -- (3.14, 4.49912) -- (3.15, 4.49975) -- (3.16, 4.5) -- (3.17, 4.49985) -- (3.18, 4.4993) -- (3.19, 4.49836) -- (3.2, 4.49701) -- (3.21, 4.49525) -- (3.22, 4.49308) -- (3.23, 4.4905) -- (3.24, 4.4875) -- (3.25, 4.48408) -- (3.26, 4.48024) -- (3.27, 4.47597) -- (3.28, 4.47127) -- (3.29, 4.46614) -- (3.3, 4.46057) -- (3.31, 4.45456) -- (3.32, 4.4481) -- (3.33, 4.4412) -- (3.34, 4.43384) -- (3.35, 4.42604) -- (3.36, 4.41777) -- (3.37, 4.40905) -- (3.38, 4.39986) -- (3.39, 4.3902) -- (3.4, 4.38007) -- (3.41, 4.36947) -- (3.42, 4.35839) -- (3.43, 4.34682) -- (3.44, 4.33478) -- (3.45, 4.32224) -- (3.46, 4.30922) -- (3.47, 4.2957) -- (3.48, 4.28168) -- (3.49, 4.26717) -- (3.5, 4.25214) -- (3.51, 4.23661) -- (3.52, 4.22057) -- (3.53, 4.20402) -- (3.54, 4.18694) -- (3.55, 4.16935) -- (3.56, 4.15122) -- (3.57, 4.13258) -- (3.58, 4.1134) -- (3.59, 4.09368) -- (3.6, 4.07343) -- (3.61, 4.05263) -- (3.62, 4.03129) -- (3.63, 4.00941) -- (3.64, 3.98697) -- (3.65, 3.96397) -- (3.66, 3.94042) -- (3.67, 3.91631) -- (3.68, 3.89163) -- (3.69, 3.86638) -- (3.7, 3.84056) -- (3.71, 3.81416) -- (3.72, 3.78719) -- (3.73, 3.75963) -- (3.74, 3.73149) -- (3.75, 3.70276) -- (3.76, 3.67344) -- (3.77, 3.64352) -- (3.78, 3.61301) -- (3.79, 3.58189) -- (3.8, 3.55017) -- (3.81, 3.51784) -- (3.82, 3.48489) -- (3.83, 3.45133) -- (3.84, 3.41715) -- (3.85, 3.38235) -- (3.86, 3.34693) -- (3.87, 3.31087) -- (3.88, 3.27418) -- (3.89, 3.23686) -- (3.9, 3.19889) -- (3.91, 3.16029) -- (3.92, 3.12104) -- (3.93, 3.08114) -- (3.94, 3.04058) -- (3.95, 2.99937) -- (3.96, 2.95751) -- (3.97, 2.91498) -- (3.98, 2.87178) -- (3.99, 2.82791) -- (4., 2.78337) -- (4.01, 2.73816) -- (4.02, 2.69226) -- (4.03, 2.64568) -- (4.04, 2.59842) -- (4.05, 2.55047) -- (4.06, 2.50182) -- (4.07, 2.45248) -- (4.08, 2.40243) -- (4.09, 2.35169) -- (4.1, 2.30024) -- (4.11, 2.24808) -- (4.12, 2.1952) -- (4.13, 2.14161) -- (4.14, 2.0873) -- (4.15, 2.03226) -- (4.16, 1.9765) -- (4.17, 1.92001) -- (4.18, 1.86279) -- (4.19, 1.80483) -- (4.2, 1.74612) -- (4.21, 1.68668) -- (4.22, 1.62649) -- (4.23, 1.56555) -- (4.24, 1.50385) -- (4.25, 1.4414) -- (4.26, 1.37819) -- (4.27, 1.31421) -- (4.28, 1.24947) -- (4.29, 1.18396) -- (4.3, 1.11767) -- (4.31, 1.05061) -- (4.32, 0.982762) -- (4.33, 0.914134) -- (4.34, 0.844719) -- (4.35, 0.774513) -- (4.36, 0.703514) -- (4.37, 0.631717) -- (4.38, 0.55912) -- (4.39, 0.485718) -- (4.4, 0.411509) -- (4.41, 0.33649) -- (4.42, 0.260656) -- (4.43, 0.184005) -- (4.44, 0.106534) -- (4.45, 0.0282376) -- (4.46, -0.0508858) -- (4.47, -0.13084) -- (4.48, -0.211629) -- (4.49, -0.293255) -- (4.5, -0.375722);
\end{tikzpicture}
}
\caption{\red{The graphic depiction of the proof of Theorem~\ref{thm:conditionalsymeq}.}\label{fig:conditionalsymeq}}
\end{figure}

\red{\textbf{Case (I): $\mathbf{U(\mu_1,\mu^\star)}$ is increasing at $\mathbf{\mu_1=\frac{\lambda}{N}}$.} If $\mu_1^\dagger >\frac{\lambda}{N}$ (see Figure~\ref{fig:conditionalsymeq-1}), $U(\mu_1,\mu^\star)$ would start out being increasing convex, reach a rising point of inflection at $\mu_1=\mu_1^\dagger$, and then become increasing concave. (Otherwise, if $\mu_1^\dagger \leq\frac{\lambda}{N}$, $U(\mu_1,\mu^\star)$ would just be increasing concave to begin with.) It would then go on to attain a (global) maximum, and finally become decreasing concave. This means that the unique stationary point of $U(\mu_1,\mu^\star)$ in this interval must be at this (interior) global maximum. Since $U'(\mu^\star,\mu^\star)=0$ (from the symmetric first order condition~(\ref{eq:MMN-SYM-FOC})), $\mu_1=\mu^\star$ must be the global maximizer of the utility function $U(\mu_1,\mu^\star)$, and hence a symmetric equilibrium.}

\red{\textbf{Case (II): $\mathbf{U(\mu_1,\mu^\star)}$ is decreasing at $\mathbf{\mu_1=\frac{\lambda}{N}}$.} Because $U(\mu^\star,\mu^\star)\geq U(\frac{\lambda}{N},\mu^\star)$, $U(\mu_1,\mu^\star)$ must eventually increase to a value at or above $U(\frac{\lambda}{N},\mu^\star)$, which means it must start out being decreasing \textit{convex} (see Figure~\ref{fig:conditionalsymeq-2}), attain a minimum, then become increasing convex. It would then follow the same pattern as in the previous case, i.e., reach a rising point of inflection at $\mu_1=\mu_1^\dagger$, and then become increasing concave, go on to attain a (global) maximum, and finally become decreasing concave. This means that it admits two stationary points -- a minimum and a maximum. Since $U'(\mu^\star,\mu^\star)=0$ (from the symmetric first order condition~(\ref{eq:MMN-SYM-FOC})) \textit{and} $U(\mu^\star,\mu^\star)\geq U(\frac{\lambda}{N},\mu^\star)$, $\mu_1=\mu^\star$ must be the (global) maximizer, and hence a symmetric equilibrium.}

\red{Note that if $U(\mu_1,\mu^\star)$ is stationary at $\mu_1=\frac{\lambda}{N}$, it could either start out increasing or decreasing in the interval $(\frac{\lambda}{N},\infty)$, and one of the two cases discussed above would apply accordingly.}

\red{Finally, to conclude the proof, note that~(\ref{eq:conditionalsymeq}) is equivalent to the inequality $U(\mu^\star,\mu^\star) \geq U(\frac{\lambda}{N},\mu^\star)$, obtained by plugging in and evaluating the utilities using~(\ref{Equation:TaggedServerUtility}) and~(\ref{eq:MMN-IDLE}). This completes the proof.}
\eProof

\subsubsection*{Proof of Theorem~\ref{lemma:MMN-SYM-FOC}.}

The symmetric first order condition~(\ref{eq:MMN-SYM-FOC}) can be rewritten as
\begin{equation*}
ErlC\left(N,\frac{\lambda}{\mu}\right)=\mu^2c'(\mu)\frac{N^2}{\lambda}+\frac{\lambda}{\mu}-N
\end{equation*}
It suffices to show that if $\lambda c'\left(\frac{\lambda}{N}\right) < 1$, then the left hand side and the right hand side intersect \change{at least once} in $\left(\frac{\lambda}{N},\infty\right)$. We first observe that the left hand side, the Erlang-C function, is shown to be convex and increasing in $\rho=\frac{\lambda}{\mu}$ (pages 8 and 11 of~\cite{WhittNotes}). This means that it is decreasing and convex in $\mu$. Moreover, $ErlC(N,N)=1$ and $ErlC(N,0)=0$, which means that the left hand side decreases from $1$ to $0$ in a convex fashion as $\mu$ runs from $\frac{\lambda}{N}$ to $\infty$. The right hand side is clearly convex in $\mu$, and is equal to $\lambda c'\left(\frac{\lambda}{N}\right)$ when $\mu=\frac{\lambda}{N}$, and approaches $\infty$ as $\mu$ approaches $\infty$. Therefore, if $\lambda c'\left(\frac{\lambda}{N}\right)<1$, \change{then the two curves must intersect at least once} in $\left(\frac{\lambda}{N},\infty\right)$.

\change{Next, it is sufficient to show that if $2\frac{\lambda}{N}c'\left(\frac{\lambda}{N}\right)+\left(\frac{\lambda}{N}\right)^2 c''\left(\frac{\lambda}{N}\right)\geq 1$, then the right hand side is non-decreasing in $\mu$. In order to do so, it suffices to show that
\begin{equation*}
\begin{split}
&\frac{\partial}{\partial\mu}\left(\mu^2c'(\mu)\frac{N^2}{\lambda}+\frac{\lambda}{\mu}-N\right)\geq 0\\
\Leftrightarrow\ &\left(\mu^2c''(\mu)+2\mu c'(\mu)\right)\frac{N^2}{\lambda} - \frac{\lambda}{\mu^2}\geq 0\\
\Leftrightarrow\ &\left(\mu^2c''(\mu)+2\mu c'(\mu)\right)\left(\frac{N\mu}{\lambda}\right)^2\geq 1
\end{split}
\end{equation*}
The left hand side is a non-decreasing function of $\mu$, therefore, in the interval $\left(\frac{\lambda}{N},\infty\right)$, we have
\begin{equation*}
\left(\mu^2c''(\mu)+2\mu c'(\mu)\right)\left(\frac{N\mu}{\lambda}\right)^2 \geq \left(\frac{\lambda}{N}\right)^2 c''\left(\frac{\lambda}{N}\right)+2\frac{\lambda}{N}c'\left(\frac{\lambda}{N}\right)\geq 1.
\end{equation*}}
This completes the proof.
\eProof

\subsubsection*{\red{Proof of Theorem~\ref{thm:MMN-larger-eq-better}.}}

\red{The utility at any symmetric point can be evaluated as $U(\mu,\mu)=1-\frac{\lambda}{N\mu}-c(\mu)$, using~(\ref{Equation:TaggedServerUtility}) and~(\ref{eq:MMN-IDLE}). Therefore, it follows that showing $U(\mu_1^\star,\mu_1^\star) > U(\mu_2^\star,\mu_2^\star)$ is equivalent to showing that
\begin{equation*}
\frac{c(\mu_1^\star)-c(\mu_2^\star)}{\mu_1^\star-\mu_2^\star} < \frac{\lambda}{N \mu_1^\star \mu_2^\star}.
\end{equation*}}
\red{The function $c$ is convex by assumption. It follows that
\begin{equation}\label{eq:tangent-inequality-thm6}
c(\mu_1^\star) - c(\mu_2^\star) \leq (\mu_1^\star - \mu_2^\star) c'(\mu_1^\star).
\end{equation}
Therefore, rearranging and substituting for $c'(\mu_1^\star)$ from the symmetric first order condition~(\ref{eq:MMN-SYM-FOC}),
\begin{equation*}
\frac{c(\mu_1^\star)-c(\mu_2^\star)}{\mu_1^\star-\mu_2^\star} \leq \frac{\lambda}{N^2(\mu_1^\star)^2}\left(N-\frac{\lambda}{\mu_1^\star}+ErlC\left(N,\frac{\lambda}{\mu_1^\star}\right)\right).
\end{equation*}
It has been shown (page 14 of~\cite{WhittNotes}, and~\cite{Harel1988}) that $ErlC\left(N,\frac{\lambda}{\mu}\right) < \frac{\lambda}{N \mu}$. Using this,
\begin{equation*}
\frac{c(\mu_1^\star)-c(\mu_2^\star)}{\mu_1^\star-\mu_2^\star} < \frac{\lambda}{N^2(\mu_1^\star)^2}\left(N-\frac{\lambda}{\mu_1^\star}\left(1-\frac{1}{N}\right)\right) < \frac{\lambda}{N^2(\mu_1^\star)^2}(N) = \frac{\lambda}{N(\mu_1^\star)^2} < \frac{\lambda}{N \mu_1^\star \mu_2^\star}.
\end{equation*}
This completes the proof.}
\eProof 

\subsection*{PROOFS FROM SECTION~\ref{Section:Staffing}}

\subsubsection*{Proof of Proposition~\ref{proposition:nonaoInd}.}
We first observe that if $f(\lambda) = \omega(\lambda)$, then
\[
\frac{C^{\star,\lambda}(N^\lambda)}{\lambda} \geq c_S \frac{N^\lambda}{\lambda}\rightarrow \infty \mbox{ as } \lambda \rightarrow \infty.
\]
Since Proposition~\ref{proposition:staffing_independent} evidences a staffing policy under which $C^{\star,\lambda}(N^\lambda)/\lambda$ has a finite limit, having $f(\lambda) = \omega(\lambda)$ cannot result in an asymptotically optimal staffing policy.

\red{Next, we consider the case $f(\lambda)=o(\lambda)$. In this case, $\lambda/N^\lambda \rightarrow \infty$.  Since any symmetric equilibrium must have $\mu^{\star,\lambda} > \lambda/N^\lambda$ from~(\ref{Equation:NashDef}), it follows that if there exists a sequence of symmetric equilibria $\{ \mu^{\star,\lambda} \}$, then $\mu^{\star,\lambda} \rightarrow \infty$ as $\lambda \rightarrow \infty$.  We conclude that such a staffing policy cannot be admissible.}
\eProof

\subsubsection*{Proof of Theorem~\ref{lemma:FOCexistence_staffingInd}.}
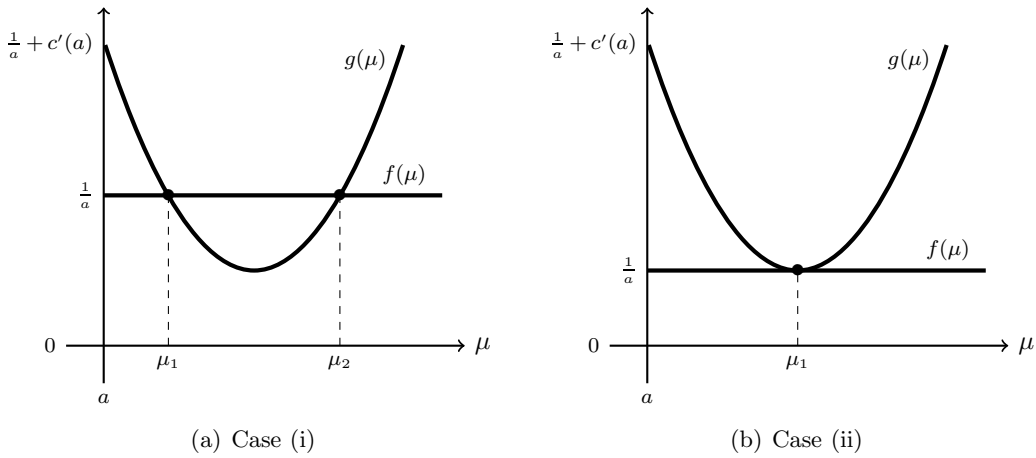
\begin{figure}[htbp]
\centering
\subfigure[Case (i)]{
\begin{tikzpicture}
\draw[thick,->] (0,-.5) -- (0,4.5);
\draw[thick,->] (-.5,0) -- (4.8,0);
\node[right] at (4.8,0) {{\small $\mu$}};
\draw[ultra thick] (0.02,4) parabola[bend at end] (2,1) parabola (3.98,4);
\draw[ultra thick] (0,2) -- (4.5,2);
\draw[dashed] (.86,0) -- (.86,2);
\node at (.86,2) {{\textbullet}};
\draw[dashed] (3.14,0) -- (3.14,2);
\node at (3.14,2) {{\textbullet}};
\node[left] at (-.5,0) {{\scriptsize $0$}};
\node[left] at (0,2) {{\scriptsize $\frac{1}{a}$}};
\node[left] at (0,4) {{\scriptsize $\frac{1}{a}+c'(a)$}};
\node[above] at (4,2) {{\scriptsize $f(\mu)$}};
\node[left] at (3.9,3.8) {{\scriptsize $g(\mu)$}};
\node[below] at (0,-.5) {{\scriptsize $a$}};
\node[below] at (.86,0) {{\scriptsize $\mu_1$}};
\node[below] at (3.14,0) {{\scriptsize $\mu_2$}};
\end{tikzpicture}
}
\subfigure[Case (ii)]{
\begin{tikzpicture}
\draw[thick,->] (0,-.5) -- (0,4.5);
\draw[thick,->] (-.5,0) -- (4.8,0);
\node[right] at (4.8,0) {{\small $\mu$}};
\draw[ultra thick] (0.02,4) parabola[bend at end] (2,1) parabola (3.98,4);
\draw[ultra thick] (0,1) -- (4.5,1);
\draw[dashed] (2,0) -- (2,1);
\node at (2,1) {{\textbullet}};
\node[left] at (-.5,0) {{\scriptsize $0$}};
\node[left] at (0,1) {{\scriptsize $\frac{1}{a}$}};
\node[left] at (0,4) {{\scriptsize $\frac{1}{a}+c'(a)$}};
\node[above] at (4,1) {{\scriptsize $f(\mu)$}};
\node[left] at (3.9,3.8) {{\scriptsize $g(\mu)$}};
\node[below] at (0,-.5) {{\scriptsize $a$}};
\node[below] at (2,0) {{\scriptsize $\mu_1$}};
\end{tikzpicture}
}
\caption{The limiting first order condition~(\ref{eq:limiting_mu_Ind}).}
\label{fig:FOC}
\end{figure}

We can rewrite~(\ref{eq:limiting_mu_Ind}) as
\[
f(\mu) = g(\mu)
\]
where
\[
f(\mu) = \frac{1}{a} \mbox{ and } g(\mu) = \frac{\mu^2}{a^2} c'(\mu) + \frac{1}{\mu}.
\]
The two cases of interest (i) and (ii) are as shown in Figure~\ref{fig:FOC}.  Our strategy for the proof is to rewrite~(\ref{eq:FOCstaffingInd}) in terms of functions $f^\lambda$ and $g^\lambda$ that are in some sense close to $f$ and $g$.  Then, in case (i), the fact that $g(\mu)$ lies below $f(\mu)$ for $\mu \in [\mu_1,\mu_2]$ implies that $f^\lambda$ and $g^\lambda$ intersect (at least) twice.  The case (ii) is more delicate, because the sign of $o(\lambda)$ determines if the functions $f^\lambda$ and $g^\lambda$ will cross (at least) twice or not at all.  (We remark that it will become clear in that part of the proof where the condition $o(\lambda) < -3$ is needed.)

The first step is to rewrite~(\ref{eq:FOCstaffingInd}) as
\[
f^\lambda(\mu) = g^\lambda(\mu)
\]
where
\begin{eqnarray*}
f^\lambda(\mu) & = & \frac{1}{\lambda} ErlC\left( N^\lambda, \frac{\lambda}{\mu} \right) + \frac{N^\lambda}{\lambda} \\
g^\lambda(\mu) & = & \mu^2 c'(\mu) \left( \frac{N^\lambda}{\lambda} \right)^2 + \frac{1}{\mu}.
\end{eqnarray*}
The function $g^\lambda$ converges uniformly on compact sets to $g$ since for any $\overline{\mu} >0$, substituting for $N^\lambda$ in~(\ref{eq:staffingInd}) shows that
\begin{equation} \label{eq:glimit}
\sup_{\mu \in [0,\overline{\mu}]} \left| g^\lambda(\mu) - g(\mu) \right| \leq \overline{\mu}^2 c'(\overline{\mu}) \left( \frac{2}{a} \left| \frac{o(\lambda)}{\lambda} \right| + \left( \frac{o(\lambda)}{\lambda}\right)^2 \right) \rightarrow 0,
\end{equation}
as $\lambda \rightarrow \infty$. Next, \red{recall $ErlC(N,\rho) \leq 1$ whenever $\rho/N <1$.} Since
\begin{equation} \label{eq:flimit}
\left| f^\lambda(\mu) - f(\mu) \right| \red{\leq} \frac{1}{\lambda} ErlC\left( \frac{1}{a}\lambda + o(\lambda), \frac{\lambda}{\mu} \right) + \red{ \left| \frac{o(\lambda)}{\lambda} \right|}
\end{equation}
and $ErlC(\lambda/a + o(\lambda), \lambda/\mu) \leq 1$ for all $\mu >a$ for all large enough $\lambda$, the function $f^\lambda$ converges uniformly to $f$ on any compact set $[a+\epsilon, \overline{\mu}]$ with $\overline{\mu} > a+\epsilon$ and $\epsilon$ arbitrarily small.  \red{The reason we need only consider compact sets having lower bound $a + \epsilon$ is that it is straightforward to see any solution to~(\ref{eq:limiting_mu_Ind}) has $\mu > a$.} It is also helpful to note that $g^\lambda$ is convex in $\mu$ because
\[
\frac{d^2}{d\mu^2} g^\lambda(\mu) = \left( 2c'(\mu) + 4\mu c''(\mu) + \mu^2 c'''(\mu) \right) + 2\frac{1}{\mu^3} >0 \mbox{ for } \mu \in (0,\infty),
\]
and $f^\lambda$ is convex decreasing in $\mu$ because $ErlC(N,\rho)$ is convex increasing in $\rho$ (pages 8 and 11 of~\cite{WhittNotes}).

We prove (i) and then (ii).

\noindent{\bf Proof of (i):}
There exists $\mu_m \in (\mu_1,\mu_2)$ for which $f(\mu_m) > g(\mu_m)$.  Then, it follows from~(\ref{eq:glimit}) and~(\ref{eq:flimit}) that $f^\lambda(\mu_m) > g^\lambda(\mu_m)$ for all large enough $\lambda$.  Also,
\[
\lim_{\mu \rightarrow \infty} f^\lambda(\mu) = \frac{1}{a} < \lim_{\mu \rightarrow \infty} g^\lambda(\mu) = \infty.
\]
and
\[
\lim_{\mu \downarrow \lambda/ N^\lambda} f^\lambda(\mu) = \frac{1}{\lambda} + \frac{N^\lambda}{\lambda} < g^\lambda\left( \frac{\lambda}{N^\lambda} \right) = c'\left( \frac{\lambda}{N^\lambda}\right) + \frac{N^\lambda}{\lambda}
\]
for all large enough $\lambda$, where the inequality follows because $c$ is strictly increasing. Since $f^\lambda$ is convex decreasing and $g^\lambda$ is convex, we conclude that there exist two solutions to~(\ref{eq:FOCstaffingInd}).

\noindent{\bf Proof of (ii):}
We prove part (a) and then part (b).  Recall that $\mu_1$ is the only $\mu>0$ for which $f(\mu_1)=g(\mu_1)$.

\noindent{\bf Proof of (ii)(a):}
For part (a),  it is enough to show that for all large enough $\lambda$,
\begin{equation} \label{eq:fminusgpos}
f^\lambda(\mu_1) - g^\lambda(\mu_1) >0.
\end{equation}
The remainder of the argument follows as in the proof of part (i).

From the definition of $f^\lambda$ and $g^\lambda$ in the second paragraph of this proof, and substituting for $N^\lambda$,
\begin{small}
\begin{eqnarray*}
\lefteqn{f^\lambda(\mu_1) - g^\lambda(\mu_1)} \\ & = &
\frac{1}{a} - \frac{1}{\mu_1} - \left( \frac{\mu_1}{a} \right)^2 c'(\mu_1)
 + \frac{1}{\lambda} ErlC \left( \frac{1}{a} \lambda + o(\lambda), \frac{\lambda}{\mu_1} \right) + \frac{o(\lambda)}{\lambda} \left( 1-\frac{2}{a} \left( \mu_1\right)^2 c'(\mu_1) \right) - \left( \mu_1 \right)^2 c'(\mu_1) \left( \frac{o(\lambda)}{\lambda} \right)^2. \nonumber
\end{eqnarray*}
\end{small}
It follows from $f(\mu_1) = g(\mu_1)$ that $1/a - 1/\mu_1 - (\mu_1/a)^2 c'(\mu_1) = 0$, and so, also noting that $ErlC(\lambda/a + o(\lambda), \lambda/\mu_1) >0$,
\begin{equation} \label{eq:lb}
f^\lambda(\mu_1) - g^\lambda(\mu_1) > \frac{o(\lambda)}{\lambda}  \left( 1-\frac{2}{a} \left( \mu_1\right)^2 c'(\mu_1) \right) - \left( \mu_1 \right)^2 c'(\mu_1) \left( \frac{o(\lambda)}{\lambda} \right)^2.
\end{equation}
Again using the fact that $f(\mu_1) = g(\mu_1)$,
\[
1-\frac{2}{a}\left( \mu_1 \right)^2 c'(\mu_1) = 1-2a\left( \frac{1}{a} - \frac{1}{\mu_1} \right) = -1 + 2 \frac{a}{\mu_1}.
\]
Then, the term multiplying $o(\lambda)/\lambda$ in~(\ref{eq:lb}) is positive if
\begin{equation} \label{eq:amuproperties}
 -1 + 2 \frac{a}{\mu_1} >0,
\end{equation}
which implies~(\ref{eq:fminusgpos}) holds for all large enough $\lambda$.

To see~(\ref{eq:amuproperties}), and so complete the proof of part (ii)(a), note that since $\mu_1$ solves~(\ref{eq:limiting_mu_Ind}), \red{ and the left-hand side of~(\ref{eq:limiting_mu_Ind}) is convex increasing while the right-hand side is concave increasing, $\mu_1$ also solves the result of differentiating~(\ref{eq:limiting_mu_Ind}), which is }
\[
\frac{1}{\mu_1^2} = \frac{1}{a^2} \left( \mu_1^2 c''(\mu_1) + 2 \mu_1 c'(\mu_1) \right).
\]
Algebra shows that
\[
\frac{1}{\mu_1} - 2\left( \frac{\mu_1^2}{a^2}c'(\mu_1) \right) = \frac{\mu_1^3}{a^2} c''(\mu_1).
\]
We next use~(\ref{eq:limiting_mu_Ind}) to substitute for $\frac{\mu_1^2}{a^2}c'(\mu_1)$ to find
\[
\frac{3}{\mu_1} - \frac{2}{a} = \frac{\mu_1^3}{a^2} c''(\mu_1).
\]
Since $c$ is convex,
\[
\frac{3}{\mu_1} - \frac{2}{a} \geq 0,
\]
and so $1.5a \geq \mu_1$, from which~(\ref{eq:amuproperties}) follows.

\noindent{\bf Proof of (ii)(b):}
Let $\mu^{\lambda} \red{\in (0,\infty)} $ be the minimizer of the function $g^\lambda$.  \red{The minimizer exists because $g^\lambda$ is convex and
  \[
  \frac{d}{d\mu}g^\lambda(\mu) = \left( \mu c'(\mu) + \mu^2 c''(\mu) \right) \left( \frac{N^\lambda}{\lambda} \right)^2 - \frac{1}{\mu^2},
  \]
  which is negative for all small enough $\mu$, and positive for all large enough $\mu$.
  }
  It is sufficient to show that for all large enough $\lambda$
\begin{equation} \label{eq:iibtoshow}
g^\lambda(\mu) - f^\lambda(\mu) >0 \mbox{ for all } \red{\mu \in [a, \mu^{\lambda}]}.
\end{equation}
This is because for all $\mu> \mu^{\lambda}$, $g^\lambda$ is increasing and $f^\lambda$ is decreasing.

Suppose we can establish that for all large enough $\lambda$
\begin{equation} \label{eq:iibneeded}
\frac{1}{\mu} \geq \frac{2}{3a} - \frac{\epsilon^\lambda}{2a}, \mbox{ for all } \red{\mu \in [a,\mu^\lambda]},
\end{equation}
where $\epsilon^\lambda$ satisfies $\epsilon^\lambda \rightarrow 0$ as $\lambda \rightarrow \infty$.  Since $g(\mu) \geq f(\mu)$ for all $\mu$, it follows that
\[
g^\lambda(\mu) = \mu^2 c'(\mu) \left( \frac{N^\lambda}{\lambda} \right)^2 + \frac{1}{\mu} \geq \left( a - \frac{a^2}{\mu} \right) \left( \frac{N^\lambda}{\lambda} \right)^2 +\frac{1}{\mu}.
\]
Substituting for $N^\lambda$ and algebra shows that
\[
\left( a - \frac{a^2}{\mu} \right) \left( \frac{N^\lambda}{\lambda} \right)^2 +\frac{1}{\mu} = \frac{1}{a} + \red{ \left( \frac{o(\lambda)}{\lambda} \right) 2\left( 1-\frac{a}{\mu} \right)} + a \left( 1-\frac{a}{\mu} \right) \left( \frac{o(\lambda)}{\lambda} \right)^2.
\]
Then, from the definition of $f^\lambda$ and the above lower bound on $g^\lambda$, also  using the fact that \red{the assumption $N^\lambda - \lambda/a <0$ implies the term $o(\lambda)$ is negative},
\begin{eqnarray*}
g^\lambda(\mu) - f^\lambda(\mu) & \geq & \red{ \left| \frac{o(\lambda)}{\lambda} \right|} \left( \frac{2a}{\mu} -1 \right) - \frac{1}{\lambda} ErlC\left(N^\lambda,\frac{\lambda}{\mu} \right) + a \left( 1-\frac{a}{\mu} \right) \left( \frac{o(\lambda)}{\lambda} \right)^2.
\end{eqnarray*}
Since $-ErlC(N^\lambda, \lambda/\mu) > -1$ and $1/a-1/\mu >0$ from~(\ref{eq:limiting_mu_Ind}) implies $1-a/\mu >0$,
\[
g^\lambda(\mu) - f^\lambda(\mu) \geq \frac{1}{\lambda} \left( \red{ \lvert o(\lambda) \rvert} \left( \frac{2a}{\mu} -1 \right) -1 \right).
\]
Next, from~(\ref{eq:iibneeded}),
\[
\frac{2a}{\mu} \geq \frac{4}{3} - \epsilon^\lambda,
\]
and so
\[
g^\lambda(\mu) - f^\lambda(\mu) \geq \frac{1}{\lambda} \left( \red{ \lvert o(\lambda) \rvert} \left( \frac{1}{3}-\epsilon^\lambda\right)-1 \right).
\]
The fact that $\lvert o(\lambda) \rvert >3$ and $\epsilon^\lambda \rightarrow 0$ then implies that for all large enough $\lambda$,~(\ref{eq:iibtoshow}) is satisfied.

Finally, to complete the proof, we show that~(\ref{eq:iibneeded}) holds. First note that $\mu^\lambda$ as the minimizer of $g^\lambda$ satisfies
\[
\left( 2 \mu c'(\mu) + \mu^2 c''(\mu) \right) \left( \frac{N^\lambda}{\lambda} \right)^2 - \frac{1}{\mu^2} = 0,
\]
and that solution is unique and continuous in $\lambda$. Hence $\mu^\lambda \rightarrow \mu_1$ as $\lambda \rightarrow \infty$.  Then,
\[
g^\lambda(\mu^\lambda) \rightarrow g(\mu_1) = \frac{1}{a} \mbox{ as } \lambda \rightarrow \infty.
\]
Furthermore, $g^\lambda(\mu^\lambda)$ approaches $g(\mu_1)$ from above; i.e.,
\[
g^\lambda(\mu^\lambda) \downarrow \frac{1}{a} \mbox{ as } \lambda \rightarrow \infty,
\]
because, \red{recalling that the term $o(\lambda)$ is negative,}
\[
g^\lambda(\mu) = \mu^2 c'(\mu) \left( \frac{1}{a} - \frac{o(\lambda)}{\lambda} \right)^2 + \frac{1}{\mu} > g(\mu)  \mbox{ for all } \mu >0.
\]
Therefore, there exists $\epsilon^\lambda \rightarrow 0$ such that
\[
g^\lambda(\mu^\lambda) = \frac{1}{a} - \frac{3}{4}\frac{\epsilon^\lambda}{a},
\]
where the $3/(4a)$ multiplier of $\epsilon^\lambda$ is chosen for convenience when obtaining the bound in the previous paragraph.  Finally,
\[
\frac{1}{\mu} \geq \frac{1}{\mu^\lambda}
\]
means that~(\ref{eq:iibneeded}) follows if
\[
\frac{1}{\mu^{\lambda}} \geq \frac{2}{3} g^\lambda(\mu^\lambda) = \frac{2}{3} \frac{1}{a} - \frac{1}{2} \frac{\epsilon^\lambda}{a}.
\]
To see the above display is valid, note that $\mu^\lambda$ solves
\[
\left( g^\lambda(\mu) \right)' = 0,
\]
which from algebra is equivalent to
\[
2 g^\lambda(\red{\mu^\lambda}) - \frac{3}{\red{\mu^\lambda}} + \red{\left(\mu^\lambda \right)^3} c''(\red{\mu^\lambda}) \left( \frac{N^\lambda}{\lambda} \right)^2 = 0.
\]
Hence
\[
2g^\lambda(\red{\mu^\lambda}) - \frac{3}{\red{\mu^\lambda}} \leq 0,
\]
as required.
\eProof

\subsubsection*{Proof of Lemma~\ref{lemma:FOCisEquilibrium}.}
\red{It is enough to show the inequality~(\ref{eq:conditionalsymeq}) of Theorem~\ref{thm:conditionalsymeq} holds. The function $c$ is convex by assumption. It follows that
\begin{equation} \label{eq:tangent-inequality}
 c(\mu^\lambda) - c\left( \frac{\lambda}{N^\lambda}\right) \leq \left( \mu^\lambda - \frac{\lambda}{N^\lambda} \right) c'(\mu^\lambda).
\end{equation}
Plugging in for $c'(\mu^\lambda)$ from the symmetric first order condition~(\ref{eq:FOCstaffingInd}) yields (after algebra)
\[
\left( \mu^\lambda - \frac{\lambda}{N^\lambda} \right) c'(\mu^\lambda) = \frac{\lambda}{\mu^\lambda N^\lambda} \left( 1 - \frac{\lambda}{\mu^\lambda N^\lambda} \right)\left( 1 - \frac{\lambda}{\mu^\lambda N^\lambda} + \frac{ErlC\left( N^\lambda, \lambda / \mu^\lambda \right)}{N^\lambda} \right).
\]
Hence, in order to show the inequality~(\ref{eq:conditionalsymeq}) is true, also substituting for $\rho^\lambda = \lambda/\mu^\lambda$, it is enough to verify that
\begin{footnotesize}
\begin{align*}
\frac{\lambda}{\mu^\lambda N^\lambda} \left( 1 - \frac{\lambda}{\mu^\lambda N^\lambda} \right)\left( 1- \frac{\lambda}{\mu^\lambda N^\lambda} + \frac{ErlC\left( N^\lambda, \lambda / \mu^\lambda \right)}{N^\lambda} \right) &\leq \left(1-\frac{\lambda}{\mu^\lambda N^\lambda}\right)\left(1+\left(1-\frac{\lambda}{\mu^\lambda N^\lambda}+\frac{ErlC(N^\lambda,\lambda/ \mu^\lambda)}{N-1}\right)^{-1}\right)^{-1}\\
\Longleftrightarrow\qquad\qquad\qquad\;\;\; 1 + \frac{1}{1-\frac{\lambda}{\mu^\lambda N^\lambda} + \frac{ErlC(N^\lambda, \lambda/ \mu^\lambda)}{N^\lambda - 1}} &\leq \frac{1}{ \left( \frac{\lambda}{\mu^\lambda N^\lambda} \right) \left(1-\frac{\lambda}{\mu^\lambda N^\lambda} + \frac{ErlC(N^\lambda, \lambda/ \mu^\lambda)}{N^\lambda} \right)}.
\end{align*}
\end{footnotesize}
Since $N^\lambda -1 < N^\lambda$, it is enough to show that
\begin{align*}
1 + \frac{1}{1-\frac{\lambda}{\mu^\lambda N^\lambda} + \frac{ErlC(N^\lambda, \lambda/ \mu^\lambda)}{N^\lambda} } &\leq \frac{1}{ \left( \frac{\lambda}{\mu^\lambda N^\lambda} \right) \left(1-\frac{\lambda}{\mu^\lambda N^\lambda} + \frac{ErlC(N^\lambda, \lambda/ \mu^\lambda)}{N^\lambda} \right)}\\
\Longleftrightarrow\qquad\qquad\qquad\qquad\qquad\qquad\qquad\qquad\qquad\quad\;\;\; 1 &\leq \frac{\left( 1 - \frac{\lambda}{\mu^\lambda N^\lambda}\right)}{\frac{\lambda}{\mu^\lambda N^\lambda}\left( 1- \frac{\lambda}{\mu^\lambda N^\lambda} + \frac{ErlC(N^\lambda, \lambda/ \mu^\lambda)}{N^\lambda} \right)}\\
\Longleftrightarrow\qquad \frac{\lambda}{\mu^\lambda N^\lambda} \left( 1 - \frac{\lambda}{\mu^\lambda N^\lambda} \right) + \frac{\lambda}{\mu^\lambda N^\lambda} \frac{ErlC(N^\lambda, \lambda/ \mu^\lambda)}{N^\lambda} &\leq \left( 1 - \frac{\lambda}{\mu^\lambda N^\lambda} \right)\\
\Longleftrightarrow\qquad\qquad\qquad\qquad\qquad\qquad\qquad\;\; \frac{ErlC(N^\lambda, \lambda/ \mu^\lambda)}{ \lambda/ \mu^\lambda} &\leq \left( \frac{N^\lambda \mu^\lambda}{\lambda} - 1 \right)^2.
\end{align*}
Since $N^\lambda \mu^\lambda / \lambda \rightarrow d >1$ by assumption, the limit of the right-hand side of the above expression is positive, and, since and $ErlC(N^\lambda, \lambda/ \mu^\lambda) \leq 1$, the limit of the left-hand side of the above expression is 0.  We conclude that for all large enough $\lambda$, the above inequality is valid.}
\eProof

\subsubsection*{Proof of Proposition~\ref{proposition:staffing_independent}.}
Let
\[
\underline{\mu}^{\star} = \argmin\{ \mu>0:~(\ref{eq:limiting_mu_Ind}) \mbox{ holds } \}.
\]
Next, recalling that $\underline{\mu}^{\star} >a$, also let
\[
    \underline{\mu} = \underline{\mu}^{\star} - \frac{1}{2} \left( \underline{\mu}^{\star} - a\right) >a,
\]
so that the system is stable if all servers were to work at rate $\underline{\mu}$ ($\lambda < \underline{\mu} N^\lambda$ for all large enough $\lambda$).  It follows from Theorem~\ref{lemma:FOCexistence_staffingInd} that, for all large enough $\lambda$, any $\mu^\lambda$ that satisfies the first order condition~(\ref{eq:FOCstaffingInd}) also satisfies $\mu^\lambda > \underline{\mu}$.  Hence any symmetric equilibrium $\mu^{\star,\lambda}$ must also satisfy $\mu^{\star,\lambda} > \underline{\mu}$ for all large enough $\lambda$, and so
\[
\overline{W}^{\star,\lambda} < \overline{W}_{\underline{\mu}}^\lambda.
\]
Therefore, also using the fact that $\overline{W}^{\star,\lambda}>0$, it follows that
\[
c_S \frac{N^\lambda}{\lambda} < \frac{C^{\star,\lambda}(N^\lambda)}{\lambda} = c_S \frac{N^\lambda}{\lambda} + \overline{w} \overline{W}^{\star,\lambda} < c_S \frac{N^\lambda}{\lambda} + \overline{w} \overline{W}_{\underline{\mu}}^\lambda.
\]
Then, since $N^\lambda / \lambda \rightarrow 1/a$ as $\lambda \rightarrow \infty$ from~(\ref{eq:staffingInd}), it is sufficient to show
\[
\overline{W}_{\underline{\mu}}^\lambda \rightarrow 0 \mbox{ as } \lambda \rightarrow \infty.
\]
This follows from substituting the staffing $N^\lambda = \lambda/a + o(\lambda)$ in~(\ref{eq:staffingInd}) into the well-known formula for the steady state mean waiting time in a $M$/$M$/$N^\lambda$ queue with arrival rate $\lambda$ and service rate $\underline{\mu}$ as follows
\begin{eqnarray*}
\overline{W}_{\underline{\mu}}^\lambda & = & \frac{1}{\lambda} \frac{\lambda/\underline{\mu}}{N^\lambda - \lambda/\underline{\mu}} ErlC\left( N^\lambda, \frac{\lambda}{\underline{\mu}} \right) \\
& = & \frac{1/\underline{\mu}}{\left(1/a-1/\underline{\mu} \right)\lambda + o(\lambda)} ErlC\left( N^\lambda, \frac{\lambda}{\underline{\mu}} \right) \\
& \rightarrow & 0, \mbox{ as } \lambda \rightarrow \infty,
\end{eqnarray*}
since $ErlC(N^\lambda, \lambda/\underline{\mu}) \in [0,1]$ for all $\lambda$.
\eProof

\subsubsection*{Proof of Lemma~\ref{lemma:Afinite}.}
It follows from the equation
\[
a(\mu - a) = \mu^3 c'(\mu)
\]
that
\[
a = \frac{\mu}{2} \left( 1 \stackrel{+}{-} \sqrt{1-4\mu c'(\mu)} \right).
\]
The condition $4\mu c'(\mu) \leq 1$ is required to ensure that there is a real-valued solution for $a$. Hence
\[
\mathcal{A} = \left\{ \frac{\mu}{2} \left( 1 \stackrel{+}{-} \sqrt{1-4\mu c'(\mu)} \right): 0 \leq 4 \mu c'(\mu) \leq 1 \right\}.
\]
Since $c'(\mu)$ is well-behaved, this implies that $\mathcal{A}$ is compact, and, in particular, closed.  We conclude that \change{$a^\star = \sup \mathcal{A} \in \mathcal{A}$}, which implies that $a^\star$ is finite.
\eProof

\subsubsection*{Proof of Theorem~\ref{theorem:staffing_independent}.}
It follows from Proposition~\ref{proposition:nonaoInd} that
\[
0 \leq \liminf_{\lambda \rightarrow \infty} \frac{N^{{opt},\lambda}}{\lambda} \leq \limsup_{\lambda \rightarrow \infty} \frac{N^{{opt},\lambda}}{\lambda} < \infty,
\]
because any staffing policy that is not asymptotically optimal also is not optimal for each $\lambda$.  Consider any subsequence $\lambda'$ on which either $\liminf_{\lambda \rightarrow \infty} N^{{opt},\lambda}/\lambda$ or $\limsup_{\lambda \rightarrow \infty} N^{{opt},\lambda}/\lambda$ is attained, and suppose that
\begin{equation} \label{eq:supposed_limit}
\frac{N^{{opt},\lambda'}}{\lambda'} \rightarrow \frac{1}{a} \mbox{ as } \lambda' \rightarrow \infty, \mbox{ where } a \in [0,\infty).
\end{equation}

The definition of asymptotic optimality requires that for each $\lambda'$, there exists a symmetric equilibrium service rate $\mu^{\star,\lambda'}$.  As in the proof of Lemma~\ref{proposition:nonaoInd}, it is enough to consider sequences $\{\mu^\lambda\}$ that satisfy the first order condition~(\ref{eq:FOCstaffingInd}).  Then, by the last sentence of Theorem~\ref{lemma:FOCexistence_staffingInd}, any sequence of solutions $\{\mu^{\lambda'}\}$ to~(\ref{eq:FOCstaffingInd}) must be such that $|\mu^{\lambda'}- \mu|$ is arbitrarily small, for $\lambda'$ large enough, for some $\mu$ that solves~(\ref{eq:limiting_mu_Ind}), given $a$ in~(\ref{eq:supposed_limit}).  In summary, the choice of $a$ in~(\ref{eq:supposed_limit}) is constrained by the requirement that a symmetric equilibrium service rate must exist.

Given that there exists at least one symmetric equilibrium service rate for all large enough $\lambda'$, it follows in a manner very similar to the proof of Proposition~\ref{proposition:staffing_independent} that
\[
\overline{W}^{\star,\lambda'} \rightarrow 0 \mbox{ as } \lambda' \rightarrow \infty,
\]
even though when there are multiple equilibria we may not be able to guarantee which symmetric equilibrium $\mu^{\star,\lambda'}$ the servers choose for each $\lambda'$.  We conclude that
\begin{equation} \label{eq:wrong_limit}
\frac{C^{\star,\lambda'}(N^{{opt},\lambda'})}{\lambda'} = c_S \frac{N^{{opt},\lambda'}}{\lambda'} + \overline{w} \overline{W}^{\star,\lambda'} \rightarrow c_S \frac{1}{a}, \mbox{ as } \lambda' \rightarrow \infty.
\end{equation}

We argue by contradiction that $a$ in~(\ref{eq:wrong_limit}) must equal $a^\star$.  Suppose not.  Then, since
\[
\frac{C^{\star,\lambda}(N^{{ao},\lambda})}{\lambda} \rightarrow c_S \frac{1}{a^\star} \mbox{ as } \lambda \rightarrow \infty
\]
by Proposition~\ref{proposition:staffing_independent} (and so the above limit is true on any subsequence), and $a^\star > a$ by its definition, it follows that
\[
C^{\star,\lambda'}(N^{{ao},\lambda'}) < C^{\star,\lambda'}(N^{{opt},\lambda'}) \mbox{ for all large enough } \lambda'.
\]
The above inequality contradicts the definition of $N^{{opt},\lambda'}$.

The previous argument did not depend on if $\lambda'$ was the subsequence on which $\liminf_{\lambda \rightarrow \infty} N^{{opt},\lambda}/\lambda$ or $\limsup_{\lambda \rightarrow \infty} N^{{opt},\lambda}/\lambda$ was attained.  Hence
\[
\lim_{\lambda \rightarrow \infty} \frac{N^{{opt},\lambda}}{\lambda} = \frac{1}{a^\star},
\]
and, furthermore,
\[
\lim_{\lambda \rightarrow \infty} \frac{C^{\star,\lambda}(N^{{opt},\lambda})}{\lambda} = c_S\frac{1}{a^\star}.
\]
Since also
\[
\lim_{\lambda \rightarrow \infty} \frac{C^{\star^,\lambda}(N^{{ao},\lambda})}{\lambda} = c_S\frac{1}{a^\star},
\]
the proof is complete.
\eProof

\subsubsection*{Proof of Lemma~\ref{lemma:ex_polynomial_cost}.}
We first observe that~(\ref{eq:limiting_mu_Ind}) is equivalently written as:
\[
0=c_E p\mu^{p+2} - a \mu + a^2.
\]
The function
\[
f(\mu) = c_E p \mu^{p+2} - a \mu + a^2
\]
attains its minimum value in $(0,\infty)$ at
\[
\underline{\mu} = \left( \frac{a}{c_E p (p+2)} \right)^{1/(p+1)}.
\]
The function $f$ is convex in $(0,\infty)$ because $f''(\mu)>0$ for all $\mu \in (0,\infty)$ and so $\underline{\mu}$ is the unique minimum.  It follows that
\[
\mbox{ if } f(\underline{\mu}) \left\{ \begin{array}{l} < \\ > \\ = \end{array} \right\} 0, \mbox{ then } \left\{ \begin{array}{l} \mbox{there are 2 non-negative solutions to~(\ref{eq:limiting_mu_Ind})} \\ \mbox{there is no non-negative solution to~(\ref{eq:limiting_mu_Ind})} \\ \mbox{there is exactly one solution to~(\ref{eq:limiting_mu_Ind})} \end{array}\right..
\]
Since
\[
f(\underline{\mu}) = a^{\frac{p+2}{p+1}} \left( a^{2-\frac{p+2}{p+1}} - \triangle \right)
\]
for
\[
\triangle := \left( \frac{1}{c_E p(p+1)} \right)^{\frac{1}{p+1}} \left( 1 - \left( \frac{1}{c_Ep} \right)^{p+1} \left( \frac{1}{p+2} \right)^{p+2} \right) >0,
\]
it follows that
\[
\mbox{ if } a^{\frac{p}{p+1}} - \triangle \left\{ \begin{array}{l} < \\ > \\ = \end{array} \right\} 0, \mbox{ then } \left\{ \begin{array}{l} \mbox{there are 2 non-negative solutions to~(\ref{eq:limiting_mu_Ind})} \\ \mbox{there is no non-negative solution to~(\ref{eq:limiting_mu_Ind})} \\ \mbox{there is exactly one solution to~(\ref{eq:limiting_mu_Ind})} \end{array}\right..
\]
The expression for $\triangle$ can be simplified so that
\[
\triangle = \frac{(p+1)}{(p+2)} \left( \frac{1}{c_E p (p+2)} \right)^{\frac{1}{p+1}}.
\]
Then, $a^\star$ follows by noting that $a^\star = \triangle^{(p+1)/p}$ and $\mu^\star$ follows by noting that $\mu^\star = \underline{\mu}$ and then substituting for $a^\star$.

To complete the proof, we must show that $a^\star$ and $\mu^\star$ are both increasing in $p$.  This is because we have already observed that any solution to~(\ref{eq:limiting_mu_Ind}) has $a < \mu$, and the fact that $\mu <1$ follows directly from the expression for $\underline{\mu}$.  We first show $a^\star$ is increasing in $p$, and then argue that this implies $\mu^\star$ is increasing in $p$.

To see that $a^\star$ is increasing in $p$, we take the derivative of $\log a^\star(p)$ and show that this is positive.  Since
\begin{eqnarray*}
\log a^\star(p) & = & \log (p+1) - \log (p+2) + \frac{1}{p} \log(p+1) \\
& & - \frac{1}{p} \log c_E - \frac{1}{p} \log p - \frac{2}{p} \log (2+p),
\end{eqnarray*}
it follows that
\begin{eqnarray*}
\left(\log a^\star(p) \right)' & = & \frac{1}{p+1} - \frac{1}{p+2} + \left( \frac{p/(p+1) - \log (p+1)}{p^2} \right) + \frac{1}{p^2} \log c_E  \\
& & - \frac{\frac{p}{p} - \log(p)}{p^2} - 2 \left( \frac{\frac{p}{p+2} - \log(p+2)}{p^2} \right).
\end{eqnarray*}
After much simplification, we have
\begin{eqnarray*}
\left(\log a^\star(p) \right)'  & = & \frac{1}{p^2} \log c_E + \frac{1}{p^2} \left( \log \left( \frac{p(p+2)^2}{p+1} \right) - \frac{p^2+p+4}{(p+1)(p+2)} \right).
\end{eqnarray*}
Hence it is enough to show that
\[
\triangle(p) =  \log \left( \frac{p(p+2)^2}{p+1} \right) - \frac{p^2+p+4}{(p+1)(p+2)} \geq 0, \mbox{ for } p \geq 1.
\]
This follows because the first term is increasing in $p$, and has a value that exceeds 1 when $p=1$; on the other hand, the second term has a value that is strictly below 1 for all $p\geq 1$.

Finally, it remains to argue that $\mu^\star$ is increasing in $p$.  At the value $\mu = \mu^\star$
\[
g(\mu) = \mu^3 c'(\mu) - a \mu + a^2 = 0.
\]
At the unique point where the minimum is attained, it is also true that
\[
g'(\mu) = \mu^3c''(\mu) + 3 \mu^2 c'(\mu) - a = 0.
\]
Since $\mu^3c''(\mu) + 3 \mu^2 c'(\mu)$ is an increasing function of $\mu$, it follows that if $a$ increases, then $\mu$ must increase.
\eProof

\subsection*{PROOFS FROM SECTION~\ref{Section:Scheduling}}

\subsubsection*{Proof of Theorem~\ref{theorem:2-server-idlerandom}.}

It is sufficient to verify the detailed balance equations.  For reference, it is helpful to refer to Figure~\ref{fig:idle-mc}, which depicts the relevant portion of the Markov chain.  We require the following additional notation.
For all $\mathcal{I}\subseteq\{1,2,\ldots,N\}$, all states $\boldsymbol s=(s_1,s_2,\ldots,s_{|\mathcal I|})$, all servers $s'\in\{1,2,\ldots,N\}\backslash\mathcal{I}$, and integers $j\in\{1,2,\ldots,|\mathcal{I}|+1\}$, we define the state $\boldsymbol s[s',j]$ by \[\boldsymbol s[s',j]\equiv(s_1,s_2,\ldots,s_{j-1},s',s_j,\ldots,s_{|\mathcal I|}).\]

\begin{figure}[b]
\begin{center}
\begin{tikzpicture}
\node[draw, circle, thick, name=s1, minimum size=.55in] at (0,0) {{\small $\boldsymbol{s}[s',1]$}};
\node[draw, circle, thick, name=t1, minimum size=.55in] at (10,0) {{\small $\boldsymbol{s}- s_1$}};
\node[draw, circle, thick, name=s2, minimum size = .55in] at (0,-3) {{\small $\boldsymbol{s}[s',2]$}};
\node[draw, circle, thick, name=t2, minimum size=.55in] at (10,-3) {{\small $\boldsymbol{s}- s_2$}};
\node[draw, circle, thick, name=si, minimum size = .55in] at (0,-6) {{\small $\boldsymbol{s}+s'$}};
\node[draw, circle, thick, name=ti, minimum size=.55in] at (10,-6) {{\small $\boldsymbol{s}- s_{|\mathcal I|}$}};
\node[draw, circle, thick, name=s, minimum size=.55in] at (5,-3) {{\small $\boldsymbol s$}};
\node at (0,-4.5) [yshift=2] {{\Large $\boldsymbol{\vdots}$}};
\node at (10,-4.5) [yshift=2] {{\Large $\boldsymbol{\vdots}$}};
\draw[thick,->] (s1) -- (s) node[midway,yshift=10] {\rotatebox{-31}{{\small $\lambda p^{\mathcal I\cup\{s'\}}(1)$}}};
\draw[thick,->] (s) -- (t1) node[midway,yshift=10] {\rotatebox{31}{{\small $\lambda p^{\mathcal I}(1)$}}};
\draw[thick,->] (s2) -- (s) node[midway,above] {{\small $\lambda p^{\mathcal I\cup\{s'\}}(2)$}};
\draw[thick,->] (s) -- (t2) node[midway,above] {{\small $\lambda p^{\mathcal I}(2)$}};
\draw[thick,->] (si) -- (s) node[midway,yshift=10] {\rotatebox{31}{{\small $\lambda p^{\mathcal I\cup\{s'\}}(|\mathcal I|+1)$}}};
\draw[thick,->] (s) -- (ti) node[midway,yshift=10] {\rotatebox{-31}{{\small $\lambda p^{\mathcal I}(|\mathcal I|)$}}};
\draw[thick,->] (s) to[out=-95,in=0] (si);
\draw[thick,->] (ti) to[out=180,in=-85] (s);
\node[below] at (2.5,-6) {{\small $\mu_{s'}$}};
\node[below] at (7.5,-6) {{\small $\mu_{s_{|\mathcal I|}}$}};
\draw[decorate, decoration={brace,amplitude=10pt}]  (1.1,-7) -- (-1.1,-7) node[midway,yshift=-20] {{\small For each $s'\not\in\mathcal I$}};
\end{tikzpicture}
\caption{Snippet of the Markov chain showing the rates into and out of state $\boldsymbol{s}=(s_1,\ldots,s_{|\mathcal I|})$.  For convenience, we use $\boldsymbol{s}-{s_j}$ to denote the state $(s_1,s_2,\ldots,s_{j-1},s_{j+1},\ldots,s_{|\mathcal I|})$ and $\boldsymbol{s}+s'$ to denote the state $\boldsymbol s[s',|\mathcal I|+1]=(s_1,s_2,\ldots,s_{|\mathcal I|},s')$. \label{fig:idle-mc}}
\end{center}
\end{figure}

We first observe that:
\begin{align*}
\mbox{Rate into state }\boldsymbol{s}\mbox{ due to an arrival}&=\lambda\sum_{s'\not\in\mathcal I}\sum_{j=1}^{|\mathcal I|+1}\pi_{\boldsymbol s[s',j]}p^{\mathcal I\cup\{s'\}}(j)\\
&=\lambda\sum_{s'\not\in\mathcal I}\sum_{j=0}^{|\mathcal I|}\frac{\mu_{s'}\pi_B}{\lambda}\prod_{s\in\mathcal I}\left(\frac{\mu_s}{\lambda}\right)p^{\mathcal I\cup\{s'\}}(j)\\
&=\sum_{s'\not\in\mathcal I}\mu_{s'}\pi_B\prod_{s\in I}\frac{\mu_s}{\lambda} = \sum_{s'\not\in\mathcal I}\mu_{s'}\pi_{\boldsymbol s}\\
&=\mbox{Rate out of state }\boldsymbol{s}\mbox{ due to a departure}.
\end{align*}
Then, to complete the proof, we next observe that for each $s' \not\in \mathcal{I}$:
\begin{align*}
\mbox{Rate into state }\boldsymbol{s}\mbox{ due to a departure}&=\mu_{s_{|\mathcal I|}}\pi_{(s_1,s_2,\ldots,s_{|\mathcal I|-1})}\\
&=\mu_{s_{|\mathcal I|}}\pi_B\prod_{s\in\mathcal I\backslash\{s_{|\mathcal I|}\}}\frac{\mu_s}{\lambda}\\
&=\lambda\pi_B\prod_{s\in\mathcal I}\frac{\mu_s}{\lambda} = \lambda\pi_{\boldsymbol s}\\
&=\mbox{Rate out of state }\boldsymbol{s}\mbox{ due to an arrival}.
\end{align*}
\eProof

\subsubsection*{Proof of Proposition~\ref{proposition:idletime}.}

In order to derive the steady state probability that a server is idle, we first solve for the steady state probabilities of the $M$/$M$/$2$ system (with arrival rate $\lambda$ and service rates $\mu_1$ and $\mu_2$ respectively) under an arbitrary probabilistic routing policy where a job that arrives to find an empty system is routed to server 1 with probability $p$ and server 2 with probability $1-p$. Then, for an $r$-routing policy, we simply substitute $p=\frac{\mu_1^r}{\mu_1^r+\mu_2^r}$.

It should be noted that this analysis (and more) for $2$ servers has been carried out by~\cite{Krishnamoorthi1963}. Prior to that,~\cite{Saaty1960} carried out a partial analysis (by analyzing an $r$-routing policy with $r=1$). However, we rederive the expressions using our notation for clarity.

The dynamics of this system can be represented by a continuous time Markov chain shown in Figure~\ref{fig:twoservermc} whose state space is simply given by the number of jobs in the system, except when there is just a single job in the system, in which case the state variable also includes information about which of the two servers is serving that job. This system is stable when $\mu_1+\mu_2>\lambda$ and we denote the steady state probabilities as follows:
\begin{itemize}
\item $\pi_0$ is the steady state probability that the system is empty.
\item $\pi_1^{(j)}$ is the steady state probability that there is one job in the system, served by server $j$.
\item For all $k\ge2$, $\pi_k$ is the steady state probability that there are $k$ jobs in the system.
\end{itemize}
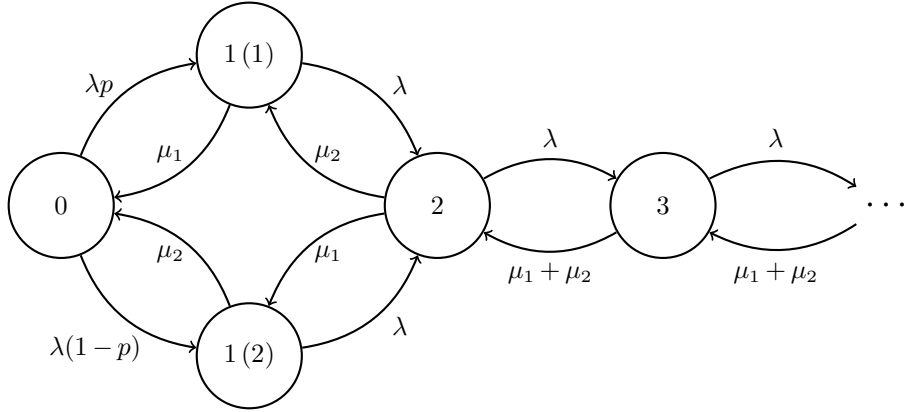
\begin{figure}[t]
\begin{center}
\begin{tikzpicture}
\node[draw, circle, thick, name=s0, minimum size=.55in] at (0,0) {{\small $0$}};
\node[draw, circle, thick, name=s11, minimum size=.55in] at (2.5,2) {{\small $1\,(1)$}};
\node[draw, circle, thick, name=s12, minimum size=.55in] at (2.5,-2) {{\small $1\,(2)$}};
\node[draw, circle, thick, name=s2, minimum size=.55in] at (5,0) {{\small $2$}};
\node[draw, circle, thick, name=s3, minimum size=.55in] at (8,0) {{\small $3$}};
\node[name=s4] at (11,0) {{\large $\cdots$}};
\draw[thick,->] (s0) to[bend left] (s11);
\draw[thick,->] (s11) to[bend left] (s0);
\draw[thick,->] (s0) to[bend right] (s12);
\draw[thick,->] (s12) to[bend right] (s0);
\draw[thick,->] (s11) to[bend left] (s2);
\draw[thick,->] (s2) to[bend left] (s11);
\draw[thick,->] (s12) to[bend right] (s2);
\draw[thick,->] (s2) to[bend right] (s12);
\draw[thick,->] (s2) to[bend left] (s3);
\draw[thick,->] (s3) to[bend left] (s2);
\draw[thick,->] (s3) to[bend left] (s4);
\draw[thick,->] (s4) to[bend left] (s3);
\node at (.5,1.6) {{\small $\lambda p$}};
\node at (.45,-1.9) {{\small $\lambda (1-p)$}};
\node at (4.5,1.6) {{\small $\lambda$}};
\node at (4.5,-1.6) {{\small $\lambda$}};
\node at (6.5,0.9) {{\small $\lambda$}};
\node at (6.5,-0.9) {{\small $\mu_1+\mu_2$}};
\node at (9.5,0.9) {{\small $\lambda$}};
\node at (9.5,-0.9) {{\small $\mu_1+\mu_2$}};
\node at (1.45,.68) {{\small $\mu_1$}};
\node at (1.45,-.65) {{\small $\mu_2$}};
\node at (3.55,.68) {{\small $\mu_2$}};
\node at (3.55,-.65) {{\small $\mu_1$}};
\end{tikzpicture}
\end{center}
\caption{The $M$/$M$/$2$ Markov chain with probabilistic routing}
\label{fig:twoservermc}
\end{figure}
We can write down the balance equations of the Markov chain as follows:
\begin{align*}
\lambda\pi_0&=\mu_1\pi_1^{(1)}+\mu_2\pi_1^{(2)}\\
(\lambda+\mu_1)\pi_1^{(1)}&=\lambda p\pi_0+\mu_2\pi_2\\
(\lambda+\mu_2)\pi_1^{(2)}&=\lambda (1-p)\pi_0+\mu_1\pi_2\\
(\lambda+\mu_1+\mu_2)\pi_2&=\lambda\pi_1^{(1)}+\lambda\pi_1^{(2)}+(\mu_1+\mu_2)\pi_3\\
\forall k\ge3\colon\quad(\lambda+\mu_1+\mu_2)\pi_k&=\lambda\pi_{k-1}+(\mu_1+\mu_2)\pi_{k+1},
\end{align*}
yielding the following solution to the steady state probabilities:
\begin{align}
\pi_0&=\frac{\mu_1\mu_2(\mu_1+\mu_2-\lambda)(\mu_1+\mu_2+2\lambda)}{\mu_1\mu_2(\mu_1+\mu_2)^2+\lambda(\mu_1+\mu_2)(\mu_2^2+2\mu_1\mu_2+(1-p)(\mu_1^2-\mu_2^2))+\lambda^2(\mu_1^2+\mu_2^2)}\label{eq:pi0}\\
\pi_1^{(1)}&=\frac{\lambda(\lambda+p(\mu_1+\mu_2))\pi_0}{\mu_1(\mu_1+\mu_2+2\lambda)}\nonumber\\
\pi_1^{(2)}&=\frac{\lambda(\lambda+(1-p)(\mu_1+\mu_2))\pi_0}{\mu_2(\mu_1+\mu_2+2\lambda)}\nonumber.
\end{align}
Consequently, the steady state probability that server $1$ is idle is given by
\begin{equation*}
I_1(\mu_1,\mu_2;p)=\pi_0+\pi_1^{(2)}=\left(1+\frac{\lambda(\lambda+(1-p)(\mu_1+\mu_2))}{\mu_2(\mu_1+\mu_2+2\lambda)}\right)\pi_0.
\end{equation*}
Substituting for $\pi_0$, we obtain
\begin{equation}\label{eq:M/M/2-p-idletime1}
I_1(\mu_1,\mu_2;p)=\frac{\mu_1(\mu_1+\mu_2-\lambda)\left[(\lambda+\mu_2)^2+\mu_1\mu_2+(1-p)\lambda(\mu_1+\mu_2)\right]}{\mu_1\mu_2(\mu_1+\mu_2)^2+\lambda(\mu_1+\mu_2)\left[\mu_2^2+2\mu_1\mu_2+(1-p)(\mu_1^2-\mu_2^2)\right]+\lambda^2(\mu_1^2+\mu_2^2)}.
\end{equation}
Finally, for an $r$-routing policy, we let $p=\frac{\mu_1^r}{\mu_1^r+\mu_2^r}$ to obtain:
\begin{align*}
I_1^r(\mu_1,\mu_2)&=I_1(\mu_1,\mu_2;p=\frac{\mu_1^r}{\mu_1^r+\mu_2^r})\\
&=\frac{\mu_1(\mu_1+\mu_2-\lambda)\left[(\lambda+\mu_2)^2+\mu_1\mu_2+\frac{\mu_2^r}{\mu_1^r+\mu_2^r}\lambda(\mu_1+\mu_2)\right]}{\mu_1\mu_2(\mu_1+\mu_2)^2+\lambda(\mu_1+\mu_2)\left[\mu_2^2+2\mu_1\mu_2+\frac{\mu_2^r}{\mu_1^r+\mu_2^r}(\mu_1^2-\mu_2^2)\right]+\lambda^2(\mu_1^2+\mu_2^2)}.
\end{align*}
By symmetry of the $r$-routing policy, it can be verified that $I^r_2(\mu_1,\mu_2)=I^r_1(\mu_2,\mu_1)$, completing the proof.
\eProof

\subsubsection*{Proof of Theorem~\ref{theorem:FSF-noeq}.}

We first highlight that when all servers operate at the same rate $\mu\in\left(\frac{\lambda}{N},\infty\right)$, both FSF and SSF are equivalent to Random routing. Henceforth, we refer to such a configuration as a symmetric operating point $\mu$. In order to prove that there does not exist a symmetric equilibrium under either FSF or SSF, we show that at any symmetric operating point $\mu$, any one server can attain a strictly higher utility by unilaterally setting her service rate to be slightly lower (in the case of FSF) or slightly higher (in the case of SSF) than $\mu$.

We borrow some notation from the proof of Proposition~\ref{proposition:idletime} where we derived the expressions for the steady state probability that a server is idle when there are only $2$ servers under any probabilistic policy, parameterized by a number $p\in[0,1]$ which denotes the probability that a job arriving to an empty system is routed to server $1$. Recall that $I_1(\mu_1,\mu_2;p)$ denotes the steady state probability that server $1$ is idle under such a probabilistic policy, and the corresponding utility function for server $1$ is $U_1(\mu_1,\mu_2;p)=I_1(\mu_1,\mu_2;p)-c(\mu_1)$. Then, by definition, the utility function for server $1$ under FSF is given by:
\begin{equation*}
U^{FSF}_1(\mu_1,\mu_2) =
\begin{cases}
U_1(\mu_1,\mu_2;p=0) &,\ \mu_1 < \mu_2\\
U_1\left(\mu_1,\mu_2;p=\frac{1}{2}\right) &,\ \mu_1 = \mu_2\\
U_1(\mu_1,\mu_2;p=1) &,\ \mu_1 > \mu_2.
\end{cases}
\end{equation*}
Similarly, under SSF, we have:
\begin{equation*}
U^{SSF}_1(\mu_1,\mu_2) =
\begin{cases}
U_1(\mu_1,\mu_2;p=1) &,\ \mu_1 < \mu_2\\
U_1\left(\mu_1,\mu_2;p=\frac{1}{2}\right) &,\ \mu_1 = \mu_2\\
U_1(\mu_1,\mu_2;p=0) &,\ \mu_1 > \mu_2.
\end{cases}
\end{equation*}
Note that while the utility function under any probabilistic routing policy is continuous everywhere, the utility function under FSF or SSF is discontinuous at symmetric operating points. This discontinuity turns out to be the crucial tool in the proof. Let the two servers be operating at a symmetric operating point $\mu$. Then, it is sufficient to show that there exists $0<\delta<\mu-\frac{\lambda}{2}$ such that
\begin{equation}\label{eq:FSF-better-response}
U_1^{FSF}(\mu-\delta,\mu)-U_1^{FSF}(\mu,\mu)>0,
\end{equation}
and
\begin{equation}\label{eq:SSF-better-response}
U_1^{SSF}(\mu+\delta,\mu)-U_1^{FSF}(\mu,\mu)>0.
\end{equation}
We show~(\ref{eq:FSF-better-response}), and~(\ref{eq:SSF-better-response}) follows from a similar argument. Note that
\begin{align*}
U_1^{FSF}(\mu-\delta,\mu)-U_1^{FSF}(\mu,\mu) = U_1(\mu-\delta,&\mu;p=0)-U_1\left(\mu,\mu;p=\frac{1}{2}\right)\\
= \big(U_1(\mu-\delta,&\mu;p=0)-U_1(\mu,\mu;p=0)\big)\\
&+ \left(U_1(\mu,\mu;p=0)-U_1\left(\mu,\mu;p=\frac{1}{2}\right)\right)
\end{align*}
Since the first difference, $U_1(\mu-\delta,\mu;p=0)-U_1(\mu,\mu;p=0)$, is zero when $\delta=0$, and is continuous in $\delta$, it is sufficient to show that the second difference, $U_1(\mu,\mu;p=0)-U_1(\mu,\mu;p=\frac{1}{2})$, is strictly positive:
\begin{align*}
U_1(\mu,\mu;p=0)-U_1\left(\mu,\mu;p=\frac{1}{2}\right)&=I_1(\mu,\mu;p=0)-I_1\left(\mu,\mu;p=\frac{1}{2}\right)\\
&=\frac{\lambda(2\mu-\lambda)}{(\mu+\lambda)(2\mu+\lambda)}>0\qquad\qquad\qquad\qquad\big(\text{using }(\ref{eq:M/M/2-p-idletime1})\big).
\end{align*}
This completes the proof.
\eProof

\subsubsection*{Proof of Theorem~\ref{theorem:unique-symeq}.}

The proof of this theorem consists of two parts. First, we show that under any $r$-routing policy, any symmetric equilibrium $\mu^\star\in(\frac{\lambda}{2},\infty)$ must satisfy the equation $\varphi(\mu^\star)=r$. This is a direct consequence of the necessary first order condition for the utility function of server $1$ to attain an interior maximum at $\mu^\star$. The second part of the proof involves using the condition $c'(\frac{\lambda}{2})<\frac{1}{\lambda}$ to show that $\varphi$ is a \textit{strictly decreasing bijection} onto $\mathbb R$, which would lead to the following implications:
\begin{itemize}
\item $\varphi$ is invertible; therefore, if an $r$-routing policy admits a symmetric equilibrium, it is unique, and is given by $\mu^\star=\varphi^{-1}(r)$.
\item $\varphi^{-1}(r)$ is strictly decreasing in $r$; therefore, so is the unique symmetric equilibrium (if it exists). Since the mean response time $\mathbb{E}[T]$ is inversely related to the service rate, this establishes that $\mathbb{E}[T]$ at symmetric equilibrium (across $r$-routing policies that admit one) is increasing in $r$.
\end{itemize}

We begin with the first order condition for an interior maximum. The utility function of server $1$ under an $r$-routing policy, from~(\ref{Equation:ServerUtility}), is given by
\begin{equation*}
U^r_1(\mu_1,\mu_2) = I^r_1(\mu_1,\mu_2) - c(\mu_1)
\end{equation*}
For $\mu^\star\in(\lambda/2,\infty)$ to be a symmetric equilibrium, the function $U^r_1(\mu_1,\mu^\star)$ must attain a global maximum at $\mu_1=\mu^\star$. The corresponding first order condition is then given by:
\begin{equation}\label{eq:symfoc-2-server}
\left.\frac{\partial I^r_1}{\partial \mu_1}(\mu_1,\mu^\star)\right|_{\mu_1=\mu^\star} = c'(\mu^\star),
\end{equation}
where $I^r_1$ is given by Proposition~\ref{proposition:idletime}. The partial derivative of the idle time can be computed and the left hand side of the above equation evaluates to
\begin{equation}\label{eq:symmetricfirstderivative}
\left.\frac{\partial I^r_1}{\partial \mu_1}(\mu_1,\mu^\star)\right|_{\mu_1=\mu^\star} = \frac{\lambda(4\lambda+4\mu^\star+\lambda r-2\mu^\star r)}{4\mu^\star(\lambda+\mu^\star)(\lambda+2\mu^\star)}.
\end{equation}
Substituting in~(\ref{eq:symfoc-2-server}) and rearranging the terms, we obtain:
\begin{equation*}
\frac{4(\lambda+\mu^\star)}{\lambda(\lambda-2\mu^\star)}\left(\mu^\star(\lambda+2\mu^\star)c'(\mu^\star)-\lambda\right)=r.
\end{equation*}
The left hand side is equal to $\varphi(\mu^\star)$, thus yielding the necessary condition $\varphi(\mu^\star)=r$.

Next, we proceed to show that if $c'(\frac{\lambda}{2})<\frac{1}{\lambda}$, then $\varphi$ is a strictly decreasing bijection onto $\mathbb R$. Note that the function
\begin{equation*}
\varphi(\mu)=\frac{4(\lambda+\mu)}{\lambda(\lambda-2\mu)}\left(\mu(\lambda+2\mu)c'(\mu)-\lambda\right)
\end{equation*}
is clearly a continuous function in $(\frac{\lambda}{2},\infty)$. In addition, it is a surjection onto $\mathbb R$, as evidenced by the facts that $\varphi(\mu)\to-\infty$ as $\mu\to\infty$ and $\varphi(\mu)\to\infty$ as $\mu\to\frac{\lambda}{2}+$ (using $c'(\frac{\lambda}{2})<\frac{1}{\lambda}$).

To complete the proof, it is sufficient to show that $\varphi'(\mu)<0$ for all $\mu\in(\frac{\lambda}{2},\infty)$.  First, observe that
\begin{equation*}
\varphi'(\mu) = \frac{4\psi(\mu)}{\lambda(\lambda-2\mu)^2},
\end{equation*}
where
\begin{equation*}
\psi(\mu) = \mu(\lambda+\mu)(\lambda^2-4\mu^2)c''(\mu)+(\lambda^3+6\lambda^2\mu-8\mu^3)c'(\mu)-3\lambda^2.
\end{equation*}
Since $c'(\frac{\lambda}{2})<\frac{1}{\lambda}$, as $\mu\to\frac{\lambda}{2}+$, $\psi(\mu)<0$.  Moreover, since $c'''(\mu)>0$, for all $\mu>\frac{\lambda}{2}$, we have
\begin{small}
\begin{equation*}
\psi'(\mu) = -4\mu(\lambda+\mu)\left(\mu^2-\left(\frac{\lambda}{2}\right)^2\right)c'''(\mu)-4\left(\mu-\frac{\lambda}{2}\right)(\lambda^2+6\lambda\mu+6\mu^2)c''(\mu)-24\left(\mu^2-\left(\frac{\lambda}{2}\right)^2\right)c'(\mu)<0.
\end{equation*}
\end{small}
It follows that $\psi(\mu)<0$ for all $\mu>\frac{\lambda}{2}$.  Since $\varphi'(\mu)$ has the same sign as $\psi(\mu)$, we conclude that $\varphi'(\mu)<0$, as desired.
\eProof

\subsubsection*{Proof of Theorem~\ref{theorem:boundedinterval}.}

From Theorem~\ref{theorem:unique-symeq}, we know that if a symmetric equilibrium exists, then it is unique, and is given by $\mu^\star=\varphi^{-1}(r)$, where $\varphi$ establishes a one-to-one correspondence between $r$ and $\mu^\star$ ($\mu^\star$ is strictly decreasing in $r$ and vice versa). Therefore, it is enough to show that there exists a finite upper bound $\overline{\mu}>\frac{\lambda}{2}$ such that no service rate $\mu>\overline{\mu}$ can be a symmetric equilibrium under \textit{any} $r$-routing policy. It would then automatically follow that for $\underline{r}=\varphi(\overline{\mu})$, no $r$-routing policy with $r\leq\underline{r}$ admits a symmetric equilibrium. We prove this by exhibiting a $\overline{\mu}$ and showing that if $\mu\geq\overline{\mu}$, then the utility function of server $1$, $U_1^r(\mu_1,\mu)$, cannot attain a global maximum at $\mu_1=\mu$ for any $r\in\mathbb{R}$.

We begin by establishing a lower bound for the maximum utility $U_1^r(\mu_1,\mu)$ that server $1$ can obtain under any $r$-routing policy:
\begin{equation}\label{eq:lowerboundonmax}
\max_{\mu_1>\frac{\lambda}{2}}U_1^r(\mu_1,\mu)\geq U_1^r\left(\frac{\lambda}{2},\mu\right)=I_1^r\left(\frac{\lambda}{2},\mu\right)-c\left(\frac{\lambda}{2}\right)\geq-c\left(\frac{\lambda}{2}\right)=U_1^r\left(\frac{\lambda}{2},\frac{\lambda}{2}\right).
\end{equation}
By definition, if $\mu^\star$ is a symmetric equilibrium under any $r$-routing policy, then the utility function of server $1$, $U_1^r(\mu_1,\mu^\star)$, is maximized at $\mu_1=\mu^\star$, and hence, using~(\ref{eq:lowerboundonmax}), we have
\begin{equation}\label{eq:necessarycondition}
U_1^r(\mu^\star,\mu^\star)\geq U_1^r(\frac{\lambda}{2},\frac{\lambda}{2}).
\end{equation}
Next, we establish some properties on $U_1^r(\mu,\mu)$ that help us translate this necessary condition for a symmetric equilibrium into an upper bound on any symmetric equilibrium service rate. We have,
\begin{equation*}
U_1^r(\mu,\mu) = 1-\frac{\lambda}{2\mu}-c(\mu),
\end{equation*}
which has the following properties:
\begin{itemize}
\item Since $c'(\frac{\lambda}{2})<\frac{1}{\lambda}$, $U_1^r(\mu,\mu)$, as a function of $\mu$, is strictly increasing at $\mu=\frac{\lambda}{2}$.
\item $U_1^r(\mu,\mu)$ is a concave function of $\mu$.
\end{itemize}
This means that $U_1^r(\mu,\mu)$ is strictly increasing at $\mu=\frac{\lambda}{2}$, attains a maximum at the unique $\mu_\dagger>\frac{\lambda}{2}$ that solves the first order condition $\mu_\dagger^2c'(\mu_\dagger)=\frac{\lambda}{2}$, and then decreases forever. This shape of the curve $U_1^r(\mu,\mu)$ implies that there must exist a unique $\overline{\mu}>\mu_\dagger$, such that $U_1^r(\overline{\mu},\overline{\mu})=U_1^r(\frac{\lambda}{2},\frac{\lambda}{2})$.

Since $U_1^r(\mu,\mu)$ is a strictly decreasing function for $\mu>\mu_\dagger$, it follows that if $\mu^\star>\overline{\mu}$, then, $U_1^r(\mu^\star,\mu^\star)< U_1^r(\overline{\mu},\overline{\mu})=U_1^r(\frac{\lambda}{2},\frac{\lambda}{2})$, contradicting the necessary condition~(\ref{eq:necessarycondition}). This establishes the required upper bound $\overline{\mu}$ on any symmetric equilibrium service rate, completing the proof.
\eProof

\subsubsection*{Proof of Theorem~\ref{theorem:guarantee-symeq}.}

A useful tool for proving this theorem is Theorem 3 from~\cite{Cheng2004}, whose statement we have adapted to our model:
\begin{theorem}\label{thm:symmetricNEresult}
A symmetric game with a nonempty, convex, and compact strategy space, and utility functions that are continuous and quasiconcave has a symmetric (pure-strategy) equilibrium.
\end{theorem}
We begin by verifying that our $2$-server game meets the qualifying conditions of Theorem~\ref{thm:symmetricNEresult}:
\begin{itemize}
\item \textit{Symmetry:} First, all servers have the same strategy space of service rates, namely, $(\frac{\lambda}{2},\infty)$. Moreover, since an $r$-routing policy is symmetric and all servers have the same cost function, their utility functions are symmetric as well. Hence, our $2$-server game is indeed symmetric.
\item \textit{Strategy space:} The strategy space $(\frac{\lambda}{2},\infty)$ is nonempty and convex, but not compact, as required by Theorem~\ref{thm:symmetricNEresult}. Hence, for the time being, we modify the strategy space to be $[\frac{\lambda}{2},\overline{\mu}+1]$ so that it is compact, where $\overline{\mu}$ is the upper bound on any symmetric equilibrium, established in Theorem~\ref{theorem:boundedinterval}, and deal with the implications of this modification later.
\item \textit{Utility function:} $U_1^r(\mu_1,\mu_2)$ is clearly continuous. From Mathematica, it can be verified that the idle time function $I_1^r(\mu_1,\mu_2)$ is concave in $\mu_1$ for $r\in\{-2,-1,0,1\}$, and since the cost function is convex, this means the utility functions are also concave. (Unfortunately, we could not get Mathematica to verify concavity for non-integral values of $r$, though we strongly suspect that it is so for the entire interval $[-2,1]$.)
\end{itemize}
Therefore, we can apply Theorem~\ref{thm:symmetricNEresult} to infer that an $r$-routing policy with $r\in\{-2,-1,0,1\}$ admits a symmetric equilibrium in $[\frac{\lambda}{2},\overline{\mu}+1]$. We now show that the boundaries cannot be symmetric equilibria. We already know from Theorem~\ref{theorem:boundedinterval} that $\overline{\mu}+1$ cannot be a symmetric equilibrium. (We could have chosen to close the interval at any $\mu>\overline{\mu}$. The choice $\overline{\mu}+1$ was arbitrary.) To see that $\frac{\lambda}{2}$ cannot be a symmetric equilibrium, observe that $c'(\frac{\lambda}{2})<\frac{1}{\lambda}$ implies that $U^r_1(\mu_1,\frac{\lambda}{2})$ is increasing at $\mu_1=\frac{\lambda}{2}$ (using the derivative of the idle time computed in~(\ref{eq:symmetricfirstderivative})), and hence server $1$ would have an incentive to deviate. Therefore, any symmetric equilibrium must be an interior point, and from Theorem~\ref{theorem:unique-symeq}, such an equilibrium must be unique. This completes the proof.
\eProof

%
%
%

\end{document}